\documentclass[aps, prd, superscriptaddress, nofootinbib, eqsecnum, preprintnumbers, showpacs]{revtex4}
\usepackage{amsmath}
\usepackage{graphicx}

\newcommand{\lesssim}{\mathrel{\mathpalette\vereq<}}

\newcommand{\chushi}[1]{}

\begin{document}

\title{%
  \hfill{\normalsize\vbox{%
  }}\\
  \vspace{-0.5cm}
  {\bf   Holographic QCD Integrated back to Hidden Local Symmetry
%~\footnote{
%Short version was given at \cite{Harada:2010iv}}
} } 
\author{Masayasu Harada}\thanks{
      {\tt harada@hken.phys.nagoya-u.ac.jp}}
       \affiliation{ Department of Physics, Nagoya University,
                    Nagoya, 464-8602, Japan.}
\author{Shinya Matsuzaki}\thanks{
      {\tt synya@pusan.ac.kr}}
      \affiliation{ Department of Physics,  
                    Pusan National University, Busan 609-735, Korea.}
\author{{Koichi Yamawaki}} \thanks{
      {\tt yamawaki@kmi.nagoya-u.ac.jp}}
      \affiliation{ Kobayashi-Maskawa Institute for the Origin of Particles and 
the Universe (KMI) \\ 
 Nagoya University, Nagoya 464-8602, Japan.}
\date{\today}

\begin{abstract}

We develop a previously proposed gauge-invariant method to integrate out 
infinite tower of Kaluza-Klein (KK) modes of 
vector and axialvector mesons 
in a class of models of holographic QCD (HQCD). 
The HQCD is reduced by our method to the chiral perturbation theory
with the hidden local symmetry (HLS) having only the lowest KK mode 
identified as the HLS gauge boson. 
We take the Sakai-Sugimoto model as a concrete HQCD, 
and completely determine the ${\cal O} (p^4)$ terms as well as the ${\cal O}(p^2)$ 
 terms from the DBI part 
and the anomaly-related (intrinsic parity odd) gauge-invariant terms  
 from the CS part. 
Effects of higher KK modes are fully included in these terms.  
To demonstrate power of our method, 
we compute momentum-dependences of several form factors such as 
the pion electromagnetic form factors, 
the $\pi^0$-$\gamma$ and $\omega$-$\pi^0$ transition form factors compared with experiment,
which was not achieved before due to complication to handle 
infinite sums.  
We also study other anomaly-related quantities like $\gamma^*$-$\pi^0$-$\pi^+$-$\pi^-$ and 
$\omega$-$\pi^0$-$\pi^+$-$\pi^-$ vertex functions. 
\end{abstract}

\maketitle

\section{Introduction}

Holography, based on 
gauge/gravity duality~\cite{Maldacena:1997re,AdS/CFT}, 
has been of late fashion to 
reveal a part of features in
strongly coupled gauge theories. 
Application to QCD, which is called holographic QCD (HQCD),  
is useful to check the validity of the holographic
correspondence.  
In some 
models~\cite{Sakai:2004cn,Erlich:2005qh,Da Rold:2005zs,Sakai:2005yt}
 which realize the chiral symmetry breaking of QCD, 
it has been shown in the large $N_c$ limit that 
some observables of low-energy QCD are consistent with the experiment. 
There are two types of holographic approaches: One is called 
``top-down" approach 
starting with a stringy setting; 
the other is called ``bottom-up" approach beginning with a
five-dimensional gauge theory defined on an AdS (anti-de Sitter space) background. 
It is a key point to notice that, whichever approaches, 
one eventually employs a five-dimensional gauge model with a
characteristic induced-metric 
and some boundary conditions on a certain brane configuration.

Holographic recipe tells us that classical solutions for boundary values of 
bulk fields serve as sources coupled to currents in the dual four-dimensional 
QCD. Green functions in QCD like current correlators are thus evaluated straightforwardly 
 from the boundary action as a generating functional 
  in the large $N_c$ limit~\cite{Erlich:2005qh,Da Rold:2005zs}. 
Equivalently, one can show that those things are calculable from the five-dimensional action 
by performing Kaluza-Klein (KK) decomposition of the bulk gauge fields 
and identifying KK fields themselves as vector and axialvector fields of 
a low-energy effective model dual to QCD~\cite{Sakai:2004cn, Sakai:2005yt}. 
In this sense, one can say that, in the low-energy region, 
any model of HQCD is reduced to a certain effective hadron model in four dimensions. 
Such effective models include vector and axialvector mesons 
as an infinite tower of KK modes 
together with the Nambu-Goldstone bosons (NGBs) 
associated with the spontaneous  chiral symmetry breaking.  
Infinite tower of KK modes 
(vector and axialvector mesons) then contributes to Green functions such as current correlators and form factors. 
Here we follow the latter approach~\cite{Sakai:2004cn,Sakai:2005yt} dealing with the bulk action as a functional of the gauge fields.

  It was pointed out~\cite{Hill:2000mu,ArkaniHamed:2001ca,Sakai:2004cn,Sakai:2005yt,Son:2003et} 
that the infinite tower of KK modes is 
interpreted as a set of gauge bosons of hidden local symmetries (HLSs)~\cite{Bando:1984ej,Bando:1987br,Harada:2003jx}. 
 Note that since the KK modes as the {\it gauge bosons of HLSs are not necessarily mass eigenstates}, 
we should distinguish them (``HLS-KK modes", $V_\mu^{(n)}$ in Eq.(\ref{Vmun:trans})) from 
the conventional KK modes ($B_\mu^{(n)}$ in Eq.(\ref{B:field})). 
Hereafter we shall call the HLS-KK modes simply the KK modes. 
Solving away higher KK modes through the equations of motion 
derived from the five-dimensional effective action, 
which is equivalent to  
integrating out KK modes in terms of functional integral, 
we showed~\cite{Harada:2006di} that, in the low-energy region,  
any holographic model can be formulated in the HLS notion to 
be reduced to the HLS model having a finite set of HLS gauge bosons 
with the lowest one identified as the $\rho$ meson and its flavor partners.

Instead of dealing with the infinite tower of KK modes,  
we demonstrated~\cite{Harada:2006di} that 
effects from the higher KK modes are {\it fully} incorporated 
into coefficients of the   ${\cal O } (p^4)$ terms in the HLS field theory extended 
from the conventional chiral perturbation theory (ChPT)~\cite{Gas:84}, 
so-called the HLS-ChPT~\cite{Tanabashi:1993sr,Harada:2003jx}. 
(Similar method of integration out was also considered in Ref.\cite{Belitsky:2010fj}.) 
 Furthermore,  since it is a manifestly HLS-gauge invariant formulation,  
one can calculate any Green function order by order in the derivative expansion, or 
loop expansion, in which higher order corrections may be identified 
with the $1/N_c$-subleading effects which are not easily figured out in HQCD. 
In fact we calculated meson-loop corrections 
as $1/N_c$-subleading effects 
in terms of the HLS-ChPT 
for the Dirac-Born-Infeld (DBI) part in the Sakai-Sugimoto (SS) model~\cite{Sakai:2004cn,Sakai:2005yt}.

In this paper, our method will be developed in details including external 
gauge fields such as photon in a class of HQCD models including the SS model. 
By construction our method is manifestly invariant under the HLS and the chiral symmetry 
including the external gauge symmetry. It will be shown that on the contrary, 
a naive truncation simply neglecting higher KK modes 
 of the HLS gauge bosons violates the HLS and the chiral symmetry including the external gauge symmetry. 
We further extend our method to the Chern-Simons (CS) part.    
In the case of the SS model 
we present a full set of the ${\cal O}(p^4)$ terms of 
the HLS Lagrangian computed from the DBI part at the leading order of $1/N_c$ expansion, 
which was partially reported in the previous work~\cite{Harada:2006di}. 
In addition, 
the anomaly-related  (intrinsic-parity odd (IP-odd)) gauge-invariant terms 
introduced in Refs.~\cite{Fujiwara:1984mp,Bando:1987br,Harada:2003jx}  
are completely determined from the CS part. 
Once the ${\cal O}(p^4)$ terms are determined, 
calculation of meson-loop corrections 
of subleading order in $1/N_c$ expansion
in terms of the HLS-ChPT can be performed.

Throughout this paper, we will confine ourselves to the large $N_c$ limit,  
leaving calculations of $1/N_c$-subleading order in future works. 
Even in the large $N_c$ limit, our method is useful 
especially for 
studying momentum-dependences of several form factors, 
which was not achieved due to complication to deal with the infinite sum. 
Actually, given a concrete holographic model not restricted to the SS model, 
our method enables us to deduce definite 
predictions of the model for any physical quantity 
to be compared with experimental data.

Here we demonstrate power of our method in the case of the SS model. 
The form factors are calculable in the general framework of the HLS model 
with its parameters determined by the SS model for 
IP-odd processes as well as IP-even ones. 
The electromagnetic (EM) gauge invariance and the chiral invariance are automatically maintained since our method 
is manifestly invariant under the external gauge symmetry as well as the HLS.   
 As to IP-even processes,      
an explicit form of 
the pion EM form factor is given 
to be compared with the experimental data.   
As to IP-odd processes, 
we also give explicit forms of the $\pi^0$-$\gamma$ and $\omega$-$\pi^0$ transition form factors 
and the related quantities such as $\gamma^*$-$\pi^0$-$\pi^+$-$\pi^-$ and 
$\omega$-$\pi^0$-$\pi^+$-$\pi^-$ vertex functions.  
To show that our formulation correctly includes contributions from infinite set of 
higher KK modes, we further derive the same results by a different method dealing with the infinite sum explicitly without using the general HLS Lagrangian.  
 It reveals the fact that infinite sum is crucial for the gauge invariance. 
Actually, the EM gauge symmetry and chiral symmetry (low-energy theorem) 
are obviously violated by  
a naive truncation simply neglecting higher KK modes
instead of taking the infinite sum. 
Note that the importance of the higher KK modes is visible only in the HLS basis: 
The higher KK modes in the mass eigenstates basis (KK modes in the usual sense) do not contribute at all, 
since our method is equivalent to setting zero the higher mass eigenstate fields $B_\mu^{(n)} =0 \,(n\ne 1)$ 
via equation of motion of $B_\mu^{(n)}$ (see Eq.(\ref{sol})). This is  in accord with the fact that the SS model may not be reliable beyond 
the scale of $M_{KK} \sim$ 1 GeV.

This paper is organized as follows:

In Sec.~\ref{sec2} we start with a class of models of HQCD including
the SS model~\cite{Sakai:2004cn,Sakai:2005yt} 
and explain our formulation integrating out 
arbitrary parts of
infinite tower of vector and axialvector mesons in a manner manifestly 
invariant under the HLS and the external gauge symmetry. 
We demonstrate that low-energy effective models of HQCD  
can be formulated by the HLS with ${\cal O}(p^4)$ terms. 
In Sec.~\ref{sec3} 
we calculate the parameters of the HLS Lagrangian from the SS model
for the IP-even and IP-odd ${\cal O}(p^4)$ terms. 
In Sec.~\ref{pheno} 
we present several applications of our method including 
the pion EM form factor and IP-odd form factors such as 
$\pi^0$-$\gamma$ and $\omega$-$\pi^0$ transition form factors. 
Sec.~\ref{summary} is devoted to summary and discussion. 
Appendix A is a proof that $\Gamma_3$ defined in the text as a part of the CS action of the SS model is HLS-invariant 
(and thereby provides the HLS-invariant terms of the IP-odd part of the HLS Lagrangian). 
In Appendix B DBI and CS terms are expanded in terms of the HLS building blocks.  
Appendix C is to demonstrate  that, as done for the IP-even processes in the text,  
the same result as that of our integrating out method for the IP-odd form factors 
is obtained by an alternative method explicitly using sum rules of the infinite tower of the HLS-KK modes.

\section{A gauge-invariant way to integrate out HQCD} 
\label{sec2}

In this section,  we develop detailed formulation of our method~\cite{Harada:2006di}:
Starting with a class of  holographic QCD (HQCD) models including  
the Sakai-Sugimoto (SS) model~\cite{Sakai:2004cn,Sakai:2005yt}, 
we introduce a way to obtain a  low-energy effective 
model in four dimensions described only by the lightest vector meson identified as the
$\rho$ meson, based on the hidden local symmetry (HLS) together with
the Nambu-Goldstone  
bosons (NGBs).  
Although most of notations adopted here follow the SS model~\cite{Sakai:2004cn,Sakai:2005yt}, 
our methodology is applicable to other types of HQCD.

\subsection{Reducing 5d-models to 4d-models with infinite tower of vector and axialvector mesons}

Suppose that the fifth direction, spanned by the coordinate $z$,  
extends from  minus infinity to plus infinity ($-\infty < z < \infty$)~\footnote{ In an application to 
another type of  HQCD~\cite{Da Rold:2005zs}, the $z$ coordinate is 
  defined on a finite interval, which is different from the $z$ 
  coordinate used here. They are related by an appropriate
  coordinate transformation as done in 
  Refs.~\cite{Sakai:2004cn,Sakai:2005yt}. 
}. The parity is introduced by imposing a reflection symmetry under an interchange 
$z \leftrightarrow -z$ along the fifth direction. 
We employ a five-dimensional gauge theory 
which has a vectorial $U(N)$ gauge symmetry 
defined on a certain background associated with 
the gauge/gravity duality.

The five-dimensional gauge field, $A_M(x^\mu, z)$ with $M=(\mu, z)$,   
transforms  inhomogeneously under the $U(N)$ gauge symmetry as 
\begin{equation}
A_M(x^{\mu},z)  \to 
  g(x^{\mu},z)  A_{M}(x^{\mu},z)  g^{\dagger}(x^{\mu},z)  
   - i \partial_{M} g(x^{\mu},z)  g^{\dagger}(x^{\mu},z) 
\,, \label{AM:trans}
\end{equation} 
where $g(x^\mu, z)$ is the transformation matrix of the gauge symmetry. 
As far as gauge-invariant sector 
such as the Dirac-Born-Infeld part of the SS model~\cite{Sakai:2004cn,Sakai:2005yt}  
is concerned, the five-dimensional action in the large $N_c$ limit  can be 
written as~\footnote{  Models of HQCD having the left- and right-bulk fields 
such as $F_L, F_R$~\cite{Da Rold:2005zs,Erlich:2005qh} 
can be described by the same action as in Eq.(\ref{5d-action}) with a suitable $z$-coordinate transformation 
prescribed. } 
\begin{eqnarray}
S_5 = N_c   
        \int d^4x dz  
         \Bigg( 
               - \frac{1}{2} K_1(z) {\rm tr}[ 
                                                   F_{\mu\nu} F^{\mu\nu} 
                                                 ]
               +      K_2(z) M_{\rm KK}^2  {\rm tr }[ 
                                                   F_{\mu z} F^{\mu z} 
                                                  ] 
         \Bigg)
\,, \label{5d-action}
\end{eqnarray} 
where 
$K_{1,2}(z)$ denote a set of metric-functions of $z$ 
constrained by the gauge/gravity duality.  
$M_{\rm KK}$ is a typical mass scale of 
the Kaluza-Klein (KK) modes of the gauge field $A_M$.

We choose the same boundary condition of the five-dimensional gauge field $A_{M}$ as done in 
Refs.~\cite{Sakai:2004cn,Sakai:2005yt}: 
\begin{equation} 
A_{M}(x^{\mu},z = \pm \infty) = 0 
\,.
\label{AM:BC}
\end{equation}
A transformation which does not change this boundary condition 
satisfies $\partial_M g(x^\mu, z)|_{z=\pm \infty}=0$. 
This implies an emergence of global chiral $U(N)_L \times U(N)_R$ symmetry in four dimensions 
characterized by the transformation matrices $g_{R, L}=g(z=\pm \infty)$.  
With the boundary condition (\ref{AM:BC}) imposed, 
the zero mode of $A_z$ is identified with the NGB 
associated with the spontaneous breaking of the chiral
symmetry. 
The chiral field 
\begin{equation}
U(x^{\mu})
   = 
   {\rm P} \exp \left[ i \int_{-\infty}^{\infty} dz' A_z(x^{\mu},z') \right] 
\, 
\end{equation}
is parameterized by the NGB field $\pi$ as 
\begin{equation} 
 U(x^{\mu}) = e^{\frac{2i \pi(x^{\mu})}{F_{\pi}}} 
\, , 
\end{equation} 
where $F_\pi$ denotes the decay constant of $\pi$. 
$U$ is divided as 
\begin{equation} 
 U(x^{\mu}) = \xi_L^{\dagger}(x^{\mu}) \cdot \xi_R(x^{\mu}) 
\,,
\end{equation} 
such that
$\xi_{R,L}$ transform as 
\begin{equation} 
\xi_{R,L} \rightarrow h(x^{\mu}) \cdot \xi_{R,L} \cdot g_{R,L}^{\dagger} 
\, , 
\end{equation} 
with $h(x^\mu)$ being the transformation of 
the hidden local symmetry (HLS)~\cite{Bando:1984ej,Bando:1987br,Harada:2003jx}. 
Here we note that we can introduce an infinite number of HLSs
by dividing $U$ into a product of an infinite number of $\xi$ fields~\cite{Bando:1984ej,Bando:1987br}.

Chiral $U(N_f)_L \times U(N_f)_R$ symmetry can be gauged by the external fields
${\mathcal L}_{\mu}$ and ${\mathcal R}_\mu$ including the photon field
through the boundary condition~\cite{Sakai:2005yt}
\begin{eqnarray} 
A_{\mu}(x^{\mu},z = +\infty) &=& 
{\mathcal R}_{\mu}(x^\mu) 
= {\cal V}_\mu (x^\mu) + {\cal A}_\mu (x^\mu) 
\, \nonumber \\ 
A_{\mu}(x^{\mu},z = -\infty) &=& 
{\mathcal L}_{\mu} (x^\mu)
= {\cal V}_\mu(x^\mu) - {\cal A}_\mu(x^\mu)
\,,
\label{Amu:BC2}
\end{eqnarray}
instead of Eq.~(\ref{AM:BC}).

Following Refs.~\cite{Sakai:2004cn,Sakai:2005yt,Harada:2006di}, 
we work in $A_z=0$ 
gauge. 
There still exists a four-dimensional gauge symmetry 
under which $A_\mu(x^\mu,z)$ transforms as 
\begin{equation} 
A_\mu(x^\mu,z) 
\to h(x^\mu) \cdot A_\mu(x^\mu, z) \cdot h^\dagger(x^\mu) 
- i \partial_\mu h(x^\mu) \cdot h^\dagger(x^\mu) 
\,. \label{Amu:trans:Az0}
\end{equation}
This gauge symmetry is identified~\cite{Sakai:2004cn,Sakai:2005yt,Harada:2006di} 
with the above HLS. 
In this gauge the NGB fields reside in the boundary condition for 
the five-dimensional gauge field $A_{\mu}$ as 
\begin{equation} 
  A_\mu (x^\mu, z=+ \infty) 
  = \alpha^{R}_\mu(x^\mu)  
  \,, \qquad 
  A_\mu (x^\mu, z=- \infty) 
  = \alpha^{L}_\mu(x^\mu)  
  \,, 
\end{equation}
where 
\begin{eqnarray}
\alpha^{R}_{\mu} (x^{\mu}) &=& 
i \xi_{R} (x^{\mu}) 
{\mathcal D}_\mu
\xi^{\dagger}_{R}(x^{\mu})
=
i \xi_{R} (x^{\mu}) 
\left( \partial_\mu - i {\mathcal R}_\mu \right)
\xi^{\dagger}_{R}(x^{\mu})
\, , 
\nonumber\\
\alpha^{L}_{\mu} (x^{\mu}) &=& i \xi_{L} (x^{\mu}) 
{\mathcal D}_\mu
\xi^{\dagger}_{L}(x^{\mu})
=
i \xi_{L} (x^{\mu}) 
\left( \partial_\mu - i {\mathcal L}_\mu \right)
\xi^{\dagger}_{L}(x^{\mu})
\, , 
\label{alphaRL:trans}
\end{eqnarray} 
which transform under the HLS in the same way as in Eq.~(\ref{Amu:trans:Az0}). 
Note that,
in this gauge, we explicitize  a single HLS among an infinite number of HLSs 
while the chiral symmetry is ``hidden".

We introduce an infinite tower of the massive KK modes of 
the vector $(V_\mu^{(n)}(x^\mu))$
and the axialvector $(A_\mu^{(n)}(x^\mu))$ meson fields. 
The vector meson fields $V_\mu^{(n)}(x^\mu)$ transform
as the HLS gauge boson: 
\begin{equation} 
   V_\mu^{(n)}(x^\mu) 
\to h(x^\mu) \cdot   V_\mu^{(n)}(x^\mu)  \cdot h^\dagger(x^\mu) 
- i \partial_\mu h(x^\mu) \cdot h^\dagger(x^\mu) 
\,, \label{Vmun:trans}
\end{equation} 
while the axialvector meson fields $A_\mu^{(n)}(x^\mu)$ transform as the matter fields: 
\begin{equation} 
A_\mu^{(n)}(x^\mu) \to h(x^\mu) \cdot 
A_\mu^{(n)}(x^\mu) \cdot  h^\dag (x^\mu)
\,. \label{Amun:trans}
\end{equation}  
It should be noted that 
the vector meson fields $V_\mu^{(n)}(x^\mu)$ are different from the mass-eigenstate fields $B_\mu^{(n)}(x^\mu)$ in Refs.~\cite{Sakai:2004cn,Sakai:2005yt} 
which transform as matter fields, 
\begin{equation} 
B_\mu^{(n)}(x^\mu) \to h(x^\mu) \cdot  B_\mu^{(n)}(x^\mu) \cdot h^\dag(x^\mu) 
\,. \label{B:field}
\end{equation}  
The five-dimensional gauge field $A_\mu(x^\mu, z)$  is now 
expanded as~\footnote{In Eq.(\ref{Amuxz}) 
we put a relative minus sign in front of 
the HLS gauge fields $V_\mu^{(n)}(x^\mu)$ 
for a convention. }
\begin{equation} 
  A_\mu(x^\mu, z) 
  = \alpha_\mu^R(x^\mu) \phi^R(z) 
  +
  \alpha_\mu^L(x^\mu) \phi^L(z) 
 + 
\sum_{n=1}^\infty \left( 
A_\mu^{(n)}(x^\mu) \psi_{2n}(z)
-    
V_\mu^{(n)}(x^\mu) \psi_{2n-1}(z) 
\right)
  \,. \label{Amuxz} 
\end{equation} 
The functions $\{ \psi_{2n-1}(z) \}$ and $\{ \psi_{2n}(z) \}$ are the eigenfunctions~\footnote{
The eigenfunction for $n=2k$ ($k=1,2,\ldots$) 
is an odd function of $z$, while that 
for $n=(2k-1)$ is an even function. } satisfying the eigenvalue 
equation obtained from the action (\ref{5d-action}): 
\begin{equation} 
-  K_1^{-1}(z) \partial_z (K_2(z) \partial_z \psi_n(z)) 
= \lambda_n \psi_n(z) 
\quad\quad  (n=0,1,2,\ldots)
\,, \label{generalKKeq}
\end{equation} 
where $\lambda_n$ denotes the $n$th eigenvalue. 
On the other hand, the gauge invariance requires 
the functions $\phi^{R,L}(z)$ to be different from the eigenfunctions: 
{} From the transformation properties in Eqs.(\ref{Amu:trans:Az0}) and (\ref{alphaRL:trans})-(\ref{Amun:trans}),  
we see that the functions, $\phi^{R,L}(z)$, $\{ \psi_{2n-1}(z) \}$ and $\{ \psi_{2n}(z) \}$ 
 are constrained as 
\begin{equation} 
  \phi^R (z)+ \phi^L (z)- \sum_{n=1}^\infty \psi_{2n-1} (z) = 1 
\,. \label{cons:general}
\end{equation}
Using this, we may rewrite Eq.(\ref{Amuxz}) to obtain 
\begin{eqnarray} 
  A_\mu(x^\mu, z) 
 &=& \alpha_{\mu ||} (x^\mu) 
  + 
  \alpha_{\mu \perp}(x^\mu) (\phi^R (z) -\phi^L (z) )
\nonumber \\ 
&&  
+ 
\sum_{n=1}^\infty 
A_\mu^{(n)}(x^\mu) \psi_{2n}(z)
+ 
\sum_{n=1}^\infty 
\left( \alpha_{\mu ||}(x^\mu) -    
V_\mu^{(n)}(x^\mu) 
\right)\psi_{2n-1}(z) 
  \,, \label{Amuxz:2} 
\end{eqnarray}
where 
\begin{equation} 
  \alpha_{\mu ||,\perp}(x^\mu) 
  =\frac{\alpha^R_\mu(x^\mu) \pm \alpha^L_\mu(x^\mu)}{2}
\, \label{alpha:para-perp} 
\end{equation}
respectively transform under the HLS as 
\begin{eqnarray} 
\alpha_{\mu ||}(x^\mu) 
&\to& 
h(x^\mu) \cdot   \alpha_{\mu ||}(x^\mu)  \cdot h^\dagger(x^\mu) 
- i \partial_\mu h(x^\mu) \cdot h^\dagger(x^\mu) 
\,, \label{alphapara:trans} \\ 
\alpha_{\mu \perp}(x^\mu) 
&\to& 
h(x^\mu) \cdot   \alpha_{\mu \perp}(x^\mu)  \cdot h^\dagger(x^\mu) 
\,. \label{alphaperp:trans} 
\end{eqnarray}
Note that $\alpha_{\mu \perp}$ includes the NGB fields as $\alpha_{\mu \perp}= \frac{1}{F_\pi} \partial_\mu \pi + \cdots$. 
The corresponding wave function $(\phi^R - \phi^L)$ should therefore be  
the eigenfunction for the zero mode ($n=0$ in Eq.(\ref{generalKKeq})), $\psi_0$: 
\begin{eqnarray} 
  \phi^R (z)-\phi^L(z)  = \psi_0(z) 
\label{eq 20}
\,. 
\end{eqnarray}
  Thus we see from Eqs.(\ref{cons:general}) and (\ref{eq 20}) 
that the wave functions $\phi^R$ and $\phi^L$ are not the eigenfunctions but are given as 
\begin{equation} 
 \phi^{R,L} (z)
 = \frac{1}{2} \left[ 
1 + \sum_{n=1}^\infty \psi_{2n-1} (z) \pm \psi_0(z) \right] 
\,. 
\end{equation}

By substituting Eq.(\ref{Amuxz:2}) into the action (\ref{5d-action}) 
with Eq.(\ref{eq 20}) taken into account, 
the five-dimensional theory is now described by the NGB fields along with 
an infinite tower of the vector and the axialvector meson fields in four dimensions:  
The action (\ref{5d-action}) is expressed as 
\begin{eqnarray} 
 S_5 
&=& 
 N_c M_{\rm KK}^2 
 \int dz d^4 x \, 
 \Bigg\{ 
K_2(z) \dot{\psi}_0^2(z) {\rm tr}[\alpha_{\mu \perp}(x^\mu)]^2 
+ 
K_2(z) \sum_{n=1}^\infty 
  \lambda_{2n} \psi_{2n}^2(z) {\rm tr}[A_\mu^{(n)}(x^\mu)]^2 
\nonumber \\ 
&& 
+ 
K_2(z) \sum_{n=1}^\infty 
\lambda_{2n-1} \psi_{2n-1}^2(z) {\rm tr}[\alpha_{\mu ||} (x^\mu) - V_\mu^{(n)}(x^\mu)]^2   
\Bigg\} 
- \frac{1}{2} N_c 
\int dz d^4 x K_1(z)  {\rm tr}[F_{\mu\nu} F^{\mu\nu}] 
\,, \label{action:Amu:in}
\end{eqnarray} 
where we have used the eigenvalue equation (\ref{generalKKeq}) and the orthogonality relation 
among the eigenfunctions.  
In the last term of Eq.(\ref{action:Amu:in}) 
the five-dimensional field strength $F_{\mu \nu}(x^\mu, z)$ can be decomposed into three parts: 
\begin{equation} 
  F_{\mu\nu}(x^\mu,z) 
  = F_{\mu\nu}^{(0)}(x^\mu,z) + \sum_{n=1}^\infty F_{\mu\nu}^{(n)}(x^\mu,z) 
+ \sum_{n=1}^\infty \sum_{m=1}^\infty F_{\mu\nu}^{(n,m)}(x^\mu, z) 
\,, \label{Fmunu:xz}
\end{equation}
where 
\begin{eqnarray} 
 F_{\mu\nu}^{(0)}(x^\mu,z)  
& =& F_{\mu\nu}(\alpha_{||}) + \psi_0(z) (D_\mu \alpha_{\nu \perp}(x^\mu) - D_\nu \alpha_{\mu \perp} (x^\mu) ) 
 -  i \psi_0^2(z)  [\alpha_{\mu \perp} (x^\mu) , \alpha_{\nu \perp} (x^\mu) ]   
\,, \nonumber \\ 
F_{\mu\nu}^{(n)}(x^\mu,z) 
&=& 
\psi_{2n} (z) (D_\mu A_\nu^{(n)} - D_\nu A_\mu^{(n)} )
- \psi_{2n-1} (z) (D_\mu \tilde{V}_\nu^{(n)} - D_\nu {\tilde V}_\mu^{(n)} ) 
\nonumber \\ 
&& 
\hspace{30pt} 
- i \psi_{2n}^2 (z) [A_\mu^{(n)} , A_\nu^{(n)}] 
- i \psi_{2n-1}^2 (z) [{\tilde V}_\mu^{(n)} , {\tilde V}_\nu^{(n)}] 
\nonumber \\ 
&& 
\hspace{30pt} 
-  i \psi_0 (z) \psi_{2n} (z) ( [\alpha_{\mu \perp}, A_\nu^{(n)}] - [\alpha_{\nu \perp}, A_\mu^{(n)}] ) 
+ i \psi_0 (z) \psi_{2n-1} (z) ( [\alpha_{\mu \perp}, {\tilde V}_\nu^{(n)}] - [\alpha_{\nu \perp}, {\tilde V}_\mu^{(n)}] ) 
\,, 
\nonumber \\ 
\nonumber \\ 
F_{\mu\nu}^{(n,m)}(x^\mu, z) 
&=& 
i  \psi_{2n-1} (z) \psi_{2m} (z) ( [A_\mu^{(m)}, {\tilde V}_\nu^{(n)}] - [A_\nu^{(m)}, {\tilde V}_\mu^{(n)}] ) 
\,, 
\end{eqnarray} 
with 
\begin{eqnarray} 
F_{\mu\nu}(\alpha_{||}) 
&\equiv & \partial_\mu \alpha_{\nu ||} (x^\mu) - \partial_\nu \alpha_{\mu ||} (x^\mu) 
- i [\alpha_{\mu ||} (x^\mu) , \alpha_{\nu ||} (x^\mu) ] 
\,, \nonumber \\
D_\mu  
&\equiv & \partial_\mu  - i [\alpha_{\mu ||}, \quad ] 
\,, \nonumber \\ 
{\tilde V}_\mu^{(n)} 
& \equiv & 
V_\mu^{(n)} - \alpha_{\mu ||} 
\,.
\end{eqnarray}

\subsection{Integrating out KK-modes of vector and axialvector mesons} 
\label{truncate} 

We are interested in constructing
a low-energy effective theory of HQCD 
written in terms of the meson fields
with their masses lower than a certain energy scale.
Suppose that those mesons are given by the KK-modes
of the HQCD at the level of 
 $m\le M$ for the axialvector mesons  $A_\mu^{(m)}(x^\mu)$
  and the level of $n\le N$ for the vector mesons $V_\mu^{(n)}(x^\mu)$.

Let us first discuss naive truncation of KK-modes 
of the HLS gauge bosons as the vector and the axialvector mesons
simply by putting $A_\mu^{(m)}(x^\mu)=0$  for $m>M$
and $V_\mu^{(n)}(x^\mu)=0$ for  
$n > N$
in Eq.~(\ref{Amuxz}):
\begin{equation} 
  A_\mu^{\rm trun}(x^\mu, z) 
  = \alpha_\mu^R(x^\mu) \phi^R(z) 
  +
  \alpha_\mu^L(x^\mu) \phi^L(z) 
 + 
\sum_{m=1}^M 
A_\mu^{(m)}(x^\mu) \psi_{2m}(z)
-    
\sum_{n=1}^N
V_\mu^{(n)}(x^\mu) \psi_{2n-1}(z) 
  \, %\label{Amuxz3} 
\end{equation}
with  the constraint in Eqs.~(\ref{cons:general}) and (\ref{eq 20}) unchanged:
\begin{eqnarray} 
&&
  \phi^R(z) + \phi^L(z) - \sum_{n=1}^{\infty} \psi_{2n-1}(z) = 1 
  \,, \label{cons:int:2}
\\
&&
 \phi^R (z) - \phi^L (z) = \psi_0(z)
 \,.
\end{eqnarray}
As a result,
$A_\mu^{\rm trun}(x^\mu,z)$ transforms under the HLS as
\begin{equation}
A_\mu^{\rm trun}(x^\mu,z) \rightarrow
h(x^\mu) \cdot A_\mu^{\rm trun}(x^\mu,z) \cdot h^\dag(x^\mu) 
- i C^{\rm trun}(z)
    \, \partial_\mu h(x^\mu) \cdot h^\dag(x^\mu)
    \, ,
    \label{Amu:trunc:trans}
\end{equation}
where
\begin{equation}
C^{\rm trun}(z) \equiv 
\phi^R(z) + \phi^L(z) - \sum_{n=1}^{N} \psi_{2n-1}(z)
= 1 + \sum_{n=N+1}^{\infty} \psi_{2n-1}(z)
\neq 1 \ .
\label{Cz}
\end{equation}
Then $A_\mu^{\rm trun}(x^\mu,z)$ no longer transforms as the gauge field. 
Since
the action in Eq.~(\ref{5d-action}) is invariant under the transformation
in Eq.~(\ref{Amu:trans:Az0}) but not in Eq.~(\ref{Amu:trunc:trans}),
then this truncation 
violates the gauge symmetry (HLS)
 and hence the chiral symmetry~\footnote{   
Some reflections of the violation of the HLS/chiral symmetry will be 
discussed  in Sec.~\ref{sec3}. }.

The violation of the HLS/chiral symmetry can also be seen  
in the expression of Eq.~(\ref{action:Amu:in})
with naive truncation,
$A_\mu^{(m)}(x^\mu)=0$  for $m>M$
and $V_\mu^{(n)}(x^\mu)=0$ for  
$n > N$:
\begin{eqnarray} 
 S_5^{\rm trun} 
& = &
 N_c M_{\rm KK}^2 
 \int dz d^4 x \, 
 \Bigg\{ 
K_2(z) \dot{\psi}_0^2(z) {\rm tr}[\alpha_{\perp \mu}(x^\mu)]^2 
+ 
K_2(z) \sum_{m=1}^M
  \lambda_{2m} \psi_{2m}^2(z) {\rm tr}[A_\mu^{(m)}(x^\mu)]^2 
\nonumber \\ 
&& 
+ 
K_2(z) \sum_{n=1}^N
\lambda_{2n-1} \psi_{2n-1}^2(z) {\rm tr}[\alpha_{\mu ||} (x^\mu)-V_\mu^{(n)}(x^\mu)]^2 
+  K_2(z) \sum_{n=N+1}^\infty
\lambda_{2n-1} \psi_{2n-1}^2(z) {\rm tr}[\alpha_{\mu ||} (x^\mu)  ]^2 
\nonumber \\ 
&& 
- \frac{1}{2}
N_c 
\int dz d^4 x K_1(z)  
{\rm tr} \left[\left( F_{\mu\nu}^{\rm trun} \right)^2 \right]
   \,,
\label{action:naive}
\end{eqnarray}
where 
$F_{\mu\nu}^{\rm trun} = \partial_\mu A_\nu^{\rm trun} - \partial_\nu 
A_\nu^{\rm trun} - i \left[ A_\mu^{\rm trun} \,,\, A_\nu^{\rm trun} \right]
$.
It is obvious that the last line is not invariant under the HLS/chiral symmetry
since $F_{\mu\nu}^{\rm trun} \not\rightarrow 
 h\cdot F_{\mu\nu}^{\rm trun} \cdot h^\dag$
 under the HLS.
 Similarly,
 one can easily see from Eq.(\ref{alphapara:trans}) that 
the last term in the second line also violates 
the chiral symmetry as well as the HLS.

Now we shall discuss a method~\cite{Harada:2006di} 
to integrate out KK modes of the HLS gauge bosons, 
or solving them away through the equations of motion
in an HLS/chiral-invariant manner.
Equivalently, our method~\cite{Harada:2006di} is nothing but 
eliminating the mass-eigenstate fields 
$B_\mu^{(n)}$ in Eq.(\ref{B:field}) 
through the equations of motion, $B_\mu^{(n)}=0$,  
which may be phrased as ``neglecting the higher mass excitation modes"~\cite{Nawa:2006gv}. 
Consider a low-energy effective theory below 
the axialvector-meson mass of $m=M+1$ level
and the vector-meson mass 
of $n=N+1$ level,
where the higher dimensional terms such as the kinetic terms
may be ignored.
Then the equations of motion for $B_\mu^{(2m)} =A_\mu^{(m)}$
with $m>M$ 
and $B_\mu^{(2n-1)} = (V_\mu^{(n)}- \alpha_{\mu ||})$ with $n > N$ read
\begin{eqnarray} 
B_\mu^{(2m)}(x^\mu) &=& A_\mu^{(m)}(x^\mu)  =  0 
 \qquad 
(m=M+1, M+2, \cdots , \infty) 
   \,, \qquad
   \nonumber\\ 
B_\mu^{(2n-1)}(x^\mu) &=& \left( V_\mu^{(n)}(x^\mu) - \alpha_{\mu ||}(x^\mu) \right) 
 = 0
 \qquad 
(n=N+1, N+2, \cdots , \infty) 
 \,. \label{sol}
\end{eqnarray}
Note that naive truncation is to eliminate
the HLS fields $V_\mu^{(n)}= A_\mu^{(m)}=0$
in contrast to integrating out as $B_\mu^{(n)}=0$.
 Putting these solutions into 
 Eq.~(\ref{Amuxz}), we obtain
 \begin{eqnarray} 
  A_\mu^{\rm integ}(x^\mu, z) 
  &=& \alpha_\mu^R(x^\mu) \phi^R(z) 
  +
  \alpha_\mu^L(x^\mu) \phi^L(z) 
+  
\sum_{m=1}^M
A_\mu^{(m)}(x^\mu) \psi_{2m}(z)
-    
\sum_{n=1}^N
V_\mu^{(n)}(x^\mu) \psi_{2n-1}(z) 
  \nonumber\\
&& {} 
-
 \sum_{n=N+1}^\infty \alpha_{\mu \parallel}(x^\mu) \psi_{2n-1}(z)
  \,. \label{Amuxz2} 
\end{eqnarray}
This 
$A_\mu^{\rm integ}$ transforms under the HLS as
\begin{equation}
A_\mu^{\rm integ}(x^\mu,z) \rightarrow
h(x^\mu) \cdot A_\mu^{\rm integ}(x^\mu,z) \cdot h^\dag(x^\mu) 
- i C^{\rm integ}(z)
    \, \partial_\mu h(x^\mu) \cdot h^\dag(x^\mu)
    \, ,
    \label{Amu:integ:trans}
\end{equation}
where $C^{\rm integ}(z)$ is identically unity from 
Eq.~(\ref{cons:int:2}):
\begin{equation}
C^{\rm integ}(z) \equiv 
\phi^R(z) + \phi^L(z) - \sum_{n=1}^{\infty} \psi_{2n-1}(z)
= 1, \label{D}
\end{equation}
in comparison with $C^{\rm trun}(z) \neq 1$ in Eq.(\ref{Cz}). 
This implies that $A_\mu^{\rm integ}$ transforms as the gauge field
in contrast to $A_\mu^{\rm trun}$
in the naive truncation,
and hence the action~(\ref{5d-action}) remains invariant under the HLS/chiral transformation 
after higher KK-modes are integrated out. 
The reason why $A_\mu^{\rm integ}$ transforms correctly is that 
the presence of the last term of Eq.(\ref{Amuxz2}) consisting of 
$\alpha_{\mu||}$ as a result of equations of motion (\ref{sol}) 
keeps the transformation property of the original higher KK fields, 
in contrast to $A_\mu^{\rm trun}$ which lacks the corresponding term. 
It is convenient to rewrite the expression in 
Eq.~(\ref{Amuxz2}) as~\cite{Harada:2006di}
\begin{equation}
 A_\mu^{\rm integ}(x^\mu, z) 
  = \alpha_\mu^R(x^\mu) \varphi^R(z) 
  +
  \alpha_\mu^L(x^\mu) \varphi^L(z) 
+  
\sum_{m=1}^M
A_\mu^{(m)}(x^\mu) \psi_{2m}(z)
-    
\sum_{n=1}^N
V_\mu^{(n)}(x^\mu) \psi_{2n-1}(z) 
  \,. \label{Amuxz3} 
\end{equation}
where
\begin{eqnarray}
&&
 \varphi^R (z)+ \varphi^L (z)- \sum_{n=1}^N \psi_{2n-1} (z) = 1 
 \,, \label{RLsum:2}
 \\
&&
 \varphi^R (z) - \varphi^L (z) = \psi_0(z)
 \,.
 \label{RLdif:2}
 \end{eqnarray}
Note the
crucial difference between 
the finite sum in Eq.~(\ref{RLsum:2})
and the infinite sum in Eq.~(\ref{cons:int:2}).
This point will be discussed in Sec.~\ref{pheno} to be
important for the HLS/chiral invariance
which includes the electromagnetic gauge invariance 
when the system is coupled to the photon as
in the pion form factor.

The invariance can also be seen by 
the action (\ref{action:Amu:in}) with the condition of integrating out KK-modes 
in Eq.(\ref{sol}):
\begin{eqnarray} 
 S_5^\textrm{integ} 
& = &
 N_c M_{\rm KK}^2 
 \int dz d^4 x \, 
 \Bigg\{ 
K_2(z) \dot{\psi}_0^2(z) {\rm tr}[\alpha_{\perp \mu}(x^\mu)]^2 
+ 
K_2(z) \sum_{m=1}^M 
  \lambda_{2m} \psi_{2m}^2(z) {\rm tr}[A_\mu^{(m)}(x^\mu)]^2 
\nonumber \\ 
&& 
+ 
K_2(z) \sum_{n=1}^N
\lambda_{2n-1} \psi_{2n-1}^2(z) {\rm tr}[\alpha_{\mu ||} (x^\mu) - V_\mu^{(n)}(x^\mu)]^2   
\Bigg\} 
\nonumber \\ 
&& 
- \frac{1}{2}
N_c 
\int dz d^4 x K_1(z)  
{\rm tr}\left[ \left(F_{\mu\nu}^{\rm integ} \right)^2 \right]
\,. \label{action:Amu:integrate}
\end{eqnarray} 
where
\begin{eqnarray}
F_{\mu\nu}^{\rm integ}(x^\mu, z)
&=&
\partial_\mu A_\nu^{\rm integ} - \partial_\nu A_\mu^{\rm integ}
- i \left[ A_\mu^{\rm integ} \,,\, A_\nu^{\rm integ} \right]
\,. 
\label{eq:F}
\end{eqnarray}
It is obvious that each term in Eq.~(\ref{action:Amu:integrate})
is invariant under the HLS.

\subsection{Integrating out HQCD back to HLS}

Let us next consider a low-energy effective model obtained by integrating 
out all the higher vector and axialvector meson fields in HQCD
except  the lowest vector meson field $V_\mu^{(1)}(x^\mu)\equiv V_\mu(x^\mu)$,
i.e. $M=0$ and $N=1$ in Eqs.~(\ref{Amuxz3})-(\ref{RLdif:2}).
Such an effective model can be described by the HLS model 
having only the NGBs and the lightest vector mesons denoted by $\rho$ 
($\rho$ meson and its flavor 
partners) plus the ${\mathcal O}(p^4)$ terms
coming from the last term in Eq.~(\ref{action:Amu:integrate}). 
Given a particular HQCD we can compute all the coefficients of
${\mathcal O}(p^4)$ terms~\cite{Harada:2006di}: 
The ${\cal O}(p^4)$ terms include the effects from infinite tower of higher KK modes 
and are completely determined as will be explicitly seen in the next sections.

Substituting Eqs.~(\ref{RLsum:2})
and
(\ref{RLdif:2})
into Eq.~(\ref{Amuxz3}) with $M=0$ and $N=1$,
we obtain
\begin{equation} 
  A_\mu^{\rm integ}(x^\mu,z) = \hat{\alpha}_{\mu \perp}(x^\mu) \psi_0(z)
  + (\hat{\alpha}_{\mu ||}(x^\mu) + V_\mu(x^\mu)  )  
  + \hat{\alpha}_{\mu ||}(x^\mu)  \psi_1(z) 
  \,, \label{generalAmu:expand3}
\end{equation} 
where 
\begin{eqnarray} 
\hat{\alpha}_{\mu \perp} 
&=& 
\frac{i}{2} \left(  \xi_R D_\mu \xi_R^\dagger - \xi_L D_\mu \xi_L^\dagger  \right)%(x^\mu) 
=
\alpha_{\mu \perp}
\,, \label{alpha:perp:cov} \\ 
\hat{\alpha}_{\mu ||}
&=& 
\frac{i}{2} \left(  \xi_R D_\mu \xi_R^\dagger + \xi_L D_\mu \xi_L^\dagger  \right)
=
- V_\mu + \alpha_{\mu ||}
\,, \label{alpha:para:cov}
\end{eqnarray} 
with
\begin{eqnarray} 
D_\mu \xi_R^\dagger 
&=& \partial_\mu \xi_R^\dagger - i {\cal R}_\mu \xi_R^\dagger  + i \xi_R^\dag V_\mu
\,, \\ 
D_\mu \xi_L^\dagger
&=& 
\partial_\mu \xi_L^\dagger - i {\cal L}_\mu \xi_L^\dagger  + i \xi_L^\dag V_\mu
\,.
\end{eqnarray}
The resultant low-energy effective theory is given by 
putting Eq.~(\ref{generalAmu:expand3}) into 
Eq.~(\ref{action:Amu:integrate})
with $M=0$ and $N=1$
through Eq.~(\ref{eq:F}).

\section{Application to Sakai-Sugimoto Model}
\label{sec3}

In this section, we apply our integrating-out method to 
the Sakai-Sugimoto (SS) model  
based on $D8/\bar{D}8/D4$ brane configuration~\cite{
Sakai:2004cn,Sakai:2005yt}. 
 As a result of integrating out higher KK modes other than 
the lowest one ($\rho$ and its flavor partners), 
we give a complete list of the ${\cal O}(p^4)$ terms of the hidden local 
symmetry (HLS) at large $N_c$ limit, which was partially reported in the previous work~\cite{Harada:2006di}. 
The anomaly-related  (intrinsic-parity odd (IP-odd)) gauge-invariant terms 
introduced in Ref.~\cite{Fujiwara:1984mp}  
are also completely determined by integrating out higher KK modes in 
the Chern-Simons (CS) term. 

We shall first summarize the action of the SS model relevant to 
our discussion, following Refs.~\cite{
Sakai:2004cn,Sakai:2005yt}. The model consists of two parts, 
the Dirac-Born-Infeld (DBI) part and the CS part.

The DBI part is given by 
\begin{eqnarray}
S_{\rm SS}^{\rm DBI} &=& N_c G  
        \int d^4x dz  
         \Bigg( 
               - \frac{1}{2} K^{-1/3}(z) {\rm tr}[ 
                                                   F_{\mu\nu} F^{\mu\nu} 
                                                 ]  
               +      K(z) M_{\rm KK}^2  {\rm tr }[ 
                                                   F_{\mu z} F^{\mu z} 
                                                  ] 
         \Bigg)
\,, \label{SSaction}
\end{eqnarray} 
where $K(z) =1 + z^2 $ is the induced metric of the five-dimensional space-time; 
the overall coupling $G$ is the rescaled 't~Hooft coupling expressed as 
$G =  N_cg_{\rm YM}^2/(108 \pi^3) $ with 
$g_{\rm YM}$ being the gauge coupling of 
the $U(N_c)$ gauge symmetry on the $N_c$ D4-branes 
the mass scale $M_{KK}$  is related to 
the scale of the compactification of the $N_c$ D4-branes onto the $S^1$. 
Comparing Eq.(\ref{SSaction}) with Eq.(\ref{5d-action}), 
we read off 
\begin{eqnarray} 
K_1(z)&= & G K^{-1/3}(z)  
\,, \nonumber \\ 
K_2(z)&=& G K(z) 
\,. 
\end{eqnarray}
Referring to  Eq.(\ref{generalKKeq}), furthermore, 
we can easily see that Eq.(\ref{SSaction}) yields the eigenvalue equation %~\cite{Sakai:2004cn,Sakai:2005yt}, 
\begin{equation} 
 -K^{1/3}(z) \partial_z  \left( K(z) \partial_z \psi_n \right) = \lambda_n \psi_n  
\, ,\label{KKeq:SS}
\end{equation}   
with the eigenvalues $\lambda_n$ and the eigenfunctions $\psi_n$ of the KK modes of 
the five-dimensional gauge field $A_\mu(x^\mu, z)$.

The CS action in the SS model is given by 
\begin{equation} 
  S_{\rm SS}^{CS}(A) =  \frac{N_c}{24\pi^2} 
  \int_{M^4 \times R} w_5(A) 
  \,, \label{CS:SS}
\end{equation} 
where 
$M^4$ and $R$ represent the four-dimensional Minkowski space-time
and the $z$-coordinate space, respectively. 
In terms of five-dimensional differential forms, 
the gauge field  and the field strength are written as 
$A=A_M dx^M = A_\mu dx^\mu + A_z dz$, $F= d A - i A^2$, 
where we use the hermitian gauge field instead of the 
anti-hermitian one used in Refs.~\cite{Sakai:2004cn,Sakai:2005yt}.   
Then the CS five-form $w_5(A)$ is expressed in terms of these five-dimensional 
differential forms as  
\begin{equation} 
 w_5(A) = {\rm tr} \left[ AF^2 + \frac{1}{2} i A^3 F  - \frac{1}{10} A^5 \right] 
 \,. \label{w5}
\end{equation} 
Crucial is to notice that 
the CS action (\ref{CS:SS}) is  not gauge-invariant under the five-dimensional gauge symmetry: Once the $A_z \equiv 0 $ gauge is realized by 
the gauge transformation 
$A \to A^g = g A g^\dagger + i g d g^\dagger$, 
the CS five-form $w_5(A)$ in Eq.(\ref{w5}) 
no longer takes the same form as in Eq.(\ref{w5}) but is modified as 
\begin{eqnarray} 
w_5(A) = 
w_5(A^g) - \frac{1}{10} {\rm tr}[g d g^{\dagger}]^5 -  d \alpha_4(id g^\dagger g, A) 
\,, \label{w5A}
\end{eqnarray}
where $\alpha_4$ is the four-form function given by 
\begin{equation} 
  \alpha_4(V,A) 
  = - \frac{1}{2}{\rm tr}[V(iAdA + idAA + A^3) - \frac{1}{2} VAVA - V^3 A] 
  \,, \label{alpha4} 
\end{equation} 
and the modified CS five-form $w_5(A^g)$ becomes 
\begin{eqnarray} 
  w_5(A^g) &=& {\rm tr}[ A^g d A^g d A^g - \frac{3}{2} i (A^g)^3 d A^g ]
  \,.  \label{w5g:SS}
\end{eqnarray}
Putting Eq.(\ref{w5A}) with Eq.(\ref{w5g:SS}) into 
the CS action (\ref{CS:SS}), we have   
\begin{eqnarray} 
S_{\rm SS}^{\rm CS} 
&=& 
 \frac{N_c}{24\pi^2} \int_{M^4 } 
 \left\{ 
\alpha_4 (id g^\dagger(+\infty) g(+\infty), {\cal R})
- \alpha_4 (id g^\dagger(-\infty) g(-\infty), {\cal L})
\right\} 
\nonumber \\ 
&& 
+ \frac{N_c}{240\pi^2} \int_{M^4 \times R} 
{\rm tr}[g d g^\dagger]^5
\nonumber \\ 
&& 
- \frac{ N_c}{24\pi^2} \int_{M^4 \times R} 
{\rm tr}[A^g d A^g d A^g - \frac{3}{2} i (A^g)^3 d A^g ]
\, \nonumber \\ 
&\equiv & 
\Gamma_1 + \Gamma_2 + \Gamma_3 
\,,  \label{123}
\end{eqnarray}
where  we have introduced 
external gauge fields ${\cal R}$ and ${\cal L}$ at the boundaries $z=\pm \infty$, and 
\begin{equation} 
g(\pm \infty) \equiv g(x^\mu , \pm \infty) 
%= \xi_{R,L}(x^\mu)
\,. 
\end{equation}   
$\Gamma_1$ and $\Gamma_2$ in Eq.(\ref{123}) 
exactly reproduce the covariantized Wess-Zumino-Witten (WZW) term~\cite{Wess:1971yu,Witten:1983tw},  
while $\Gamma_3$ is HLS-gauge invariant as shown in Appendix~\ref{proof} 
and provides 
the IP-odd interactions involving vector and axialvector mesons.

\subsection{Dirac-Born-Infeld part}  
\label{DBI-part}

In this subsection, we shall integrate out higher KK modes in 
the DBI part of the SS model given in Eq.(\ref{SSaction}): 
\begin{equation} 
  S_{\rm DBI}^{\rm SS}(A) \to S_{\rm DBI\, integ}^{\rm SS}(A^{\rm integ}) 
  \,, 
\end{equation}
where $A^{\rm integ}$ is the integrated-out five-dimensional gauge field given in 
Eq.(\ref{generalAmu:expand3}).  
The integrated-out action $S_{\rm DBI\, integ}^{\rm SS} = \int d^4x {\cal L}$ is expanded 
in terms of derivatives: 
The leading order terms counted as $\mathcal{O}(p^2)$ arise from $(F^{\rm integ}_{\mu z})^2$ term 
together with the kinetic term of the HLS gauge field $V_{\mu}$ from $(F^{\rm integ}_{\mu\nu})^2$ term. 
On the other hand, the $\mathcal{O}(p^4)$ terms come from the remainder of $(F^{\rm integ}_{\mu\nu})^2$ term.  
The Lagrangian ${\cal L}$ 
thus takes the form of the HLS Lagrangian~\cite{Bando:1984ej,Bando:1987br,Harada:2003jx}:  
\begin{eqnarray}
\mathcal{L}
      &=&
      F_{\pi}^2 {\rm tr}[ \widehat{\alpha}_{\mu\perp} \widehat{\alpha}^{\mu}_{\perp} ] 
      + a F_\pi^2 {\rm tr}[ \widehat{\alpha}_{\mu||} \widehat{\alpha}^{\mu}_{||} ] 
      - \frac{1}{2g^2}{\rm tr}[ V_{\mu\nu} V^{\mu\nu} ] 
      + \mathcal{L}_{(4)} 
\,, \label{HLS Lag} 
\end{eqnarray}
where $a$ is a parameter and 
$V_{\mu\nu}$ the field strength of the HLS gauge field defined as  
$V_{\mu\nu}= \partial_\mu V_\nu - \partial_\nu V_\mu - i [V_\mu, V_\nu]$.  
$\mathcal{L}_{(4)}$ includes the ${\mathcal O}(p^4)$ terms~\cite{Tanabashi:1993sr,Harada:2003jx} given by  
\begin{eqnarray}
{\mathcal{L}}_{(4)}
&=&
y_1 \, 
{\rm tr}[{\hat{\alpha}}_{\mu\perp}{\hat{\alpha}}^{\mu}_{\perp}
{\hat{\alpha}}_{\nu\perp}{\hat{\alpha}}^{\nu}_{\perp}]
+
y_2 \, 
{\rm tr}[{\hat{\alpha}}_{\mu\perp}{\hat{\alpha}}_{\nu\perp}
{\hat{\alpha}}^{\mu}_{\perp}{\hat{\alpha}}^{\nu}_{\perp}]
\nonumber \\
&&
+y_3 \, 
{\rm tr}[{\hat{\alpha}}_{\mu||}{\hat{\alpha}}^{\mu}_{||}
{\hat{\alpha}}_{\nu||}{\hat{\alpha}}^{\nu}_{||}] 
+y_4 \, 
{\rm tr}[{\hat{\alpha}}_{\mu||}{\hat{\alpha}}_{\nu||}
{\hat{\alpha}}^{\mu}_{||}{\hat{\alpha}}^{\nu}_{||}]
\nonumber \\
&&
+y_5 \, 
{\rm tr}[{\hat{\alpha}}_{\mu\perp}{\hat{\alpha}}^{\mu}_{\perp}
{\hat{\alpha}}_{\nu||}{\hat{\alpha}}^{\nu}_{||}]
+y_6 \, 
{\rm tr}[{\hat{\alpha}}_{\mu\perp}{\hat{\alpha}}_{\nu\perp}
{\hat{\alpha}}^{\mu}_{||}{\hat{\alpha}}^{\nu}_{||}]
+y_7 \, 
{\rm tr}[{\hat{\alpha}}_{\mu\perp}{\hat{\alpha}}_{\nu\perp}
{\hat{\alpha}}^{\nu}_{||}{\hat{\alpha}}^{\mu}_{||}]
\nonumber \\
&&
+y_8 \, 
\left\{ 
{\rm tr}[{\hat{\alpha}}_{\mu\perp}{\hat{\alpha}}^{\mu}_{||}
{\hat{\alpha}}_{\nu\perp}{\hat{\alpha}}^{\nu}_{||}]
+
{\rm tr}[{\hat{\alpha}}_{\mu\perp}{\hat{\alpha}}^{\nu}_{||}
{\hat{\alpha}}_{\nu\perp}{\hat{\alpha}}^{\mu}_{||}]
\right\}
+y_9 \, 
{\rm tr}[{\hat{\alpha}}_{\mu\perp}{\hat{\alpha}}^{\nu}_{||}
{\hat{\alpha}}_{\mu\perp}{\hat{\alpha}}^{\nu}_{||}]
\nonumber \\ 
&& + \sum_{i=10}^{18} y_i {\cal L}_i 
\nonumber \\ 
&&
+z_1 \, 
{\rm tr}[{\hat{\mathcal{V}}}_{\mu\nu}{\hat{\mathcal{V}}}^{\mu\nu}]
+z_2 \, 
{\rm tr}[{\hat{\mathcal{A}}}_{\mu\nu}{\hat{\mathcal{A}}}^{\mu\nu}]
+z_3 \, 
{\rm tr}[{\hat{\mathcal{V}}}_{\mu\nu}V^{\mu\nu}]
\nonumber \\
&&
+iz_4 \, 
{\rm tr}[V_{\mu\nu}{\hat{\alpha}}^{\mu}_{\perp}
{\hat{\alpha}}^{\nu}_{\perp}]
+iz_5 \, 
{\rm tr}[V_{\mu\nu}{\hat{\alpha}}^{\mu}_{||}
{\hat{\alpha}}^{\nu}_{||}]
\nonumber \\
&&
+iz_6 \, 
{\rm tr}[{\hat{\mathcal{V}}}_{\mu\nu}
{\hat{\alpha}}^{\mu}_{\perp}{\hat{\alpha}}^{\nu}_{\perp}]
+iz_7 \, 
{\rm tr}[{\hat{\mathcal{V}}}_{\mu\nu}
{\hat{\alpha}}^{\mu}_{||}{\hat{\alpha}}^{\nu}_{||}]
-iz_8 \, 
{\rm tr}\left[{\hat{\mathcal{A}}}_{\mu\nu}
\left({\hat{\alpha}}^{\mu}_{\perp}{\hat{\alpha}}^{\nu}_{||}
+{\hat{\alpha}}^{\mu}_{||}{\hat{\alpha}}^{\nu}_{\perp}
\right)\right] 
\,, \label{order p4 HLS Lagrangian}
\end{eqnarray} 
 where the explicit form~\cite{Tanabashi:1993sr,Harada:2003jx} of ${\cal L}_i$ ($i=10$-$18$)  is irrelevant to discussions here, and 
\begin{eqnarray} 
\hat{\cal V}_{\mu\nu} 
= \frac{1}{2} \left( 
\xi_R {\cal R}_{\mu\nu} \xi^\dag_R + \xi_L {\cal L}_{\mu\nu} \xi_L^\dag \right)  
\,, \qquad 
\hat{\cal A}_{\mu\nu} 
=  \frac{1}{2} \left( 
\xi_R {\cal R}_{\mu\nu} \xi^\dag_R - \xi_L {\cal L}_{\mu\nu} \xi_L^\dag 
\right)
\,.  
\end{eqnarray} 
In the SS model all the HLS parameters in ${\cal L}$ 
are calculated as~\footnote{In Ref.~\cite{Harada:2006di}, 
the overall sign of $z_4$ and the expression of $y_1$ should be corrected.  
In Eqs.(\ref{y1}) and (\ref{z4}) these corrections have been made properly. } 
[For details, see Appendix~\ref{derivation}]
\begin{eqnarray} 
F_{\pi}^2 
      &=& 
      N_c GM_{KK}^2  \int dz K(z) \left[ \dot{\psi}_0(z) \right]^2
\,, \label{Fpi} \\
a F_{\pi}^2
      &=& 
      N_c GM_{KK}^2  \lambda_1 \langle \psi^2_1 \rangle \quad (\lambda_1\simeq0.669) 
\, , \label{Fsigma} \\
\frac{1}{g^2}
      &=& N_cG \langle \psi_1^2 \rangle 
\,, \label{g} \\
y_1
&=&-y_2 =
-N_cG \cdot 
         \langle (1 + \psi_1 - \psi_0^2)^2 \rangle 
\,, \label{y1} \\
y_3
&=& -y_4=
-N_cG \cdot 
\langle \psi^2_1 \left( 
                   1 + \psi_1 
            \right)^2 
\rangle 
\,, \label{y3} \\ 
y_5
&=&2 y_8=-y_9=
-2N_cG \cdot 
\langle \psi_1^2 \psi_0^2 
\rangle 
\, , \label{y5} \\
y_6
&=&
-y_5-y_7 \,, \label{y6} \\
y_7
&=&
2N_cG \cdot 
\langle \psi_1 \left ( 
                  1 + \psi_1 
          \right) 
          \left( 
                 1 + \psi_1 - \psi_0^2 
          \right)
\rangle 
\,, \label{y7} \\ 
y_i  
&=& 0 \, (i=10{\rm -}18) \,
\,, 
\\
z_1
&=&
-\frac{1}{2}N_cG 
\langle ( 1+{\psi}_1)^2 \rangle 
\, , \label{z1} \\
z_2
&=&
-\frac{1}{2}N_cG
\langle \psi_0^2 \rangle 
\,, \label{z2} \\
z_3
&=&
 N_cG
\langle {\psi}_1 
( 1+ {\psi}_1) \rangle 
\,, \label{z3} \\
z_4
&=&
2N_cG
\langle {\psi}_1 (1+{\psi}_1 -\psi_0^2 ) \rangle 
\,, \label{z4} \\
z_5
&=&
-2N_cG
\langle {\psi}_1^2 ( 1+{\psi}_1 ) \rangle 
\,, \label{z5} \\
z_6
&=&
-2N_cG
\langle 
( 1+{\psi}_1-\psi_0^2 )( 1+{\psi}_1 ) 
\rangle 
\, ,\label{z6} \\
z_7
&=&
2N_cG
\langle {\psi}_1 ( 1+{\psi}_1 )^2 
\rangle 
\,, \label{z7} \\
z_8
&=&
-2N_cG
\langle 
{\psi}_1\psi_0^2 
\rangle 
\,, \label{z8}
\end{eqnarray}
where $\lambda_1\simeq0.669$ is the eigenvalue obtained from Eq.(\ref{KKeq:SS})~\cite{Sakai:2004cn,Sakai:2005yt} 
and we defined 
\begin{equation} 
\langle A \rangle \equiv \int_{-\infty}^{\infty}  dz K^{-1/3}(z) A(z)
\end{equation} 
   for a function $A(z)$. 
In obtaining Eq.(\ref{Fsigma}) we used the following identity 
\begin{equation}
\int dz K(z) \dot{\psi}_1^2(z) = \lambda_1 \int dz K^{-1/3}(z) \psi_1^2(z) 
\,. 
\end{equation} 
Note that the result $y_i = 0$ ($i=10$-18) reflects the fact that 
the SS model picks up only the large $N_c$ limit, since  
${\cal L}_i$ such as 
${\cal L}_{10}={\rm tr}[\hat{\alpha}_{\perp \mu} \hat{\alpha}^\mu_\perp] 
{\rm tr}[\hat{\alpha}_{\perp \nu} \hat{\alpha}^\nu_\perp]$
are of order $1/N_c$ subleading.

The 't~Hooft coupling $G$ and the mass scale $M_{KK}$ are free parameters 
of the SS model to be fixed by physical inputs, e.g.,  experimental values of 
$F_\pi$ and $m_\rho$. 
In the holographic QCD 
the normalization of the eigenfunction $\psi_1$ is usually 
taken to be $N_cG \langle \psi_1^2 \rangle  = 1$ 
corresponding to the canonical normalization of kinetic term of the HLS gauge field $V_\mu$ 
in Eq.(\ref{HLS Lag}). 
In that case, the corresponding HLS gauge coupling $g$ is moved over to the expression 
$\hat{\alpha}_{\mu||}$, etc., such as 
$\hat{\alpha}_{\mu||}= - V_\mu + \alpha_{\mu||} \to 
\hat{\alpha}_{\mu||}= - g V_\mu + \alpha_{\mu||}$.    
 Thus the HLS gauge coupling $g$ is not determined by the holography. 
However,   
as far as 
tree-level computation including ${\cal O}(p^4)$ terms is concerned, it turns out that 
physical quantities are independent of $g$, and thus of normalization 
of $\psi_1$~\footnote{ 
It has been shown~\cite{Tanabashi:1995nz,Harada:2003jx} that some of the ${\cal O}(p^4)$ terms 
can be absorbed into the redundancy of $g$ through the redefinition of 
the HLS gauge field.  
 This redundancy corresponds to the fact that physical quantities 
in the holography do not depend on 
the normalization of $\psi_1$.  
This redundancy is no longer true at loop level, 
i.e., physical quantities should depend on $g$, or normalization of 
$\psi_1$~\cite{Harada:2006di}. }.

\subsection{Chern-Simons part}

We shall next turn to the CS part in the SS model. 
 In this subsection we integrate out higher KK modes 
of the HLS gauge bosons in 
the CS part of the SS model to determine 
the anomaly-related IP-odd terms  
in the HLS model~\cite{Bando:1984ej,Bando:1987br,Fujiwara:1984mp,Harada:2003jx}. 
In Refs.~\cite{Sakai:2004cn,Sakai:2005yt} it was shown that  
$\Gamma_1$ and $\Gamma_2$ in Eq.(\ref{123}) 
exactly reproduce 
the covariantized WZW term~\cite{Wess:1971yu,Witten:1983tw}.   
On the other hand, 
the HLS-gauge invariant portion $\Gamma_3$ produces the IP-odd interactions 
involving vector and axialvector mesons. 
After integrating out higher KK modes in $\Gamma_3$, 
we obtain the four HLS-gauge invariant  
IP-odd terms introduced in Refs.~\cite{Fujiwara:1984mp,Bando:1987br,Harada:2003jx}~\footnote{
The same result follows in a different approach at tree level, see Ref.~\cite{Kaymakcalan:1983qq}. 
}: 
\begin{eqnarray} 
  \Gamma_{\rm IP-odd}^{\rm HLS} 
 &=&  \frac{N_c}{16\pi^2} \int_{M^4} \Bigg\{  
  \hspace{10pt}
c_1 i {\rm tr} [\hat{\alpha}_L^3 \hat{\alpha}_R - \hat{\alpha}_R^3 \hat{\alpha}_L ] 
 + c_2 i {\rm tr}[\hat{\alpha}_L \hat{\alpha}_R \hat{\alpha}_L \hat{\alpha}_R ] 
 \nonumber \\ 
 && 
 + c_3 {\rm tr}[F_V(\hat{\alpha}_L \hat{\alpha}_R - \hat{\alpha}_R \hat{\alpha}_L) ] 
 + c_4 {\rm tr}[\hat{F}_V (\hat{\alpha}_L \hat{\alpha}_R - \hat{\alpha}_R \hat{\alpha}_L)] 
 \Bigg\} 
 \,, \label{HLS:anomaly-action}
\end{eqnarray}
where the normalization of $c_1$-$c_4$ terms followed Ref.~\cite{Harada:2003jx}, 
and  
\begin{eqnarray}  
\hat{\alpha}_{R,L} 
&=& \hat{\alpha}_{||} \pm \hat{\alpha}_{\perp}  
\,, \nonumber \\  
F_V &=& d V  - i V^2  
\,, \nonumber \\ 
\hat{F}_V &=& \frac{\hat{F}_{L} + \hat{F}_{R}}{2}  
\,, \nonumber \\ 
\hat{F}_{L,R} &=& 
\xi_{L,R}^\dagger \cdot F_{L,R} \cdot \xi_{L,R} 
\,, \nonumber \\ 
F_{L,R} &=& d {\cal L}({\cal R}) -  i {\cal L}^2({\cal R}^2)  
\,.  
\end{eqnarray} 
The coefficients of the IP-odd terms are determined as 
(Detail of the derivation is given in Appendix~\ref{derivation}. )
\begin{eqnarray} 
c_1 
&=& \left\langle \left\langle  \dot{\psi}_0 \psi_1 
\left( 
   \frac{1}{2} \psi_0^2 + \frac{1}{6} \psi_1^2 - \frac{1}{2} 
\right)   \right\rangle \right\rangle
\,, \label{c1} \\ 
c_2 
&=&
\left\langle  \left\langle \dot{\psi}_0 \psi_1 
\left( 
- \frac{1}{2} \psi_0^2 + \frac{1}{6} \psi_1^2 + \frac{1}{2} \psi_1 + \frac{1}{2} 
\right)   \right\rangle \right\rangle
\,, \label{c2} \\ 
c_3 
&=&
\left\langle \left\langle \dot{\psi}_0 \psi_1 \left( \frac{1}{2} \psi_1 \right)     
\right\rangle \right\rangle
\,, \label{c3} \\ 
c_4 
&=& 
\left\langle \left\langle \dot{\psi}_0 \psi_1 
\left(  - \frac{1}{2} \psi_1 - 1 \right)   \right\rangle \right\rangle
\,, \label{c4}
\end{eqnarray} 
 where we have introduced, for a function $A(z)$,  
\begin{equation} 
\langle\langle A \rangle\rangle \equiv  \int_{-\infty}^\infty dz A(z) 
\,. 
\end{equation} 
  Thus, after integrating out higher KK modes, 
the CS action (\ref{123}) is reduced to the covariantized WZW term $\Gamma_{\rm cov.}^{\rm WZW}$ and the IP-odd 
HLS-gauge invariant terms: 
\begin{equation} 
S_{\rm CS}^{\rm SS} \to 
 S_{{\rm CS} \, {\rm integ}}^{\rm SS} = \Gamma_{\rm cov.}^{\rm WZW}(U, {\cal R}, {\cal L}) 
+ \Gamma^{\rm HLS}_{\rm IP-odd}(\hat{\alpha}_{\perp}, \hat{\alpha}_{||}, V) 
\,. \label{CS:tot}
\end{equation}

\section{Applications}
\label{pheno}

Given a concrete holographic model, our method presented in Sec.~II  
enables us to deduce definite 
predictions of the model for any physical quantity 
to be compared with experimental data. 
In this section we demonstrate power of our method in the case of 
the Sakai-Sugimoto (SS) model~\cite{Sakai:2004cn,Sakai:2005yt}. 
 Physical quantities are written in terms of the generic hidden local symmetry (HLS) model with ${\cal O}(p^4)$ 
 terms, with the Lagrangian parameters being determined by the SS model.  
Since we  have integrated out the higher KK modes of the HLS gauge bosons
keeping only the lowest one, the $\rho$ meson (and its flavor partners), 
the applicable momentum range should be restricted to  
$0 \le Q^2 \ll \{m_{\rho'}^2, m_{\rho^{\prime\prime}}^2, \cdots\}$. 
We compute momentum dependence of several form factors in 
the low-energy region ($\lesssim$ 1 GeV), 
including the pion electromagnetic (EM) form factor (Sec.~\ref{EMformfactor}) 
and intrinsic-parity (IP)-odd form factors such as $\pi^0$-$\gamma$ 
and $\omega$-$\pi^0$ transition form factors (Sec.~\ref{anomalous}).  
In Sec.~\ref{anomalous} we also calculate 
anomaly-related vertex functions such as $\gamma^*$-$\pi^0$-$\pi^+$-$\pi^-$ vertex function. 
Such results were not obtained in the original formulation of the SS model~\cite{Sakai:2005yt} 
due to complication to handle infinite sum. 
We further confirm that 
our method correctly incorporates contributions from 
higher KK modes of the HLS gauge bosons, in a different way 
starting with the original expressions of form factors 
in the SS model written in terms of infinite sum of KK modes. 
We perform a low-energy expansion of those form factors  
in a way consistent with our formalism which integrates out higher KK modes into ${\cal O}(p^4)$ terms of the HLS Lagrangian.  
This reproduces the same results as those obtained from our integrating-out method.

Hereafter we will take $N_f=3$ in which case 
the vector meson fields $(\rho_\mu^{\pm,0}, \cdots)$ and the photon ($A_\mu$) field 
are embedded as follows:  
\begin{eqnarray} 
  V_\mu &=& g \rho_\mu =  \frac{g}{\sqrt{2}} \left(
\begin{array}{ccc} 
  \frac{1}{\sqrt{2}} (\rho^0_\mu + \omega_\mu) & \rho_\mu^+ & {K_\mu^{*}}^+ \\ 
  \rho_\mu^- &  - \frac{1}{\sqrt{2}} (\rho_\mu^0 - \omega_\mu) & K_\mu^{*0} \\ 
{K_\mu^{*}}^- & \bar{K}_\mu^{* 0} & \phi_\mu  
\end{array}
\right) 
\,, \nonumber \\ 
{\cal V}_\mu &=& e A_\mu \left( 
\begin{array}{ccc} 
 \frac{2}{3} & 0 &0 \\ 
 0& - \frac{1}{3} & 0 \\ 
 0 & 0 &  - \frac{1}{3} 
\end{array}
\right)
\,. \label{parametrization}
\end{eqnarray}

\subsection{Pion electromagnetic form factor} 
\label{EMformfactor}

We shall first derive an expression of the pion EM form factor 
  from the general HLS Lagrangian given in Eq.(\ref{HLS Lag}).  
Taking unitary gauge of the HLS, we expand ${\hat \alpha}_{|| \mu}$ and ${\hat \alpha}_{\perp \mu}$ 
defined in Eqs.(\ref{alpha:perp:cov}) and (\ref{alpha:para:cov}) in terms of the pion fields $\pi$ as~\cite{Harada:2003jx}  
\begin{eqnarray} 
{\hat \alpha}_{|| \mu} &=& \frac{1}{F_\pi} \partial_\mu \pi + {\cal A}_\mu - \frac{i}{F_\pi} [ {\cal V}_\mu, \pi] 
+ \cdots 
\,,  \label{para:expand}\\ 
{\hat \alpha}_{\perp \mu} &=& - V_\mu + {\cal V}_\mu - \frac{i}{2F_\pi^2} [\partial_\mu \pi, \pi] + \cdots  
\,. \label{perp:expand} 
 \end{eqnarray} 
  Substituting these expansion forms into the Lagrangian (\ref{HLS Lag}),    
we have 
\begin{eqnarray} 
{\cal L}_{\rho\rho} 
&=& - \frac{1}{2} {\rm tr}[(\partial_\mu \rho_\nu - \partial_\nu \rho_\mu)^2] + 
ag^2 F_\pi^2 {\rm tr}[\rho_\mu \rho^\mu] 
\,, \nonumber \\  
{\cal L}_{\gamma \pi \pi} 
&=& 
  2 ie \left(1 - \frac{a}{2}\right) 
{\rm tr}[A_\mu [ \partial^\mu \pi, \pi]]  + \frac{ ie z_6}{F_\pi^2} {\rm tr}[\partial_\mu A_\nu [\partial^\mu \pi, \partial^\nu \pi]] 
\,, \nonumber \\
{\cal L}_{\rho \pi \pi}  
&=& 
 ia g {\rm tr}[\rho_\mu [\partial^\mu \pi, \pi]] + \frac{ iz_4}{F_\pi^2} {\rm tr}[\partial_\mu \rho_\nu [\partial^\mu \pi, \partial^\nu \pi]] 
\,, \nonumber \\
{\cal L}_{\gamma \rho} 
&=& 
- 2 eag F_\pi^2 {\rm tr}[\rho_\mu A^\mu] + 2 eag z_3 {\rm tr}[\partial_\mu A_\nu (\partial^\mu \rho^\nu - \partial^\nu \rho^\mu)] 
\,.  \label{IPeven:int}  
\end{eqnarray}
 From these, we read off the $\rho$ mass $m_\rho$, 
the direct $\gamma$-$\pi$-$\pi$ vertex $g_{\gamma \pi\pi}(Q^2)$, 
$\rho$-$\pi$-$\pi$ vertex $g_{\rho \pi\pi}(Q^2)$ 
and $\rho$-$\gamma$ mixing strength $g_\rho(Q^2)$: 
\begin{eqnarray} 
m_{\rho}^2  & =& a g^2 F_{\pi}^2 
\,, \\ 
 g_{\gamma \pi\pi} (Q^2) 
 &=& 
 \left(1 - \frac{a}{2} \right) + \frac{a g^2z_6}{4} \frac{Q^2}{m_\rho^2} 
\,, \label{ggammapipi} \\ 
g_\rho (Q^2) 
&=& 
 \frac{m_\rho^2}{g} \left(1 + g^2 z_3 \frac{Q^2}{m_\rho^2} \right)  
 \,, \label{grho:off}\\ 
g_{\rho \pi \pi} (Q^2) 
&=& 
 \frac{1}{2} a g \left( 1 + \frac{g^2z_4}{2} \frac{Q^2}{m_\rho^2} \right) 
 \,,  \label{grhopipi:off}
\end{eqnarray}  
and the on-shell $g_\rho$ and $g_{\rho\pi\pi}$ couplings:~\footnote{In Ref.~\cite{Harada:2006di} 
the plus sign in front of $g^2 z_4$ term in the expression of $g_{\rho\pi\pi}$ 
should read minus as in Eq.(\ref{grhopipi}).  } 
\begin{eqnarray} 
g_\rho &\equiv& g_\rho(Q^2=-m_\rho^2) 
= agF_\pi^2 (1- g^2 z_3) 
\,, \label{grho} \\ 
g_{\rho\pi\pi} 
&\equiv& 
g_{\rho\pi\pi}(Q^2=-m_\rho^2)
= \frac{1}{ 2}ag \left( 1 -  \frac{1}{2} g^2 z_4  \right) 
\,, \label{grhopipi} 
\end{eqnarray} 
where $Q^2 = -p^2$ is space-like momentum squared.

\begin{figure} 

\begin{center} 
\includegraphics[scale=0.6]{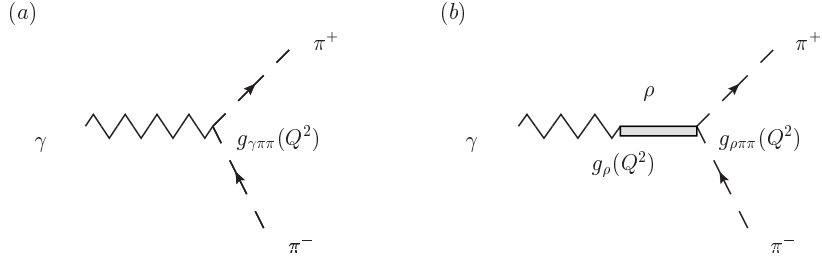}
\end{center}
\caption{Diagrams relevant to the pion EM form factor $F_V^{\pi^\pm}$ 
in the general HLS Lagrangian (\ref{HLS Lag}). 
} 
\label{graph-fv}
\end{figure}

The pion EM form factor $F_V^{\pi^\pm}$ is thus 
constructed from two contributions %from diagrams 
illustrated in Fig.~\ref{graph-fv}, one from $\rho$-mediated 
diagram (graph $(b)$), and the rest (graph $(a)$). 
  By using quantities in Eqs.(\ref{ggammapipi})-(\ref{grhopipi:off}), 
$F_V^{\pi^\pm}(Q^2)$ can be written as  
\begin{equation} 
  F_V^{\pi^\pm}(Q^2) %|_{\rm HLS} 
  = g_{\gamma \pi\pi}(Q^2) + \frac{g_\rho(Q^2) g_{\rho\pi\pi}(Q^2)}{m_\rho^2 + Q^2} 
  \,. \label{FV:HLS1}
\end{equation} 
We may rewrite this expression as %Eq.(\ref{FV:HLS1}) as 
\begin{equation} 
  F_V^{\pi^\pm} (Q^2) %|_{\rm HLS}
  = 
  \left( 1 - \frac{1}{2} \tilde{a} \right) + \tilde{z} \frac{Q^2}{m_\rho^2} 
  + \frac{{\tilde a}}{2} \frac{m_\rho^2}{m_\rho^2 + Q^2} 
\,, \label{FV}
\end{equation} 
where 
\begin{eqnarray} 
  \tilde{a} 
&=& 
 a \left( 1 - \frac{g^2 z_4}{2} 
-g^2z_3 + \frac{(g^2z_3)(g^2z_4)}{2} 
\right) 
\,, \label{atilde:HLS} \\
 \tilde{z} 
 &=&  \frac{1}{4} a \left(  g^2 z_6 + (g^2 z_3)(g^2 z_4) \right) 
\,. \label{ztilde:HLS}
\end{eqnarray} 
Note that our form factor (\ref{FV}) automatically 
ensures the EM gauge invariance no matter what values 
$\tilde a$ and $\tilde z$ may take,  
\begin{equation}
 F_V^{\pi^\pm}(0) %|_{\rm HLS}
 = 
\left(  1 - \frac{\tilde a}{2}\right) + \frac{\tilde a}{2}
=1 
 \,. \label{EMg:HLS}
\end{equation}

Here we note that the ``$\rho$ meson dominance" is defined as 
\begin{equation} 
  F_V^{\pi^\pm} (Q^2) 
= \frac{m_\rho^2}{m_\rho^2 + Q^2}
\,, \label{rhoD}
\end{equation} 
which is equivalent to taking 
\begin{equation} 
{\tilde a}=2 
\,, \qquad 
\tilde{z}=0 
\,, 
\end{equation} 
and 
is different from taking $g_{\gamma \pi\pi}(Q^2)=0$ in Eq.(\ref{ggammapipi}) 
(absence of graph $(a)$ of Fig.~\ref{graph-fv}). 
Were it not for the ${\cal O}(p^4)$ terms, 
the definition  of the $\rho$ meson dominance would be the same as $g_{\gamma \pi\pi}(Q^2)
= (1-a/2) = 0$.

\begin{figure}%[!h]
\begin{center}
\includegraphics[scale=1.0]{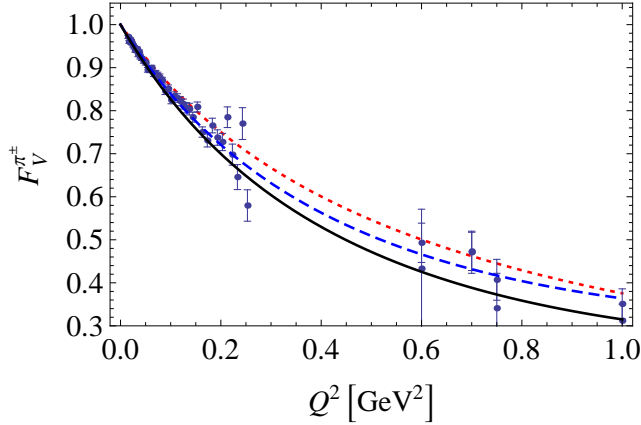} 
\caption{ 
The prediction (black solid curve) of the pion EM form factor 
$F_V^{\pi^\pm}$ fitting the experimental data~\cite{Amendolia:1986wj,Volmer:2000ek,Tadevosyan:2007yd,Horn:2006tm} 
with $\chi^2/{\rm d.o.f}=147/53= 2.8$. 
The red dotted curve corresponds to 
the form factor in the $\rho$ meson dominance 
hypothesis with ${\tilde a}=2$ and ${\tilde z}=0$. 
($\chi^2/{\rm d.o.f}=226/53 = 4.3$). 
The blue dashed curve is the best fit to experimental data 
with ${\tilde a}|_{\rm best}=2.44$, $\tilde{z}|_{\rm best} = 0.08$ 
($\chi^2/{\rm d.o.f}=81/51= 1.6$). }
\label{Fv-holo} 
\end{center}
\end{figure}

Let us now evaluate the form (\ref{FV}) in the SS model.  
By using Eqs.(\ref{g}), (\ref{z3}), (\ref{z4}), (\ref{z6}), 
with Eqs.(\ref{Fpi}) and (\ref{Fsigma}),  
$\tilde{a}$ and $\tilde{z}$ are determined in the SS model as 
\begin{eqnarray} 
   \tilde{a}_{_{\rm SS}} &=&  
\frac{\pi}{4} \lambda_1 
\frac{\langle \psi_1  \rangle  \langle  \psi_1 (1-\psi_0^2) \rangle}{\langle \psi_1^2 \rangle} \simeq 2.62
\,, \label{atilde} \\ 
 \tilde{z}_{_{\rm SS}} 
 &=& \frac{\pi}{8} \lambda_1 \left( 
\frac{\langle \psi_1  \rangle  \langle  \psi_1 (1-\psi_0^2) \rangle}{\langle \psi_1^2 \rangle} 
- \langle 1- \psi_0^2 \rangle \right) 
= \frac{\tilde a}{2} - \frac{\pi}{8} \lambda_1 \langle  1- \psi_0^2 \rangle 
\simeq 0.08
\,, \label{ztilde}
\end{eqnarray} 
where we have used $\lambda_1 \simeq 0.669$ in Eq.(\ref{Fsigma}). 
Substituting the values in Eqs.(\ref{atilde}) and (\ref{ztilde}) 
into Eq.(\ref{FV}), 
we evaluate the momentum-dependence of $F_V^{\pi^\pm}$ 
as a definite prediction of the SS model for the pion EM form 
 factor. 
See Fig.~\ref{Fv-holo} (black solid curve).  
Here we used the experimental input of $m_\rho$, $m_\rho = 775$ MeV~\cite{Amsler:2008zz}.   
(We do not need an experimental input of $F_\pi$ for this quantity.)   
The experimental data from Refs.~\cite{Amendolia:1986wj,Volmer:2000ek,Tadevosyan:2007yd,Horn:2006tm} 
are also shown. 
The $\chi^2$-fit results in good agreement with 
the data ($\chi^2/{\rm d.o.f}=147/53 =  2.8$).

 For comparison, we have also shown 
the best fit curve (denoted by a red dotted line) 
resulted from fitting the parameters $(\tilde{a}, \tilde{z})$ in the general HLS model (\ref{FV}) 
to the experimental data, which yields the best fit values of ${\tilde a}$ and $\tilde{z}$, 
${\tilde a}|_{\rm best}=2.44$, $\tilde{z}|_{\rm best} = 0.08$ 
($\chi^2/{\rm d.o.f}=81/51 =1.6$).  
It is interesting to note that the best fit values of $\tilde{a}$ and $\tilde{z}$ 
are quite close to those in the predicted curve, which  
reflects  that the predicted curve fits well the experimental data. 
Comparison with  the $\rho$ meson dominance 
with ${\tilde a}=2$ and ${\tilde z}=0$ in Eq,(\ref{FV}) 
is also shown by a blue dashed curve 
($\chi^2/{\rm d.o.f}=226/53 =4.3$).

Given any holographic model, our method can make a definite prediction of 
the model for the pion EM form factor in terms of two parameters $\tilde a$ and $\tilde z$ of the 
general HLS model which 
are determined by integrating out higher KK modes of the HLS gauge bosons.  
The best fit values in the above are the reference values to be
compared with 
those of the holographic models.  
In comparison with the $\rho$ meson dominance, the deviation from
${\tilde a}=2$ and ${\tilde z}=0$  
represents the contributions of higher KK modes in the generic
holographic model as  
will be shown below.

We shall next discuss an implication of our method from 
a different point of view. 
We start with the expression of the form factor originally studied in the SS model, 
which takes the form of the infinite sum
of KK modes of the HLS gauge fields:
\begin{equation} 
  F_V^{\pi^\pm}(Q^2) \Bigg|_{\rm SS} 
  = \sum_{k=1}^\infty \frac{g_{\rho_k} g_{\rho_k\pi\pi}}{m_{\rho_k}^2 + Q^2}
\,. \label{FVSS:1}
\end{equation}
Our method is to integrate out the higher KK mode effects into 
the ${\cal O}(p^4)$ terms of the HLS Lagrangian having only 
the $\rho$ meson as a dynamical degree of freedom. 
To be consistent with our method, 
we expand this form factor as 
\begin{eqnarray} 
  F_V^{\pi^\pm}(Q^2) \Bigg|_{\rm SS} 
 & = & 
 \frac{g_\rho g_{\rho\pi\pi}}{m_\rho^2 + Q^2} 
+ 
\sum_{k=2}^\infty \frac{g_{\rho_k} g_{\rho_k\pi\pi}}{m_{\rho_k}^2 + Q^2}
\, \nonumber \\ 
&=& 
 \frac{g_\rho g_{\rho\pi\pi}}{m_\rho^2 + Q^2} 
+ 
  \left(\sum_{k=2}^\infty \frac{g_{\rho_k} g_{\rho_k\pi\pi}}{m_{\rho_k}^2}  \right) 
+ \left(- \sum_{k=2}^\infty \frac{g_{\rho_k} g_{\rho_k\pi\pi}}{m_{\rho_k}^2} \frac{Q^2}{m_{\rho_{k}}^2} \right)   
\,, \label{FVSS:2}
\end{eqnarray} 
up to ${\cal O}(Q^4/m_{\rho_k}^4)$ ($k \ge 2$). 
Note that the ${\cal O}(p^4)$ terms of the Lagrangian 
correspond to ${\cal O}(Q^2)$ terms in the expansion of the pion EM form factor. 
 Using the sum rules~\cite{Sakai:2005yt}, 
\begin{eqnarray} 
\sum_{k=1}^\infty 
  \frac{g_{\rho_k} g_{\rho_k\pi\pi}}{m_{\rho_k}^2} 
&=& 1 
  \,, \label{sum1}\\
 \sum_{k=1}^\infty \frac{g_{\rho_k} g_{\rho_k\pi\pi}}{m_{\rho_k}^4}  
&=& \frac{\pi}{8 m_\rho^2} \lambda_1 \langle  1 - \psi_0^2  \rangle
\, , \label{sum2}
\end{eqnarray} 
we have 
 \begin{eqnarray} 
\sum_{k=2}^\infty \frac{g_{\rho_k} g_{\rho_k\pi\pi}}{m_{\rho_k}^2}
&=& 
1 - \frac{g_\rho g_{\rho\pi\pi}}{m_\rho^2} 
\,, \label{contact1:SS} \\ 
- \sum_{k=2}^\infty \frac{g_{\rho_k} g_{\rho_k\pi\pi}}{m_{\rho_k}^4}    
&=& 
\frac{g_\rho g_{\rho\pi\pi}}{m_\rho^4} - 
\frac{\pi}{8m_\rho^2} \lambda_1 \langle  1 - \psi_0^2  \rangle 
\,. \label{contact2:SS}
\end{eqnarray} 
From Ref.~\cite{Sakai:2005yt}, we read off  
$g_{\rho}$ and $g_{\rho\pi\pi}$ as well as $m_\rho$~\footnote{ 
In Ref.~\cite{Harada:2006di} 
the expression corresponding to Eq.(\ref{grhopipi:5d})  
has a typo. } 
\begin{eqnarray} 
m_\rho^2 &=& \lambda_1 M_{KK}^2 
\,, \label{Mrho:5d} \\
g_\rho &=& 
\sqrt{N_c G} \lambda_1 M_{KK}^2 \sqrt{\frac{\langle \psi_1 \rangle^2}{\langle \psi_1^2 \rangle}}
\,, \label{grho:5d}\\
g_{\rho\pi\pi} 
      &=&
       \frac{\pi}{8} \frac{\lambda_1 }{\sqrt{N_c G}} 
       \sqrt{ \frac{ \langle \psi_1(1-\psi_0^2) \rangle^2 }{ \langle \psi_1^2 \rangle}} 
\,, \label{grhopipi:5d} 
\end{eqnarray} 
where we retained $\langle \psi_1^2 \rangle$ to make explicit the ambiguity of 
the normalization of $\psi_1$ in contrast to Ref.~\cite{Sakai:2005yt} where 
the normalization of $\psi_1$ is fixed as $N_c G \langle \psi_1^2 \rangle = 1$.  
Substituting these into the right hand sides of Eqs.(\ref{contact1:SS}) and (\ref{contact2:SS}), 
we find 
\begin{eqnarray} 
 \sum_{k=2}^\infty \frac{g_{\rho_k} g_{\rho_k\pi\pi}}{m_{\rho_k}^2} 
&=& 
 1 - \frac{\pi}{8} \lambda_1 
\frac{\langle \psi_1  \rangle  \langle  \psi_1 (1-\psi_0^2) \rangle}{\langle \psi_1^2 \rangle} 
\,, \nonumber \\ 
- \sum_{k=2}^\infty \frac{g_{\rho_k} g_{\rho_k\pi\pi}}{m_{\rho_k}^4}  
&=& 
\frac{\pi}{8m_\rho^2} \lambda_1 \left( 
\frac{\langle \psi_1  \rangle  \langle  \psi_1 (1-\psi_0^2) \rangle}{\langle \psi_1^2 \rangle} 
- \langle 1- \psi_0^2 \rangle \right) 
\,. \label{relation:0}
\end{eqnarray}
 Comparing Eqs.(\ref{atilde}) and (\ref{ztilde}), we arrive at 
\begin{eqnarray} 
 \sum_{k=2}^\infty \frac{g_{\rho_k} g_{\rho_k\pi\pi}}{m_{\rho_k}^2} 
&=& 
 1 - \frac{\tilde{a}_{_{\rm SS}}}{2} 
\,, \nonumber \\ 
 - \sum_{k=2}^\infty \frac{g_{\rho_k} g_{\rho_k\pi\pi}}{m_{\rho_k}^4}  
&=& 
\frac{{\tilde z}_{_{\rm SS}}}{m_\rho^2}
\,. \label{relation:1}
\end{eqnarray}
and hence at the same result as that obtained 
by our method integrating out higher KK modes (Eq.(\ref{EMg:HLS}) with Eqs.(\ref{atilde}) and (\ref{ztilde})):  
\begin{equation} 
  F_V^{\pi^\pm} (Q^2) |_{\rm SS}
  = 
  \left( 1 - \frac{1}{2} \tilde{a}_{_{\rm SS}} \right) + \tilde{z}_{_{\rm SS}} 
\frac{Q^2}{m_\rho^2} 
  + \frac{{\tilde a}_{_{\rm SS}}}{2} \frac{m_\rho^2}{m_\rho^2 + Q^2} 
\,. \label{FV:1}
\end{equation} 

The deviation from the $\rho$ meson dominance is parameterized as
\begin{eqnarray} 
\Delta \tilde{a} &\equiv& \left( \frac{\tilde{a}- 2 }{2} \right) 
 \,, \\ 
 \Delta \tilde{z} &\equiv&  (\tilde{z}-0) 
\,. 
\end{eqnarray}  
  From Eq.(\ref{relation:1}) and referring to Ref.~\cite{Sakai:2005yt}, 
we may numerically read off these quantities in the SS model as 
\begin{eqnarray} 
 \Delta \tilde{a}_{_{\rm SS}}  
 &=& \frac{{\tilde a}_{_{\rm SS}}}{2} - 1 
 \simeq 0.31 
 \,,\nonumber \\  
&=& - \sum_{k=2}^\infty \frac{g_{\rho_k} g_{\rho_k \pi\pi}}{m_{\rho_k}^2} 
= (0.346)_{\rho'} + (-0.0505)_{\rho^{\prime\prime}} + (0.0128)_{\rho^{\prime\prime\prime}} + \cdots 
\,, \\ 
 \Delta \tilde{z}_{_{\rm SS}} 
 &=& - m_\rho^2 \sum_{k=2}^\infty \frac{g_{\rho_k} g_{\rho_k \pi\pi}}{m_{\rho_k}^4}
= (0.0806)_{\rho'} + (-0.0051)_{\rho^{\prime\prime}} + (0.0007)_{\rho^{\prime\prime\prime}} + \cdots 
\, . 
\end{eqnarray}
 This implies that the deviation from the $\rho$ meson dominance 
 (${\tilde a}\simeq 2.62, {\tilde z} \simeq 0.08$) 
comes dominantly from the $\rho'$-meson contribution.

Since the result is identical to that of our method 
which is manifestly EM-gauge invariant by construction (See Eqs.(\ref{Amu:integ:trans}),(\ref{D}) and (\ref{action:Amu:integrate})), 
 the resultant form factor (\ref{FV:1}) should be EM gauge invariant. 
 In fact, we have  
\begin{equation} 
  F_V^{\pi^\pm}(0) |_{\rm  SS} 
  = \left( 1 - \frac{\tilde{a}_{_{\rm SS}}}{2} \right)+ \frac{\tilde{a}_{_{\rm SS}}}{2} 
= 1 
  \,. \label{sum1:HLS}
\end{equation} 
In contrast, a naive truncation corresponding to 
the Lagrangian in Eq.(\ref{action:naive}) would read 
\begin{equation} 
   F_V^{\pi^\pm}(Q^2) \Bigg|_{\rm SS}^{\rm trun} 
=  
 \frac{g_\rho g_{\rho\pi\pi}}{m_\rho^2 + Q^2} 
\,, \label{trunFV}
\end{equation}
which corresponds to ignoring the last two terms in Eq.(\ref{FVSS:2}) 
coming from higher KK modes to both ${\cal O}(p^2)$ and 
${\cal O}(p^4)$ terms. Since the Lagrangian (\ref{action:naive}) is
gauge non-invariant, so is the form factor above,  
\begin{equation} 
   F_V^{\pi^\pm}(0) \Bigg|_{\rm SS}^{\rm trun} 
=  
 \frac{g_\rho g_{\rho\pi\pi}}{m_\rho^2} 
 = \frac{\tilde{a}_{_{\rm SS}}}{2}
\simeq 1.31 \neq 1  
\,,
\end{equation} 
where we used Eq.(\ref{atilde}) 
together with Eqs.~(\ref{Mrho:5d}), (\ref{grho:5d}) and
(\ref{grhopipi:5d}).
 Note that {\it the truncation (\ref{trunFV}) is different from the $\rho$ 
 meson dominance (\ref{rhoD})} which is gauge invariant.

\subsection{Anomaly-related intrinsic parity-odd processes} 
\label{anomalous}

In this subsection we calculate momentum dependence of 
the IP-odd form factors, $\pi^0$-$\gamma$ and $\omega$-$\pi^0$ transition form factors. 
We also study several IP-odd vertex functions such as 
$\pi^0$-$\gamma^*$-$\gamma^*$ (Sec.~\ref{pi2g:sec}), 
$\omega$-$\pi^0$-$\gamma^*$ (Sec.~\ref{opig:sec}), $\gamma^*$-$\pi^0$-$\pi^+$-$\pi^-$ (Sec.~\ref{pi3g:sec}), 
and $\omega \to \pi^0 \pi^+ \pi^-$ decay (Sec.~\ref{o3pi:sec}), and discuss  
the $\rho/\omega$ meson dominance. 
Discussion of an alternative method, which leads to 
the same results as ours shown in this subsection 
as done in Sec.~\ref{EMformfactor} (See Eq.(\ref{FVSS:2})), 
will be given in Appendix~\ref{alternative}.

The IP-odd interactions in the general HLS model are read off from Eq.(\ref{HLS:anomaly-action}) together with the WZW term as 
~\cite{Harada:2003jx} 
\begin{eqnarray} 
 {\cal L}_{VV\pi} 
 &=& - \frac{g^2N_c}{4 \pi^2 F_\pi} c_3 \epsilon^{\mu\nu\lambda\sigma} {\rm tr}[\partial_\mu \rho_\nu \partial_\lambda \rho_\sigma \pi] 
\,, \nonumber \\ 
{\cal L}_{V A \pi} 
&=& - \frac{egN_c}{8\pi^2 F\pi} (c_4 - c_3) \epsilon^{\mu\nu\lambda\sigma} {\rm tr}[\{ \partial_\mu \rho_\nu, \partial_\lambda A_\sigma\} \pi] 
\,, \nonumber \\ 
{\cal L}_{A A \pi}
&=& - \frac{e^2N_c}{4\pi^2 F\pi} (1 - c_4) \epsilon^{\mu\nu\lambda\sigma} {\rm tr}[\partial_\mu A_\nu \partial_\lambda A_\sigma \pi] 
\,, \nonumber \\ 
{\cal L}_{A \pi^3} 
&=& 
- i \frac{eN_c}{3\pi^2 F_\pi^3}  \left( 1 - \frac{3(c_1 - c_2 + c_3)}{4} \right) 
\epsilon^{\mu\nu\lambda\sigma} {\rm tr}[A_\mu \partial_\nu \pi \partial_\lambda \pi \partial_\sigma \pi] 
\,, \nonumber \\ 
 {\cal L}_{V \pi^3} 
&=& 
- i \frac{g N_c}{4\pi^2 F_\pi^3}  (c_1 - c_2  - c_3)  
\epsilon^{\mu\nu\lambda\sigma} {\rm tr}[\rho_\mu \partial_\nu \pi \partial_\lambda \pi \partial_\sigma \pi] 
\,. \label{IPodd:int}
\end{eqnarray} 
 For relevant Feynman graphs for each IP-odd process, see Figs.4, 5, 6, and 7 in Ref.~\cite{Harada:2003jx}.

\subsubsection{$\pi^0$-$\gamma^*$-$\gamma^*$ vertex function and $\pi^0$-$\gamma$ transition form factor} 
\label{pi2g:sec}

 We start with an expression of effective $\pi^0$-$\gamma^*$-$\gamma^*$ vertex function from the general 
 HLS Lagrangian (Eqs.(\ref{HLS Lag}) and (\ref{IPodd:int})):  
\begin{eqnarray} 
\Gamma^{\mu\nu} [\pi^0, \gamma^*_\mu (q_1), \gamma_\nu^* (q_2)] %\Bigg|_{\rm HLS}
&=& 
\frac{e^2N_c}{12\pi^2 F_\pi} 
\epsilon^{\mu\nu\alpha \beta} q_{1\alpha} q_{2 \beta} 
\cdot 
\Bigg[
(1-c_4) 
+ \frac{(c_4-c_3)}{4} \times 
\nonumber \\ 
&& 
\left\{ 
D_\rho(q_1^2) \left( 1 - g^2 z_3 \frac{q_1^2}{m_\rho^2} \right)  
+  
D_\omega(q_1^2) \left( 1 - g^2 z_3 \frac{q_1^2}{m_\omega^2} \right)   + (q_1^2 \to q_2^2)  
\right\}  
\nonumber \\ 
&& 
+ \frac{ c_3}{2} 
\left\{ 
D_\rho(q_1^2) \left( 1 - g^2 z_3 \frac{q_1^2}{m_\rho^2} \right)  
\cdot 
D_\omega(q_2^2) \left( 1 - g^2 z_3 \frac{q_2^2}{m_\omega^2} \right)  
+ (q_1^2 \leftrightarrow q_2^2 )  
\right\} 
\Bigg]
\,, \nonumber \\ 
\label{pi2g:form1}
\end{eqnarray} 
where 
$q_{1,2}$ represent outgoing four-momenta of virtual photons $\gamma^*$ and 
\begin{eqnarray} 
 D_{\rho,\omega}(p^2) &\equiv & \frac{m_{\rho,\omega}^2}{m_{\rho,\omega}^2 - p^2}
\,, \\ 
g_{\rho,\omega} (p^2) &\equiv & \frac{m_{\rho,\omega}^2}{g} \left(  1 - g^2 z_3 \frac{p^2}{m_{\rho,\omega}^2} \right)
\,. 
\end{eqnarray} 
We further rewrite Eq.(\ref{pi2g:form1}) as follows:  
\begin{eqnarray} 
\Gamma^{\mu\nu} [\pi^0, \gamma^*_\mu (q_1), \gamma_\nu^* (q_2)] 
&=& 
\frac{e^2N_c}{12\pi^2 F_\pi} 
\epsilon^{\mu\nu\alpha \beta} q_{1\alpha} q_{2 \beta} 
\Bigg[
A^{\pi 2\gamma} + \frac{B^{\pi 2\gamma}}{4}  
\left\{  
D_\rho(q_1^2) + D_\omega(q_1^2) + (q_1^2 \to q_2^2)  
\right\}  
\nonumber \\ 
&& 
+ \frac{C^{\pi 2\gamma}}{2}
\left\{  
D_\rho(q_1^2) \cdot D_\omega(q_2^2) + (q_1^2 \leftrightarrow q_2^2)  
\right\} 
\Bigg]
\,, \label{pi2g:form2} 
\end{eqnarray} 
where we have used an identity for an arbitrary coefficient $C$,  
\begin{equation} 
D_{\rho, \omega}(p^2) \left( 1 - C \cdot \frac{p^2}{m_{\rho, \omega}^2} \right)  
 = (1-C) D_{\rho,\omega}(p^2) + C  
 \,, \label{ident:1}
\end{equation}
and defined 
\begin{eqnarray} 
A^{\pi 2\gamma}   &=& 
  1- 
  (1- g^2 z_3) (c_4 + c_3\cdot g^2 z_3) 
  \,, \\ 
    B^{\pi 2\gamma} 
  &=& 
   (1- g^2 z_3) \left[ 
(c_4 + c_3\cdot g^2 z_3) - c_3 (1-g^2z_3) 
\right] 
  \,, \\ 
    C^{\pi 2\gamma} 
  &=& 
 c_3 (1- g^2 z_3)^2 
  \,.
\end{eqnarray} 
Note that these parameters satisfy 
\begin{equation} 
 A^{\pi 2 \gamma} +  B^{\pi 2 \gamma} +  C^{\pi 2 \gamma} = 1 
 \label{LET:pi2g}
\end{equation} 
which reproduces the low-energy theorem: 
\begin{equation} 
\Gamma^{\mu\nu} [\pi^0, \gamma^*_\mu (q_1), \gamma_\nu^* (q_2)] 
\stackrel{{q_1^2, q_2^2 \to 0}}{\longrightarrow} 
\frac{e^2N_c}{12\pi^2 F_\pi} 
\epsilon^{\mu\nu\alpha \beta} q_{1\alpha} q_{2 \beta} 
\,. \label{LET:pi2g1} 
\end{equation} 
The $\rho/\omega$ meson dominance~\cite{GellMann:1962jt} for this process is defined 
in a way similar to Eq.(\ref{rhoD}) by 
taking $A^{\pi 2 \gamma}=B^{\pi 2 \gamma}=0$ 
($C^{\pi 2\gamma}=1$) 
\begin{equation} 
 \Gamma^{\mu\nu} [\pi^0, \gamma^*_\mu (q_1), \gamma_\nu^* (q_2)] %\Bigg|_{\rm }
= 
\frac{e^2N_c}{24\pi^2 F_\pi} 
\epsilon^{\mu\nu\alpha \beta} q_{1\alpha} q_{2 \beta} 
\left(  
D_\rho(q_1^2) \cdot D_\omega(q_2^2) + (q_1^2 \leftrightarrow q_2^2)  
\right) 
\,. \label{GSW}
\end{equation}

The $\pi^0$-$\gamma$ transition form factor $F_{\pi^0 \gamma}$ is 
obtained from Eq.(\ref{pi2g:form2}) by setting one of the photon-momentum squared ($q_1^2$ or $q_2^2$) 
to be zero: 
\begin{equation} 
  F_{\pi^0 \gamma}(Q^2) 
  = 
\left(1 -  
{\tilde c}
\right)
 + \frac{{\tilde c}}{2} [D_\rho(Q^2) + D_\omega(Q^2)] 
  \,, \label{Fpig:form}
\end{equation} 
where $Q^2=-q_1^2$ (or $-q_2^2$) and we defined 
\begin{equation} 
{\tilde c} \equiv 1 - (A^{\pi2\gamma} +  B^{\pi2\gamma}) 
\,. \label{cbar}
\end{equation} 
The $\rho/\omega$ meson dominance defined by Eq.(\ref{GSW}) reads     
\begin{eqnarray}  
  F_{\pi^0 \gamma}(Q^2) 
  &=&  \frac{1}{2} \left( \frac{m_\rho^2}{m_\rho^2 + Q^2} +
  \frac{m_\omega^2}{m_\omega^2 + Q^2}\right) 
\, , \nonumber \\ 
{\tilde c}&=&1  
\,. 
\end{eqnarray}

\begin{figure}%[!h]
\begin{center}
\includegraphics[scale=1.0]{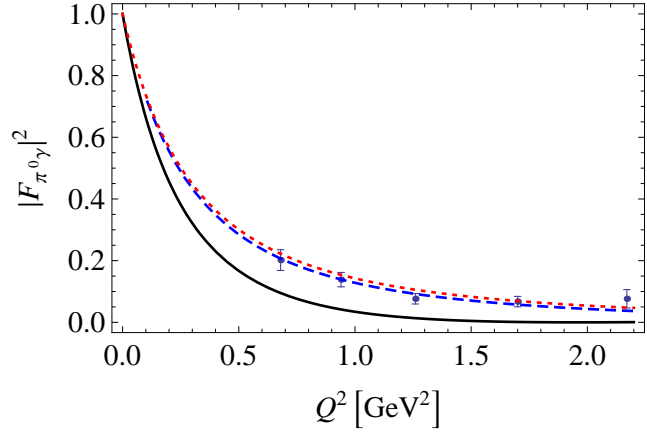} 
\vspace{10pt}  
\caption{The prediction (black solid curve) of the $\pi^0$-$\gamma$ transition form factor 
$F_{\pi^0 \gamma}$ with respect to space-like momentum-squared $Q^2$. 
Comparison with the experimental data~\cite{Behrend:1990sr} yields 
$\chi^2/{\rm d.o.f} = 63/5 =13$.  
The blue dashed and red dotted curves respectively correspond to the best fit curve with 
${\tilde c}_{\rm best}=1.03$ ($\chi^2/{\rm d.o.f}= 3.0/4 = 0.7$) and 
the $\rho/\omega$ meson dominance with ${\tilde c}=1$ ($\chi^2/{\rm d.o.f}= 4.8/5 =1.0$).  
}
\label{pi0gfig} 
\end{center}
\end{figure}

We shall now evaluate the parameters 
in Eqs.(\ref{pi2g:form2}) and (\ref{Fpig:form}) 
in the SS model. 
  Using Eqs.(\ref{g}), (\ref{z3}), (\ref{c3}), and (\ref{c4}), we calculate 
$A^{\pi 2\gamma}$, $B^{\pi 2\gamma}$, and $C^{\pi 2\gamma}$ to get    
\begin{eqnarray} 
A^{\pi 2\gamma}_{_{\rm SS}} 
&=& 1 - \left[ 
\frac{\langle  \psi_1 \rangle \langle \langle \dot{\psi}_0 \psi_1 \rangle \rangle}
{\langle \psi_1^2 \rangle} 
- \frac{1}{2} 
\frac{\langle  \psi_1 \rangle^2 \langle \langle \dot{\psi}_0 \psi_1^2   \rangle \rangle}
{\langle \psi_1^2 \rangle^2} 
\right] 
\simeq  1-(0.61) = 0.39 
\,, \label{Api2g} \\ 
B^{\pi 2\gamma}_{_{\rm SS}}  
&=&   
\frac{\langle \psi_1 \rangle \langle \langle \dot{\psi}_0 \psi_1  \rangle \rangle}
{\langle \psi_1^2 \rangle} 
- 
\frac{\langle \psi_1 \rangle^2  \langle \langle \dot{\psi}_0 \psi_1^2  \rangle \rangle}
{\langle \psi_1^2 \rangle^2} 
\simeq - 0.09
\,, \label{Bpi2g} \\ 
C^{\pi 2\gamma}_{_{\rm SS}}  
&=& \frac{1}{2}\left[ 
\frac{\langle \psi_1 \rangle^2 \langle  \langle \dot{\psi}_0 \psi_1^2  \rangle \rangle}
{\langle \psi_1^2 \rangle^2} 
\right] 
\simeq  0.50
\,. \label{Cpi2g}  
\end{eqnarray} 
   By using these values and Eq.(\ref{cbar}), the value of ${\tilde c}$ of the SS model 
is calculated as  
\begin{equation} 
{\tilde c}_{_{\rm SS}} \simeq 1.31  
\,, \label{cbar:SS}
\end{equation} 
which implies that the form factor $ F_{\pi^0 \gamma}$ in the SS model 
violates (about 30\%) the $\rho/\omega$ meson dominance.  
 Putting the value in Eq.(\ref{cbar:SS}) into Eq.(\ref{Fpig:form}), we 
 evaluate momentum dependence of $F_{\pi^0\gamma}$ as a definite prediction of the SS model 
 for the $\pi^0$-$\gamma$ transition form factor. 
 See Fig.~\ref{pi0gfig} (black solid curve). 
Here use has been made of the experimental inputs of 
 $m_\rho$ and $m_\omega$, $m_\rho=775$ MeV and $m_\omega =783$ MeV~\cite{Amsler:2008zz}.  
The experimental data are from Ref.~\cite{Behrend:1990sr} 
which is the only experiment in the space-like region~\footnote{
The experiment of Ref.~\cite{Behrend:1990sr} yields 
the linear coefficient of $F_{\pi^0 \gamma}$  
consistent with the current average value~\cite{Amsler:2008zz}, 
${\rm a}|_{\rm Ave.} = 0.032 \pm 0.004$, within $1\sigma$ error.}. 
 Figure~\ref{pi0gfig} shows that the momentum dependence of $ F_{\pi^0 \gamma}$ in the SS model disagrees with the experiment 
($\chi^2/{\rm d.o.f} = 63/5 = 13$).

 For a comparison, in Fig.~\ref{pi0gfig} we plot a curve drawn by a blue dashed line 
obtained by fitting the parameter $\tilde c$ of the general HLS model 
to the experimental data, 
which yields the best fit value of $\tilde{c}$, $\tilde{c}|_{\rm best} = 1.03$ 
($\chi^2/{\rm d.o.f} =3/4 = 0.7$). 
This best fit value is very close to ${\tilde c}=1$ of 
the $\rho/\omega$ meson dominance (red dotted curve in the figure).

\subsubsection{$\omega$-$\pi^0$-$ \gamma^*$ vertex function and $\omega$-$\pi^0$ transition form factor}  
\label{opig:sec}

  We begin with an expression of effective $\omega$-$\pi^0$-$ \gamma^*$ vertex function 
  obtained from the general HLS Lagrangian (Eqs.(\ref{HLS Lag}) and (\ref{IPodd:int})): 
\begin{eqnarray} 
\Gamma^{\mu\nu} [\omega_\mu(p), \pi^0, \gamma^*_\nu (k)] 
&=& 
\frac{e N_c}{8 \pi^2 F_\pi} 
\epsilon^{\mu\nu\alpha \beta} p_{\alpha} k_{\beta} 
\cdot g 
\left[ 
 \left( \frac{c_4-c_3}{2} \right) + c_3 \cdot 
D_\rho(k^2) \left( 1 - g^2 z_3 \frac{k^2}{m_\rho^2} \right) 
\right]
\,, 
\label{opig:form1}  
\end{eqnarray} 
where $p$ and $k$ respectively denote incoming four-momentum of $\omega$ and 
outgoing momentum of $\gamma^*$.  
Using the identity (\ref{ident:1}), we rewrite 
Eq.(\ref{opig:form1}) into the following form: 
\begin{eqnarray} 
\Gamma^{\mu\nu} [\omega_\mu(p), \pi^0, \gamma^*_\nu (k)] 
&=& 
\frac{e N_c}{8 \pi^2 F_\pi} 
\epsilon^{\mu\nu\alpha \beta} p_{\alpha} k_{\beta} 
\cdot \left[ 
A^{\omega\pi \gamma} + B^{\omega \pi\gamma}  D_\rho(k^2) 
\right]
\,, 
\label{opig:form2} 
\end{eqnarray} 
 where 
\begin{eqnarray} 
A^{\omega\pi \gamma} 
&=& 
\frac{1}{2} g \left[ (c_4 + c_3 \cdot g^2 z_3)  - c_3 (1-g^2z_3) \right] 
\,, \\ 
B^{\omega\pi \gamma} 
&=& 
g c_3 (1-g^2 z_3) 
\,. 
\end{eqnarray} 
Note that $(A^{\omega \pi \gamma} + B^{\omega \pi \gamma})$ is 
not constrained, which is consistent with the fact that 
there is no low-energy theorem for Eq.(\ref{opig:form2}) 
in the low-energy limit:  
 \begin{eqnarray} 
\Gamma^{\mu\nu} [\omega_\mu(p), \pi^0, \gamma^*_\nu (k)] 
&\stackrel{k^2 \to 0}{\longrightarrow}& 
\frac{e N_c}{8 \pi^2 F_\pi} 
\epsilon^{\mu\nu\alpha \beta} p_{\alpha} k_{\beta} 
\cdot \left[ 
A^{\omega\pi \gamma} + B^{\omega \pi\gamma} 
\right]
\,,  
\end{eqnarray} 
in contrast to Eq.(\ref{LET:pi2g}) for the $\pi^0$-$\gamma^*$-$\gamma^*$ vertex function.

The $\omega$-$\pi^0$ transition form factor $F_{\omega \pi^0}$ can be extracted 
from the $\omega \to \pi^0 l^+l^-$ decay width $\Gamma(\omega \to \pi^0 l^+ l^-)$ 
($l^\pm = e^\pm, \mu^\pm$) which is calculated through 
the effective $\omega$-$\pi^0$-$ \gamma^*$ vertex function as 
\begin{eqnarray} 
\Gamma(\omega \to \pi^0 l^+l^-) 
&=& 
  \int^{(m_\omega - m_{\pi^0})^2}_{4 m_l^2} dq^2 
  \frac{\alpha}{3\pi} \frac{\Gamma(\omega \to \pi^0 \gamma)}{q^2}
  \left(  1+ \frac{2 m_l^2}{q^2} \right) \sqrt{\frac{q^2-4 m_l^2}{q^2}} 
\nonumber \\ 
&& \times \left[ 
 \left(1 + \frac{q^2}{m_\omega^2-m_{\pi^0}^2} \right)^2 
 - \frac{4 m_\omega^2 q^2}{(m_\omega^2 - m_{\pi^0}^2)^2}
\right]^{3/2} \cdot |F_{\omega \pi^0} (q^2) |^2
\,, 
\end{eqnarray}  
where $\Gamma(\omega \to \pi^0 \gamma)$ denotes the $\omega \to \pi^0 \gamma$ decay width~\footnote{\label{foot} 
  The SS model predicts~\cite{Sakai:2005yt}  
   $g_{\omega \pi \gamma} = g_{\rho\pi\pi}$.  
   This leads to 
$\frac{\Gamma(\omega \to \pi^0 \gamma)}{\Gamma(\rho^0 \to \pi^+ \pi^-)} \Bigg|_{\rm SS} 
  = ( 5.37 \pm 0.03 ) \times 10^{-3}
$ which is compared with the experimental value~\cite{Amsler:2008zz} 
$  \frac{\Gamma(\omega \to \pi^0 \gamma)}{\Gamma(\rho^0 \to \pi^+ \pi^-)}
  \Bigg|_{\rm exp}   = ( 5.07 \pm 0.12 ) \times 10^{-3}$, 
although values for each decay width deviate (about 40\%) from the experimental values. }, 
\begin{equation} 
\Gamma(\omega \to \pi^0 \gamma) 
= 
\frac{3 \alpha}{64\pi^4 F_\pi^2} 
\, \, g_{\omega \pi \gamma}^2 \,  
\left(  
  \frac{m_\omega^2 - m_{\pi^0}^2 }{2 m_\omega}
\right)^3 
\,,  \label{decay:wpg}
\end{equation}
with  
\begin{equation} 
g_{\omega \pi \gamma} = A^{\omega \pi \gamma}+B^{\omega \pi \gamma} = \frac{g(c_3+c_4)}{2}
\,. \label{g:opig}
\end{equation}
The transition form factor $F_{\omega \pi^0}$ 
is then expressed as 
\begin{equation} 
  F_{\omega \pi^0} (q^2) 
= 
(1 - \tilde{r})
 + 
 \tilde{r} D_\rho(q^2) 
\,, \label{Fopi:HLS}
\end{equation}
where 
\begin{equation} 
\tilde{r} = 
\frac{B^{\omega \pi \gamma}}{A^{\omega \pi \gamma}+B^{\omega \pi \gamma}}
\,. \label{bar-r}
\end{equation} 
The $\rho$ meson dominance~\cite{GellMann:1962jt} for this transition form factor 
is defined in a way similar to Eq.(\ref{rhoD}) by taking $\tilde{r}=1$
as 
\begin{eqnarray} 
  F_{\omega \pi^0} (q^2) 
&=& \frac{m_\rho^2}{m_\rho^2 - q^2} 
\,, \nonumber \\ 
\tilde{r} &=& 1 
\,. 
\label{Fopi:rhoD} 
\end{eqnarray}

\begin{figure}%[!h]
\begin{center}
\includegraphics[scale=1.0]{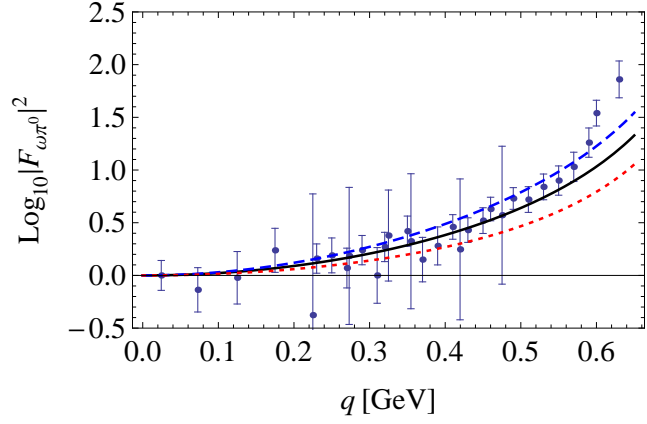} 
\vspace{10pt}  
\caption{The prediction (black solid curve) of the $\omega$-$\pi^0$ transition form factor 
$F_{\omega\pi^0}(q^2)$ with respect to time-like momentum $q$.  
Comparison with the experimental data~\cite{Akhmetshin:2005vy,:2009wb} yields 
$\chi^2/{\rm d.o.f} = 45/31 = 1.5$. 
The best fit curve with ${\tilde r}_{\rm best}=2.08$  ($\chi^2/{\rm d.o.f} = 24/30 =0.8$) 
and curve corresponding to the $\rho$ meson dominance with ${\tilde r}=1$ 
($\chi^2/{\rm d.o.f}= 124/31 = 4.0$) 
are drawn by blue dashed and red dotted lines, respectively. 
}
\label{opi0fig} 
\end{center}
\end{figure}

Let us now evaluate the parameter ${\tilde r}$ in Eq.(\ref{Fopi:HLS}) 
in the SS model.  
Using Eqs.(\ref{g}), (\ref{z3}), (\ref{c3}), and (\ref{c4}), 
we have 
\begin{eqnarray} 
 A_{_{\rm SS}}^{\omega\pi \gamma} 
&=& 
- \frac{1}{2\sqrt{N_c G}} 
\left[ 
\frac{\langle \langle \dot{\psi}_0 \psi_1 \rangle \rangle}{\sqrt{\langle \psi_1^2 \rangle}} 
- 
\frac{\langle  \psi_1 \rangle  \langle \langle \dot{\psi}_0 \psi_1^2 \rangle \rangle}{\langle \psi_1^2 \rangle^{3/2}} 
\right] 
\,, \label{Aopg:5d} \\ 
 B_{_{\rm SS}}^{\omega\pi \gamma} 
&=& 
- \frac{1}{2\sqrt{N_c G}} \left[ 
\frac{\langle \psi_1 \rangle  \langle \langle \dot{\psi}_0 \psi_1^2 \rangle \rangle}{\langle \psi_1^2 \rangle^{3/2}} 
\right] 
\,, \label{Bopg:5d}
\end{eqnarray}
and calculate $\tilde r$ to get   
\begin{equation} 
{\tilde r}_{_{\rm SS}} \simeq 1.53 
\,, 
\end{equation}   
which implies that 
the form factor $ F_{\omega \pi^0} $ in the SS model violates 
(about 50\%) the $\rho$ meson dominance with ${\tilde r}=1$. 
The predicted curve in time-like momentum region 
is shown in Fig.~\ref{opi0fig} as a black solid line  
together with the experimental data~\cite{Akhmetshin:2005vy,:2009wb}. 
 Figure.~\ref{opi0fig}  shows that the momentum dependence of 
$ F_{\omega \pi^0} $ in the SS model 
 is consistent with the experimental data ($\chi^2/{\rm d.o.f} = 45/31 = 1.5$).

 The predicted curve is compared with 
a blue dashed curve obtained by fitting the 
parameter $\tilde r$ of the general HLS model to the experimental data, 
which gives the best fit value of ${\tilde r}$, ${\tilde r}_{\rm best}=2.08$   
($\chi^2/{\rm d.o.f} = 24/30 =0.8$).  
Comparison with the $\rho$ meson dominance with ${\tilde r}=1$ (red dotted curve) is also given  
($\chi^2/{\rm d.o.f}= 124/31 = 4.0$).

\subsubsection{$\gamma^*$-$\pi^0$-$\pi^+$-$\pi^-$ vertex function} 
\label{pi3g:sec}

We start with an expression of effective  $\gamma^*$-$\pi^0$-$\pi^+$-$\pi^-$ vertex function 
  obtained from the general HLS Lagrangian (Eqs.(\ref{HLS Lag}) and (\ref{IPodd:int})):  
 \begin{eqnarray}  
&&
\Gamma_{\mu} [\gamma^*_\mu (p), \pi^0(q_0), \pi^+(q_+), \pi^-(q_-)] 
\nonumber \\ 
&=&
- \frac{e N_c}{12\pi^2 F_\pi^3} 
\epsilon_{\mu\nu\alpha \beta} q_0^\nu q_+^\alpha q_-^\beta  
\Bigg[
1 - \frac{3}{4} (c_1-c_2 + c_4) 
+ \frac{3}{4} (c_1-c_2-c_3) D_\omega(p^2) \left(1 - g^2 z_3 \frac{p^2}{m_\omega^2}\right) 
\nonumber \\ 
&& 
+ 
\left\{ 
 \frac{c_4-c_3}{4} + \frac{c_3}{2} D_\omega (p^2) \left(1 - g^2 z_3 \frac{p^2}{m_\omega^2}\right) 
\right\} 
\times 
\Bigg\{
D_\rho((q_++q_-)^2) \left( 1 - \frac{1}{2}g^2 z_4 \frac{(q_+ + q_-)^2}{m_\rho^2} \right) 
\nonumber \\ 
&& 
+ 
D_\rho((q_-+q_0)^2) \left( 1 - \frac{1}{2}g^2 z_4 \frac{(q_- + q_0)^2}{m_\rho^2} \right) 
+ 
D_\rho((q_0+q_+)^2) \left( 1 - \frac{1}{2}g^2 z_4 \frac{(q_0 + q_+)^2}{m_\rho^2} \right) 
\Bigg\}
\Bigg]
\,, 
\label{g3pi:form1}
\end{eqnarray} 
where $p$ and $q_{\pm,0}$ respectively stand for incoming four-momentum of $\gamma^*$ 
and outgoing four-momenta of $\pi^\pm$ and $\pi^0$. 
By using the identity (\ref{ident:1}), the expression (\ref{g3pi:form1}) may be 
rewritten as  
 \begin{eqnarray} 
&& 
\Gamma_{\mu} [\gamma^*_\mu (p), \pi^0(q_0), \pi^+(q_+), \pi^-(q_-)] 
\nonumber \\ 
&=& 
- \frac{e N_c}{12\pi^2 F_\pi^3} 
\epsilon_{\mu\nu\alpha \beta} q_0^\nu q_+^\alpha q_-^\beta  
\Bigg[
A^{\gamma 3\pi} + B^{\gamma 3\pi} \cdot D_\omega(p^2) 
+ 
\frac{C^{\gamma 3 \pi}}{3}  \cdot 
\Bigg\{  
D_\rho((q_++q_-)^2) 
+ 
D_\rho((q_-+q_0)^2) 
+ 
D_\rho((q_0+q_+)^2) 
\Bigg\}  
\nonumber \\ 
&& 
+ \frac{D^{\gamma 3\pi}}{3} \cdot 
 D_\omega(p^2) 
 \cdot 
 \Bigg\{ 
D_\rho((q_++q_-)^2) 
+ 
D_\rho((q_-+q_0)^2) 
+ 
D_\rho((q_0+q_+)^2) 
\Bigg\}  
\Bigg]
\,, 
\label{g3pi:form2}
\end{eqnarray} 
where 
\begin{eqnarray} 
 A^{\gamma 3\pi} 
 &=& 
 1 - \frac{3}{4} \Bigg[
(1-g^2 z_3) (c_1 - c_2 + c_3) 
- 2 c_3 (1-g^2 z_3) \left( 1 - \frac{1}{2} g^2 z_4 \right) 
+ \left(1 - \frac{1}{2} g^2 z_4 \right) (c_3+c_4) 
\Bigg] 
\,, \\
  B^{\gamma 3\pi} 
 &=& 
\frac{3}{4} \left[
(1-g^2 z_3) (c_1 - c_2 + c_3) 
- 2 c_3 (1-g^2 z_3) \left( 1 - \frac{1}{2} g^2 z_4 \right) 
\right] 
\,, \\
 C^{\gamma 3\pi} 
 &=& 
\frac{3}{4} \left[
- 2 c_3 (1-g^2 z_3) \left( 1 - \frac{1}{2} g^2 z_4 \right) 
+ \left(1 - \frac{1}{2} g^2 z_4 \right) (c_3+c_4) 
\right] 
\,, \\
 D^{\gamma 3\pi} 
 &=& 
\frac{3}{4} \left[ 
2 c_3 (1-g^2 z_3) \left( 1 - \frac{1}{2} g^2 z_4 \right) 
\right] 
\,. 
\end{eqnarray} 
Note that these parameters satisfy 
\begin{equation} 
 A^{\gamma 3 \pi} + B^{\gamma 3 \pi} + C^{\gamma 3 \pi} +  D^{\gamma 3 \pi} =1 
\label{LET:g3pi1}
\end{equation}
which reproduces the low-energy theorem:  
\begin{equation} 
  \Gamma_{\mu} [\gamma^*_\mu (p), \pi^0(q_0), \pi^+(q_+), \pi^-(q_-)] 
\stackrel{q_{\pm,0}^2 \to 0}{\longrightarrow} 
  - \frac{e N_c}{12\pi^2 F_\pi^3} 
\epsilon_{\mu\nu\alpha \beta} q_0^\nu q_+^\alpha q_-^\beta  
\,. \label{LET:g3pi} 
\end{equation} 
The $\rho/\omega$ meson dominance %~\cite{GellMann:1962jt} 
for this process is defined in a way similar to Eq.(\ref{rhoD}) by taking $A^{\gamma 3 \pi}=B^{\gamma 3 \pi}=C^{\gamma 3 \pi}=0$ 
($D^{\gamma 3 \pi}=1$) as 
 \begin{eqnarray} 
&& 
\Gamma_{\mu} [\gamma^*_\mu (p), \pi^0(q_0), \pi^+(q_+), \pi^-(q_-)] 
\nonumber \\ 
&=& 
- \frac{e N_c}{36 \pi^2 F_\pi^3} 
\epsilon_{\mu\nu\alpha \beta} q_0^\nu q_+^\alpha q_-^\beta  
\, \frac{m_\omega^2}{m_\omega^2 - p^2} 
 \left(   
\frac{m_\rho^2}{m_\rho^2 -  (q_++q_-)^2 }
+ 
(q_+ \leftrightarrow q_0) 
+ 
(q_- \leftrightarrow q_0)  
\right)    
\,. \label{rho-omega-D:g3pi}
\end{eqnarray}

We shall now evaluate the parameters in Eq.(\ref{g3pi:form2}) in the SS model. 
  Using Eqs.(\ref{g}), (\ref{z3}), (\ref{z4}), and (\ref{c1})-(\ref{c4}),  
we have 
\begin{eqnarray} 
 A_{_{\rm SS}}^{\gamma 3 \pi} 
 &=& 
 1 - \frac{3}{4} \Bigg[ 
 \frac{\langle  \psi_1 \rangle \langle \langle  \dot{\psi}_0 \psi_1 (1-\psi_0^2) \rangle \rangle}{\langle \psi_1^2 \rangle}
+ 
 \frac{\langle \langle \dot{\psi}_0 \psi_1 \rangle \rangle 
\langle \psi_1(1-\psi_0^2)  \rangle}{\langle \psi_1^2 \rangle}
- 
 \frac{\langle \langle \dot{\psi}_0 \psi_1^2 \rangle \rangle 
\langle \psi_1 \rangle \langle  \psi_1(1-\psi_0^2)  \rangle}
{\langle \psi_1^2 \rangle^2} 
 \Bigg] 
 \,, \label{Ag3pi} \\ 
  B_{_{\rm SS}}^{\gamma 3 \pi} 
 &=& 
\frac{3}{4} \Bigg[ 
 \frac{\langle  \psi_1 \rangle 
\langle \langle  \dot{\psi}_0 \psi_1 (1-\psi_0^2)  \rangle\rangle}
{\langle  \psi_1^2 \rangle}
- 
 \frac{\langle \langle \dot{\psi}_0 \psi_1^2  \rangle \rangle 
\langle \psi_1 \rangle 
\langle  \psi_1(1-\psi_0^2)  \rangle}
{\langle \psi_1^2 \rangle^2} 
 \Bigg] 
 \,, \label{Bg3pi} \\  
C_{_{\rm SS}}^{\gamma 3 \pi} 
 &=& 
\frac{3}{4} \Bigg[ 
 \frac{\langle \langle \dot{\psi}_0 \psi_1  \rangle \rangle 
\langle \psi_1(1-\psi_0^2)  \rangle}
{\langle \psi_1^2 \rangle}
- 
 \frac{\langle \langle \dot{\psi}_0 \psi_1^2  \rangle \rangle 
\langle \psi_1 \rangle 
\langle \psi_1(1-\psi_0^2)  \rangle}
{\langle \psi_1^2 \rangle^2} 
 \Bigg] 
 \,, \label{Cg3pi}\\  
D_{_{\rm SS}}^{\gamma 3 \pi} 
&=& 
\frac{3}{4} \left[ 
  \frac{\langle \langle \dot{\psi}_0 \psi_1^2 \rangle \rangle 
\langle \psi_1 \rangle 
\langle \psi_1(1-\psi_0^2)  \rangle}
{\langle \psi_1^2 \rangle^2} 
\right]  
\,. \label{Dg3pi}
\end{eqnarray}
They are calculated in the SS model as
\begin{eqnarray} 
 A_{_{\rm SS}}^{\gamma 3\pi} &\simeq & 1- (1.03)   
= -0.03
\,, \nonumber  \\ 
 B_{_{\rm SS}}^{\gamma 3\pi} &\simeq&  
0.04
\,, \nonumber \\  
 C_{_{\rm SS}}^{\gamma 3\pi} &\simeq& 
-0.51 
\,, \nonumber \\ 
D_{_{\rm SS}}^{\gamma 3\pi} &\simeq&  
1.50
\,, \label{A-Dvals:g3pi}
\end{eqnarray}
in which $D_{_{\rm SS}}^{\gamma 3 \pi} \neq 1$ 
implies that the  effective $\gamma^*$-$\pi^0$-$\pi^+$-$\pi^-$ vertex function 
in the SS model violates (about 50\%) the $\rho/\omega$ meson dominance.

\subsubsection{$\omega \to \pi^0 \pi^+ \pi^-$ decay} 
\label{o3pi:sec}

The $\omega \to \pi^0 \pi^+ \pi^-$ decay width 
is given by 
\begin{equation} 
 \Gamma(\omega \to \pi^0 \pi^+ \pi^-) 
 = \frac{m_\omega}{192 \pi^3} \int\int dE_+ dE_- 
\left(   |\vec{q}_+|^2 |\vec{q}_- |^2 - (\vec{q}_+ \cdot \vec{q}_- )^2   \right) 
|  F_{\omega \to 3\pi}  |^2 
\,, \label{width:w3pi}
\end{equation}
where $E_\pm$ and $\vec{q}_\pm$ are respectively energies and three-momenta of $\pi^\pm$ 
in the rest frame of $\omega$. 
 We construct the $\omega \to 3 \pi$ form factor $F_{\omega \to 3\pi}$ from the general HLS 
Lagrangian (Eqs.(\ref{HLS Lag}) and (\ref{IPodd:int})): 
\begin{eqnarray} 
 F_{\omega \to 3\pi} 
&=& - \frac{N_c }{4\pi^2 F_\pi m_\rho^2} 
\Bigg[ 
 \frac{3 ag }{4} g^2(c_1-c_2-c_3)  
+ \frac{ag^3}{2} c_3 \Bigg\{  
\left( 1 - \frac{g^2 z_4}{2} \frac{(q_+ + q_-)^2}{m_\rho^2} \right)
\cdot 
D_\rho ((q_+ + q_-)^2) 
\nonumber \\ 
&& 
+ 
( q_+ \leftrightarrow q_0) 
+ (q_- \leftrightarrow q_0) 
\Bigg\} 
\Bigg]
\,, \label{Fw3pi:1}
\end{eqnarray} 
where  $q_{0, \pm}$ are four-momenta of $\pi^{0,\pm}$, respectively. 
Using the identity in Eq.(\ref{ident:1})  
we may rewrite $F_{\omega \to 3 \pi}$ into the following form: 
\begin{eqnarray} 
 F_{\omega \to 3\pi} 
 = - \frac{N_c }{4\pi^2 F_\pi m_\rho^2} g_{\rho\pi\pi} 
\Bigg[A^{\omega 3 \pi}  + \frac{B^{\omega 3 \pi}}{3}  \Bigg\{  D_\rho ((q_+ + q_-)^2) 
+ ( q_+ \leftrightarrow q_0) 
+ (q_- \leftrightarrow q_0) 
\Bigg\} 
\Bigg]
\,, \label{o3pi:HLS}
\end{eqnarray} 
where  
\begin{eqnarray} 
A^{\omega 3 \pi} 
&=& 
 \frac{3}{2}  \frac{g^2(c_1-c_2+c_3)}{1-g^2z_4/2}  
 - 3 g^2 c_3 
 \,, \\ 
 B^{\omega 3 \pi} 
&= & 
3 g^2 c_3 
 \,. \label{B}
\end{eqnarray} 
Note that $(A^{\omega 3 \pi}+ B^{\omega 3 \pi})$ is not constrained because there is no low-energy theorem 
   for Eq.(\ref{o3pi:HLS}) in the low-energy limit: 
\begin{eqnarray} 
 F_{\omega \to 3\pi} %\Bigg|_{\rm HLS}
\stackrel{q^2_{\pm,0} \to 0}{\longrightarrow}  
 - \frac{N_c }{4\pi^2 F_\pi m_\rho^2} g_{\rho\pi\pi} 
\left( A^{\omega 3 \pi}  + B^{\omega 3 \pi} \right)
\,, %\label{o3pi:HLS}
\end{eqnarray} 
in contrast to Eqs.(\ref{LET:pi2g}) and (\ref{LET:g3pi}) for the $\pi^0$-$\gamma^*$-$\gamma^*$ and 
$\gamma^*$-$\pi^0$-$\pi^+$-$\pi^-$ vertex functions, respectively.  
The $\rho$ meson dominance~\cite{GellMann:1962jt} for this process is defined 
in a way similar to Eq.(\ref{rhoD}) by taking $A^{\omega 3 \pi}=0$ in the form factor $F_{\omega \to 3 \pi}$ 
as 
\begin{eqnarray} 
 F_{\omega \to 3\pi} 
 = - \frac{N_c }{12\pi^2 F_\pi m_\rho^2} g_{\rho\pi\pi} \cdot B^{\omega 3 \pi} 
\cdot \left[ 
\frac{m_\rho^2}{m_\rho^2 - (q_+ + q_-)^2} 
+ ( q_+ \leftrightarrow q_0) 
+ (q_- \leftrightarrow q_0) 
\right] 
\,.
\end{eqnarray}

  We shall evaluate the parameters in Eq.(\ref{o3pi:HLS}) in the SS model. 
 Using Eqs.(\ref{g}), (\ref{z4}), (\ref{c1}), (\ref{c2}), and (\ref{c3}), 
 we have
\begin{eqnarray} 
A_{_{\rm SS}}^{\omega 3 \pi}
&=& 
\frac{3}{2N_c G}\left( 
\frac{ \langle \langle \psi_1 \dot{\psi}_0 (1-\psi_0^2) \rangle \rangle}
{ \langle \psi_1 (1-\psi_0^2) \rangle }
- 
\frac{ \langle \langle \dot{\psi}_0 \psi_1^2 \rangle \rangle}
{ \langle \psi_1^2 \rangle}
\right) 
\,, \label{A5dim}\\ 
B_{_{\rm SS}}^{\omega 3 \pi}
&=&  
\frac{3}{2N_c G } \frac{ \langle \langle \dot{\psi}_0 \psi_1^2 \rangle\rangle}
{\langle \psi_1^2 \rangle}
\,, \label{B5dim}
\end{eqnarray}
which are calculated in the SS model as   
\begin{eqnarray} 
  A_{_{\rm SS}}^{\omega 3 \pi}&\simeq & 2.31   
\,, \nonumber \\ 
  B_{_{\rm SS}}^{\omega 3 \pi}&\simeq & 81.01
\,, \label{val:AandB}  
\end{eqnarray}
where we used the experimental inputs of $m_\rho$ and $F_\pi$, $m_\rho=775$ MeV and $F_\pi = 92.4$ MeV~\cite{Amsler:2008zz},   
to determine the value of $(N_cG)$ ($N_c G \simeq 0.01$). 
An amount of deviation from the $\rho$ meson dominance 
is estimated independently of value of $(N_cG)$ as  
\begin{equation} 
  \frac{B_{_{\rm SS}}^{\omega 3 \pi}}{A_{_{\rm SS}}^{\omega 3 \pi} + B_{_{\rm SS}}^{\omega 3 \pi}} \simeq 0.97 
  \,, 
\end{equation}
which implies that the form factor $F_{\omega \to 3 \pi}$ in the SS model is well 
approximated by the $\rho$ meson dominance.

Let us now calculate the decay width $\Gamma(\omega \to \pi^0 \pi^+\pi^-)$ in the SS model. 
To do this, we first estimate the value of $g_{\rho\pi\pi}$ which appears in Eq.(\ref{o3pi:HLS}) 
as the overall coefficient. 
    Using Eq.(\ref{ksrf2}) and 
the experimental values of $m_\rho$ and $F_\pi$, we get 
$g_{\rho \pi\pi}|_{\rm SS} \simeq 4.84 $. 
  Second, we evaluate the phase space integral using experimental inputs for 
  $m_{\pi^{\pm,0}}$ and $m_\omega$, $m_{\pi^\pm} = 140$ MeV, $m_{\pi^0} = 135$ MeV and $m_\omega = 783$ MeV~\cite{Amsler:2008zz}. 
Thus we obtain 
\begin{eqnarray} 
 \Gamma(\omega \to \pi^0 \pi^+ \pi^-) \Bigg|_{\rm SS}
&\simeq & 
2.78 
\, {\rm MeV} 
\,. \label{SS:omega3pi}
\end{eqnarray} 
This is the first full result obtained by our method which includes 
effects from {\it infinite} tower of higher KK modes. 
The result is compared with the experimental value~\cite{Amsler:2008zz}
$ \Gamma(\omega \to \pi^0\pi^+\pi^-) |_{\rm exp}  = 7.57 \pm 0.09 \, {\rm MeV}$.

  It is interesting to note that the estimate in Eq.(\ref{SS:omega3pi}) 
is different by about 7\% from the value obtained in 
Ref.~\cite{Sakai:2005yt} 
where higher KK modes of the HLS gauge bosons are truncated at the
level of $n=4$:   
\begin{equation} 
 \Gamma(\omega \to \pi^0 \pi^+ \pi^-) \Bigg|^{{\rm trun}\,(n\le4)}_{\rm SS} \simeq 2.58 \, {\rm MeV}  
\,.  
 \end{equation} 
In order to study this difference, 
let us discuss the original form of the form factor~\cite{Sakai:2005yt}: 
\begin{eqnarray} 
 F_{\omega \to 3 \pi} \Bigg|_{\rm SS} 
= - \frac{N_c}{4\pi^2 F_\pi} 
\sum_{k=1}^\infty 
\left[ 
\frac{g_{\omega \rho_k \pi} g_{\rho_k \pi\pi}}{m_{\rho_k}^2 - (q_+ + q_-)^2}
+ (q_+ \to q_0) + (q_- \to q_0) 
\right] 
\,. \label{o3pi:F:SS}
\end{eqnarray}
To be consistent with our method which integrates out higher KK modes into ${\cal O}(p^4)$ terms of the general HLS Lagrangian,  
we expand Eq.(\ref{o3pi:F:SS}) as 
\begin{eqnarray} 
 F_{\omega \to 3 \pi} \Bigg|_{\rm SS} 
&= & - \frac{N_c}{4\pi^2 F_\pi m_\rho^2}  g_{\rho \pi\pi}
\Bigg[ 
\left( 
3 \frac{m_\rho^2}{g_{\rho\pi\pi}} 
\sum_{k=2}^\infty 
\frac{g_{\omega \rho_k \pi} g_{\rho_k \pi\pi}}{m_{\rho_k}^2}
\right) 
\nonumber \\ 
&& 
\hspace{100pt}
+ 
\frac{1}{3} \left( 3g_{\omega \rho \pi} \right)
\{ D_\rho((q_+ + q_-)^2) 
+ (q_+ \to q_0) + (q_- \to q_0) 
\} 
\Bigg] 
\,, \label{o3pi:SS:expand}
\end{eqnarray} 
up to ${\cal O}(q_{\pm,0}^2/m_{\rho_k}^2)$ ($k \ge 2$) which corresponds to terms higher than ${\cal O}(p^4)$ in the Lagrangian. 
The coefficient of the second term, $g_{\omega \rho \pi}$, 
is read off from Ref.~\cite{Sakai:2005yt} as  
\begin{equation} 
 g_{\omega \rho \pi} 
= \frac{1}{2N_c G } \frac{\langle \langle \dot{\psi_0} \psi_1^2 \rangle \rangle}{\langle \psi_1^2  \rangle}  
\,. \label{gorhopi:5d}
\end{equation}
Note that the expression of ($3g_{\omega \rho \pi}$)  
is exactly the same as that of $B_{_{\rm SS}}^{\omega 3 \pi }$ in Eq.(\ref{B5dim}). 
 We may therefore write  
\begin{eqnarray} 
 F_{\omega \to 3\pi} \Bigg|_{\rm SS}
 &=& - \frac{N_c }{4\pi^2 F_\pi m_\rho^2} g_{\rho\pi\pi} 
\Bigg[
\left( 
3 \frac{m_\rho^2}{g_{\rho\pi\pi}} 
\sum_{k=2}^\infty 
\frac{g_{\omega \rho_k \pi} g_{\rho_k \pi\pi}}{m_{\rho_k}^2}
\right) 
\nonumber \\ 
&&
 + \frac{B_{_{\rm SS}}^{\omega 3 \pi}}{3}  \Bigg\{  D_\rho ((q_+ + q_-)^2) 
+ ( q_+ \leftrightarrow q_0) 
+ (q_- \leftrightarrow q_0) 
\Bigg\} 
\Bigg]
\,. \label{o3pi:SS}
\end{eqnarray}  
Identifying the first term of Eq.(\ref{o3pi:SS}) with $A^{\omega 3 \pi}_{_{\rm SS}}$ in Eq.(\ref{A5dim}), 
\begin{eqnarray}
3 \frac{m_\rho^2}{g_{\rho\pi\pi}} 
\sum_{k=2}^\infty 
\frac{g_{\omega \rho_k \pi} g_{\rho_k \pi\pi}}{m_{\rho_k}^2} 
&=& 
\frac{3}{2N_c G}\left( 
\frac{ \langle \langle \psi_1 \dot{\psi}_0 (1-\psi_0^2) \rangle \rangle}
{ \langle \psi_1 (1-\psi_0^2) \rangle }
- 
\frac{ \langle \langle \dot{\psi}_0 \psi_1^2 \rangle \rangle}
{ \langle \psi_1^2 \rangle}
\right) 
\,, 
\nonumber \\ 
&=& 
\frac{3}{2N_c G}\left( 
\frac{ \langle \langle \psi_1 \dot{\psi}_0 (1-\psi_0^2) \rangle \rangle}
{ \langle \psi_1 (1-\psi_0^2) \rangle } 
\right) 
- 
3 g_{\omega \rho \pi} 
\,,  
\end{eqnarray}
and using the expression of $g_{\rho\pi\pi}$ in 
Eq.(\ref{grhopipi:5d}) and that of $g_{\omega \rho \pi}$ in Eq.(\ref{gorhopi:5d}),  
we may read off %the infinite sum in Eq.(\ref{S}) to get the result 
\begin{equation} 
 \sum_{k=1}^\infty 
\frac{g_{\omega \rho_k \pi} g_{\rho_k \pi\pi}}{m_{\rho_k}^2}
= 
\frac{1}{2 m_\rho^2 (N_c G)^{3/2}}
\frac{\langle \langle \psi_0 \dot{\psi}_1  \rangle \rangle 
\langle \langle  \psi_1 \dot{\psi}_0  (1 - \psi_0^2) \rangle \rangle}
{\langle \psi_1 \rangle^{1/2} \langle  \psi_1 (1-\psi_0^2)  \rangle}
\,. \label{newSR}
\end{equation}
This is a new sum rule which was not obtained
in Ref.~\cite{Sakai:2005yt}.  
This sum rule shows that the form factor (\ref{o3pi:SS}) 
includes effects of {\it full set} of the infinite tower of the vector mesons.
In contrast, 
in Ref.~\cite{Sakai:2005yt} some parts of the contributions are examined 
by naively truncating the infinite tower as in Eq.(\ref{action:naive}).

\section{Summary and discussion} 
\label{summary}

In this paper, we developed our method of integrating out 
higher KK modes of the HLS gauge bosons identified as
vector and axialvector mesons 
in a class of HQCD models including the SS model. 
Our method is to integrate out higher KK modes 
through their equations of motion for the HLS gauge bosons
$A_\mu^{(m)}$ and $V_\mu^{(n)}$ in Eq.(\ref{sol}):
$A_\mu^{(m)}=0$ and 
$V_\mu^{(n)} = \alpha_{\mu ||}$.
Thus the higher vector mesons are replaced by
$\alpha_{\mu ||}= \frac{i}{2F_\pi^2} 
[\pi, \partial_\mu \pi] + \cdots$ (``pion cloud'')
which generates the ${\cal O}(p^4)$ terms as well as the ${\cal O}(p^2)$ terms 
of the HLS Lagrangian.  
Since $\alpha_{\mu ||}$ keeps the same HLS transformation property 
as that of the fields of the integrated-out KK modes, our method 
is manifestly invariant under the HLS and chiral symmetry including 
the external gauge symmetry.  
On the contrary, 
a naive truncation corresponds to simply putting fields of higher KK modes to be zero  
which does not reproduce the correct transformation property as shown in Eq.(\ref{Amu:trunc:trans}), and hence violates the HLS and the external gauge symmetry.

Given a concrete HQCD not restricted to the SS model, 
our method enables us to deduce definite 
predictions for any physical quantity which can always be 
written in terms of the parameters of the general HLS model, thus 
can be compared with experimental data once those parameters are 
determined from the HQCD.

 To show the power of our method,  
we took the SS model as an example. 
The SS model is thought to be valid only below the $M_{KK}$ scale, 
so higher-KK (mass eigenstate) fields should not contribute in the low-energy physics. 
 In our integrating out method  this was reflected by setting higher mass eigenstate 
 fields $B_\mu^{(n)}=0$ through the equations of motion: 
In terms of  the HLS basis ($A_\mu^{(m)}$ and $V_\mu^{(n)}$) are no longer independent degrees of freedom 
but simply generate ${\cal O}(p^4)$ terms and modify ${\cal O}(p^2)$ terms as well. 
We presented a full set of the ${\cal O}(p^4)$ terms of 
the HLS Lagrangian computed from the DBI part and the CS part 
at the leading order of $1/N_c$ expansion. 
Once the parameters of the HLS model are determined by the SS model, 
we can compute the form factors which are always given in the general framework of the HLS model.  
The EM gauge invariance and the chiral invariance are automatically maintained since our method is manifestly invariant under the external gauge symmetry as well as the HLS. 
The result of the pion EM form factor was 
compared with the experimental data together with 
the best fit within the general HLS model and 
the result of the $\rho$ meson dominance (See Fig.~\ref{Fv-holo}). 
It turned out that the SS model agrees with the experiment.

In the same fashion, 
we evaluated the $\pi^0$-$\gamma$ (Fig.\ref{pi0gfig}) and $\omega$-$\pi^0$ (Fig.~\ref{opi0fig}) transition form factors, which were compared with experimental data 
together with the best fit within the general HLS model and 
the result of the $\rho/\omega$ meson dominance. 
It turned out that in the SS model  
the $\pi^0$-$\gamma$ transition form factor disagrees with the experimental data, 
while the $\omega$-$\pi^0$ transition form factor is consistent with the data. 
We also presented the results for 
the related quantities such as $\gamma^*$-$\pi^0$-$\pi^+$-$\pi^-$ and 
$\omega$-$\pi^0$-$\pi^+$-$\pi^-$ vertex functions.

We further derived the same form factors 
by a different method dealing with the infinite sum explicitly without using the general HLS Lagrangian.  
This confirms that our formulation correctly includes contributions from infinite set of 
higher KK modes and that infinite sum is crucial for the gauge invariance. 
Actually, the EM gauge symmetry and chiral symmetry (low-energy theorem) in the 
 form factors are obviously violated by a naive truncation simply neglecting higher KK modes instead of taking the infinite sum.

Our method was used to deduce predictions of the SS model which were not 
available before. 
Summarizing the SS model prediction to be compared with the experiment:

\begin{itemize} 
\item[(I)] 
The pion EM form factor (Fig.~\ref{Fv-holo}) agrees with 
the experiment ($\chi^2/{\rm d.o.f} = 147/53 = 2.8$)    
compared with the best fit of the general HLS model 
($\chi^2/{\rm d.o.f} = 81/51 = 1.6$) 
and the $\rho$ meson dominance 
($\chi^2/{\rm d.o.f} = 226/53 = 4.3$). 

\item[(II)] 
The $\pi^0$-$\gamma$ transition form factor (Fig.~\ref{pi0gfig}) disagrees with the 
experiment ($\chi^2/{\rm d.o.f} = 63/5 = 13$) 
compared with the best fit of the general HLS model 
($\chi^2/{\rm d.o.f} = 3/4 = 0.7$) 
and the $\rho/\omega$ meson dominance 
($\chi^2/{\rm d.o.f} = 4.8/5 = 1.0$).

\item[(III)]
 
The $\omega$-$\pi^0$ transition form factor (Fig.~\ref{opi0fig}) 
is consistent with the experiment ($\chi^2/{\rm d.o.f} = 45/31=1.5$) 
compared with the best fit of the general HLS model 
($\chi^2/{\rm d.o.f} = 24/30 = 0.8$) 
and the $\rho$ meson dominance 
($\chi^2/{\rm d.o.f} = 124/31 = 4.0$).

\end{itemize} 
The item (I) implies no obvious 
need for corrections 
as to space-like momentum region, 
while in the time-like region the DBI part of the SS model yields for the KSRF I and II~\cite{Sakai:2005yt},  
\begin{eqnarray}  
\frac{g_{\rho}}{2 g_{\rho \pi\pi} F_\pi^2} \Bigg|_{\rm SS} &\simeq& 2.0 
\qquad 
\left(\frac{g_{\rho}}{2 g_{\rho \pi\pi} F_\pi^2} \Bigg|_{\rm exp} \simeq 1.0 \right) 
\nonumber \\  
\frac{m_\rho^2}{g_{\rho\pi\pi}^2 F_\pi^2} \Bigg|_{\rm SS} &\simeq& 3.0 
\qquad  
\left(\frac{m_\rho^2}{g_{\rho\pi\pi}^2 F_\pi^2} \Bigg|_{\rm exp} \simeq 2.0 \right) 
\, ,  
\end{eqnarray}
which would need some corrections such as $1/N_c$ subleading corrections~\cite{Harada:2006di}. 
The item (II) implies that the SS model would need corrections for the CS term 
which may arise as $1/N_c$ subleading terms. 
Our formulation can also be used to test other HQCD and 
suggest possible corrections.

Throughout this paper, we confined ourselves to the leading order in 
the $1/N_c$ expansion. 
We demonstrated that as far as the $1/N_c$-leading order form factors  
are concerned, the same results as those of our method can also be 
obtained by other methods using sum rules for infinite sum of KK modes
instead of the HLS Lagrangian.  
As far as the tree level is concerned,
our method setting fields of higher mass eigenstates $B_\mu^{(n)}=0$ 
as in Eq.~(\ref{sol}) in the HLS basis
is obviously equivalent to 
the same setting in the Lagrangian 
written in terms of $B_\mu^{(n)}$~\cite{Nawa:2006gv}
without explicit use of the HLS gauge basis. 
However, in our method based on the HLS formalism,
the systematic chiral perturbation 
can straightforwardly incorporate the $1/N_c$ subleading 
effects through loop calculations. 
  Further studies along this line will be done in future.

Our focus in this paper has been on the $1/N_c$-leading action 
and its derivative expansion. 
There could be another source, which affects coefficients of 
${\cal O}(p^4)$ terms, arising from $1/{\lambda}$ expansion.  
Further development of our method 
incorporating such another source will be pursued in future.

In the end, we emphasize that our formulation  
can be applicable to several types of HQCD models~\cite{Erlich:2005qh,Da Rold:2005zs,rev} 
and models including baryons~\cite{Nawa:2006gv,baryons}.
It will also be interesting to apply our method to
HQCD models in hot and/or dense matter~\cite{hotdenseQCD},
 and furthermore 
so-called holographic (walking) technicolor models~\cite{HTC} and
Higgsless models~\cite{Belyaev:2009ve}.

\section*{Acknowledgments}

We would like to thank D. K. Hong, 
T. Kugo, M. Rho, T. Sakai, S. Sugimoto and H. U. Yee 
for useful comments and fruitful discussions. 
This work was supported in part 
by the Grant-in-Aid for Nagoya University Global COE Program, ``Quest
for Fundamental Principles in the Universe: from Particles to the
Solar System and the Cosmos'', from the 
Ministry of Education, Culture, Sports, Science and Technology of
Japan 
and the JSPS Grant-in-Aid for Scientific Research (S) \#22224003 
(K.Y. and M.H.), 
M.H. was supported in part by 
the JSPS Grant-in-Aid for Scientific Research (c) 20540262 and
Grant-in-Aid for Scientific Research on Innovative Areas (No. 2104) 
``Quest on New Hadrons with Variety of Flavors'' from MEXT. 
S.M. was supported by the Korea Research 
Foundation Grant funded by the Korean Government
(KRF-2008-341-C00008).

\appendix 
\renewcommand\theequation{\Alph{section}.\arabic{equation}}

\section{HLS-gauge invariance of $\Gamma_3$} 
 \label{proof}

In this section we give a proof for  
the HLS-gauge invariance of the $\Gamma_3$ term 
in Eq.(\ref{123}).

 We begin by decomposing the five-dimensional gauge field 
$A=A_Mdx^M$ ($M=\mu, z$)~\footnote{
In this section we take $A_M$ to be anti-hermitian.   
} in Eq.(\ref{Amuxz:2}) into two parts including infinite tower of 
vector and axialvector meson fields: 
\begin{eqnarray} 
A &=& v +a 
\,,  \\ 
v &=& v_0 + \tilde{v} = \alpha_{||} + \sum_n^\infty \hat{\alpha}^{(n)}_{||} \psi_{2n-1}  
\,, \\ 
a &=& a_0 + \tilde{a} = \alpha_{\perp} \psi_0 + \sum^\infty_n A^{(n)} \psi_{2n}
\,,  
\end{eqnarray} 
where $\hat{\alpha}^{(n)} \equiv \alpha_{||} - V^{(n)}$. 
They transform under the HLS as 
\begin{eqnarray} 
  \alpha_{||} 
&\to& h \cdot \alpha_{||} \cdot h^\dag + h \cdot d \cdot h^\dag 
\,,  \\ 
\hat{\alpha}^{(n)}_{||} &\to& h \cdot \alpha_{||}^{(n)} \cdot h^\dag
\,, \\ 
\hat{\alpha}_{\perp} &\to& h \cdot \hat{\alpha}_{||} \cdot h^\dag 
\,, \\ 
A^{(n)} &\to& h \cdot A^{(n)} \cdot h^\dag
\,,
\end{eqnarray} 
 so that $v \to h \cdot v \cdot h^\dag + h \cdot d \cdot h^\dag$ and 
 $a \to h \cdot a \cdot h^\dag$.

 In $A_z \equiv 0$ gauge, 
the action $\Gamma_3$ in Eq.(\ref{123}) takes the form: 
\begin{eqnarray} 
\Gamma_3 
= \frac{N_c}{24\pi^2} \int_{M^4 \times R} 
{\rm tr} \Bigg[
 3 a dv dv + a da da + 3 (v^2a + a v^2 + a^3) dv 
+ [3 avada ]_{\rm non-zero}  \Bigg]
\,, \label{gamma3:SS}
\end{eqnarray}
which can be separated into two portions,   
\begin{eqnarray}
\Gamma_3 &\equiv& \Gamma_{31} + \Gamma_{32} \ , 
\\
&&
\Gamma_{31} = 
\int_{M^4 \times R} \, \mbox{tr}
\Biggl(
  3 a d v d v + 3 (v^2 a + a v^2 ) d v
\Biggr)
\ ,
\\
&&
\Gamma_{32} = 
\int_{M^4 \times R} \, \mbox{tr}
\Biggl(
  a d a d a + 3 a^3 d v
  + 3 \left[ a v a d a \right]_{\rm non-zero}
\Biggr)
\,. 
\end{eqnarray}

As to $\Gamma_{31}$, we calculate it as 
\begin{eqnarray}
\Gamma_{31} &=&
3 \int d z \, \int_{M^4} \, \mbox{tr}
\Biggl(
  - a (\partial_z v) d v - a d v (\partial_z v )
  - (v^2 a + a v^2 ) \partial_z v
\Biggr)
\nonumber\\
&=&
- 3 \int d z \, \int_{M^4} \, \mbox{tr}
\Biggl(
  \left[ d v a + a d v + (v^2 a + a v^2 ) \right]
  \partial_z v
\Biggr)
\nonumber\\
&=&
- 3 \int d z \, \int_{M^4} \, \mbox{tr}
\Biggl(
  \left\{ d v + v^2 \right\} a + a \left\{ d v + v^2 \right\} 
  \partial_z v
\Biggr)
\,. 
\end{eqnarray}
 From the transformation properties of $v$ and $a$, and noting 
that $\partial_z v =  \partial_z \tilde{v}$ transforms homogeneously under the HLS, 
we see that $\Gamma_{31}$ is HLS-gauge invariant.

As to $\Gamma_{32}$, we first consider the first term in it: 
\begin{eqnarray}
\int_{M^4 \times R} \, \mbox{tr}
\Biggl( 
  a d a d a
\Biggr)
&=&
\int_{M^4 \times R} \, \mbox{tr}
\Biggl( 
  a_0 d a_0 d a_0
  + \left[ a d a d a \right]_{\rm non-zero}
\Biggr)
\ .\label{ft:32}
\end{eqnarray}  
 The first term of Eq.(\ref{ft:32}) is calculated to be zero:  
\begin{eqnarray}
\int_{M^4 \times R} \, \mbox{tr}
\Biggl( 
  a_0 d a_0 d a_0
\Biggr)
&=&
- \int d z \, \int_{M^4} \, \mbox{tr}
\Biggl( 
  a_0 (\partial_z a_0) d a_0
  + a_0 d a_0 \partial_z a_0
\Biggr)
\nonumber\\
&=&
- \int d z \, \int_{M^4} \, \mbox{tr}
\Biggl( 
  ( d a_0 a_0 + a_0 d a_0 ) \partial_z a_0
\Biggr)
\nonumber\\
&=&
- \frac{1}{8} \int d z \, \int_{M^4} \, \mbox{tr}
\Biggl( 
\left[
  ( d \hat{\alpha}_\perp) \hat{\alpha}_\perp+ \hat{\alpha}_\perp ( d \hat{\alpha}_\perp )
\right]
\hat{\alpha}_\perp
\Biggr)
\psi_0^2 \dot{\psi}_0
\nonumber\\
&=&
- \frac{1}{24} \int d z \, 
\frac{\partial}{\partial z} (\psi_0^3)
\int_{M^4} \, \mbox{tr}
\Biggl( 
\left[
  ( d \hat{\alpha}_\perp ) \hat{\alpha}_\perp + \hat{\alpha}_\perp ( d \hat{\alpha}_\perp )
\right]
\hat{\alpha}_\perp
\Biggr)
\nonumber\\
&=&
- \frac{1}{12} 
\int_{M^4} \, \mbox{tr}
\Biggl( 
\left[
  ( d \hat{\alpha}_\perp ) \hat{\alpha}_\perp + \hat{\alpha}_\perp ( d \hat{\alpha}_\perp )
\right]
\hat{\alpha}_\perp
\Biggr)
\nonumber\\
&=&
- \frac{1}{12}
\int_{M^4} \, \mbox{tr}
\Biggl( 
  ( d \hat{\alpha}_\perp ) (\hat{\alpha}_\perp)^2 - ( d \hat{\alpha}_\perp ) (\hat{\alpha}_\perp)^2 
\Biggr)
\nonumber\\
&=&
0
\ .
\end{eqnarray}
Thus we have 
\begin{eqnarray}
\int_{M^4 \times R} \, \mbox{tr}
\Biggl( 
  a d a d a
\Biggr)
&=&
\int_{M^4 \times R} \, \mbox{tr}
\Biggl( 
  \left[  a d a d a \right]_{\rm non-zero}
\Biggr)
\,. 
\end{eqnarray}
  From this and noting that the second term in $\Gamma_{32}$ does not include 
  zero modes since $\partial_z v = \partial_z \tilde{v}$, we may write 
\begin{eqnarray}
\Gamma_{32} &= &
\int_{M^4 \times R} \, \mbox{tr}
\Biggl(
  \left[
    a d a d a + 3 a^3 d v
    + 3 a v a d a 
  \right]_{\rm non-zero}
\Biggr)\,. 
\end{eqnarray} 
We further rewrite this $\Gamma_{32}$ as follows: 
\begin{eqnarray}
\Gamma_{32} &= &
-
\int_{M^4 \times R} \, \mbox{tr}
\Biggl(
  \left[
    a d a \partial_z a + a \partial_z a d a + 3 a^3 \partial_z v
    + 3 a v a \partial_z a 
  \right]_{\rm non-zero}
\Biggr)
\nonumber\\
&=&
-
\int_{M^4 \times R} \, \mbox{tr}
\Biggl(
  \left[
    \left\{
      a d a + d a a + 3 a v a 
    \right\} \partial_z a 
    + 3 a^3 \partial_z v
  \right]_{\rm non-zero}
\Biggr)
\nonumber\\
&=&
-
\int_{M^4 \times R} \, \mbox{tr}
\Biggl(
  \left[
    \left\{
      a ( d a + v a + a v ) + ( d a + v a + a v ) a 
      + a v a - a^2 v - v a^2 
    \right\} \partial_z a 
    + 3 a^3 \partial_z v
  \right]_{\rm non-zero}
\Biggr)
\nonumber\\
&=&
-
\int_{M^4 \times R} \, \mbox{tr}
\Biggl(
  \left[
    \left\{
      a ( d a + v a + a v ) + ( d a + v a + a v ) a 
    \right\} \partial_z a 
  \right]_{\rm non-zero}
\Biggr)
\nonumber\\
&& {}
-
\int_{M^4 \times R} \, \mbox{tr}
\Biggl(
  \left[
    a\partial_z a a v +\partial_z a  a^2 v + a^2 \partial_z a v
    + 3 a^3 \partial_z v
  \right]_{\rm non-zero}
\Biggr)
\nonumber\\
&=&
-
\int_{M^4 \times R} \, \mbox{tr}
\Biggl(
  \left[
    \left\{
      a ( d a + v a + a v ) + ( d a + v a + a v ) a 
    \right\} \partial_z a 
    + 2 a^3 \partial_z v
  \right]_{\rm non-zero}
\Biggr)
\nonumber\\
&& {}
-
\int_{M^4 \times R} \, \mbox{tr}
\partial_z
\Biggl( 
  \left[ a^3 v 
  \right]_{\rm non-zero}
\Biggr)
\label{gam 32}
\end{eqnarray}
Note that the last term includes at least one non-zero mode/normalizable-mode 
which vanishes by the boundary condition at $z= \pm \infty$. 
Thus we find it goes to zero after integration with respect to $z$:  
\begin{eqnarray}
\int_{M^4 \times R} \, \mbox{tr}
\partial_z
\Biggl( 
  \left[ a^3 v 
  \right]_{\rm non-zero}
\Biggr)
= 0\,. 
\end{eqnarray}
On the other hand, we can easily see that 
the remaining terms in the first line of Eq.(\ref{gam 32}) are HLS-gauge invariant.

Thus it has been proven that the action $\Gamma_3$ is HLS-gauge invariant.

\section{Expanding Dirac-Born-Infeld and Chern-Simons parts in terms of HLS-building blocks} 
\label{derivation} 

In this section we derive 
Eqs.(\ref{Fpi})-(\ref{z8}) in the DBI part and Eqs.(\ref{c1})-(\ref{c4}) 
in the CS part.

\subsection{Dirac-Born-Infeld part}

Taking into account $A_z\equiv 0$ gauge and 
substituting Eq.(\ref{alpha:perp:cov}) into the field strength $F_{\mu z}$, we have 
\begin{eqnarray}
F_{\mu z}&=& 
\partial_{\mu}A_z - \partial_z A_{\mu} - i [A_{\mu} , A_z] 
\nonumber \\ 
&=& 
-\partial_z \left(
                  \hat{\alpha}_{\mu \perp} \psi_0 + V_{\mu} 
                      + \hat{\alpha}_{\mu ||} \left( 1 + \psi_1 \right) 
            \right)
\nonumber \\ 
&=& 
- \hat{\alpha}_{\mu \perp} \dot{\psi}_0 - 
\hat{\alpha}_{\mu ||} \dot{\psi}_1  
\,. \label{Fmuz}
\end{eqnarray} 
Similarly for $F_{\mu\nu}$ one can calculate it as 
\begin{eqnarray} 
 F_{\mu\nu}
&=& 
\partial_\mu A_\nu - \partial_\nu A_\mu 
 - i [A_\mu, A_\nu]   
\nonumber \\ 
&=&
\partial_{\mu} \left( 
                      \hat{\alpha}_{\nu \perp} \psi_0 + V_{\nu} 
                      + \hat{\alpha}_{\nu ||} (1 + \psi_1) 
               \right) 
- \partial_{\nu} \left ( 
                         \hat{\alpha}_{\mu \perp} \psi_0 + V_{\mu} 
                      + \hat{\alpha}_{\mu ||} (1 + \psi_1) 
                 \right) 
\nonumber \\
&&
- i \left[ 
              \hat{\alpha}_{\mu \perp} \psi_0 + V_{\mu} 
                      + \hat{\alpha}_{\mu ||} (1 + \psi_1) \, , \, 
              \hat{\alpha}_{\nu \perp} \psi_0 + V_{\nu} 
                      + \hat{\alpha}_{\nu ||} (1 + \psi_1) 
    \right] 
 \nonumber \\ 
&=& 
 \left( 
        D_{\mu} \hat{\alpha}_{\nu \perp} - D_{\nu} \hat{\alpha}_{\mu \perp} 
 \right) \psi_0 
 + 
 \left( 
        D_{\mu} \hat{\alpha}_{\nu ||} - D_{\nu} \hat{\alpha}_{\mu ||} 
 \right) (1 + \psi_1) 
 + V_{\mu\nu} 
\nonumber \\ 
&& 
 - i \left[ 
           \hat{\alpha}_{\mu \perp} \, , \, \hat{\alpha}_{\nu \perp} 
     \right] \psi_0^2 
 - i \left( 
            \left[ 
                  \hat{\alpha}_{\mu \perp} \, , \, \hat{\alpha}_{\nu ||} 
            \right] 
            + 
             \left[ 
                  \hat{\alpha}_{\mu ||} \, , \, \hat{\alpha}_{\nu \perp} 
            \right] 
     \right) (1 + \psi_1) \psi_0 
\nonumber \\ 
&& 
 - i \left[ 
            \hat{\alpha}_{\mu ||} \, , \, \hat{\alpha}_{\nu ||} 
     \right] (1 + \psi_1)^2 
\,, 
\end{eqnarray}
where we have defined 
\begin{eqnarray}
D_{\mu} \hat{\alpha}_{\nu\perp}
&=&
   \partial_{\mu} \hat{\alpha}_{\nu\perp} 
    - i [V_{\mu} , \hat{\alpha}_{\nu\perp}] 
\,, \nonumber \\ 
D_{\mu} \hat{\alpha}_{\nu||}
&=& 
   \partial_{\mu} \hat{\alpha}_{\nu||} 
    - i [V_{\mu} , \hat{\alpha}_{\nu||}] 
\,. \label{HLS derivative acting alpha hats}
\end{eqnarray} 
Using the identities, 
\begin{eqnarray}
D_{\mu} \hat{\alpha}_{\nu||} - D_{\nu} \hat{\alpha}_{\mu||} 
&=&
  i [ \hat{\alpha}_{\mu||} , \hat{\alpha}_{\nu||} ] 
  + i [ \hat{\alpha}_{\mu\perp} , \hat{\alpha}_{\nu\perp}] 
  + \hat{\mathcal{V}}_{\mu\nu} - V_{\mu\nu} 
\,, \label{HLS identity 1} \\
D_{\mu} \hat{\alpha}_{\nu\perp} - D_{\nu} \hat{\alpha}_{\mu\perp} 
&=&
  i [\hat{\alpha}_{\mu||} , \hat{\alpha}_{\nu\perp} ] 
  + i [\hat{\alpha}_{\mu\perp} , \hat{\alpha}_{\nu||} ] 
  + \hat{\mathcal{A}}_{\mu\nu} 
\,, \label{HLS identity 2}
\end{eqnarray} 
we obtain 
\begin{eqnarray} 
F_{\mu\nu} 
&=& 
   -V_{\mu\nu} \psi_1 + \hat{\mathcal{V}}_{\mu\nu} (1+\psi_1) 
   + \hat{\mathcal{A}}_{\mu\nu} \psi_0 
\nonumber \\ 
&& 
   - i [ \hat{\alpha}_{\mu||} \,, \, \hat{\alpha}_{\nu||} ] (1+\psi_1)\psi_1 
   + i [ \hat{\alpha}_{\mu\perp} \,,\, \hat{\alpha}_{\nu\perp}] 
   \left( 
          1 + \psi_1 - \psi_0^2 
   \right) 
\nonumber \\
&&
   - i \left( 
              [ \hat{\alpha}_{\mu||} \,,\, \hat{\alpha}_{\nu\perp} ] 
              + [ \hat{\alpha}_{\mu\perp} \,,\, \hat{\alpha}_{\nu||}] 
       \right) \psi_1\psi_0 
\,. \label{Fmunu}
\end{eqnarray} 
Substituting the final expressions in Eqs.(\ref{Fmuz}) and (\ref{Fmunu}) into the DBI part (\ref{SSaction}), 
we are readily led to Eq.(\ref{Fpi})-Eq.(\ref{z8}).

\subsection{Chern-Simons part: $\Gamma_3$}

We start with an expression of $\Gamma_3$ given in Eq.(\ref{gamma3:SS}): 
\begin{eqnarray} 
\Gamma_3 
&=& \frac{N_c}{24\pi^2} \int_5 
{\rm tr} \Bigg[
 3 a dv dv + a da da + 3 (v^2a + a v^2 + a^3) dv 
+ [3 avada ]_{\rm non-zero}  
\Bigg]
 \nonumber \\ 
 &=&
  - \frac{3 N_c}{24\pi^2} \int dz \int_{M^4} 
{\rm tr} \Bigg[
a \partial_z v d v + a d v \partial_z v 
+ 
v^2 a \partial_z v + a v^2 \partial_z v + a^3 \partial_z v 
+ 
[a v a \partial_z a ]_{\rm non-zero}  
\Bigg] 
\,,  \label{G3}
\end{eqnarray} 
where vector ($v$) and axialvector ($a$) fields are taken to be anti-hermitian for convenience. 
 After integrating out higher KK modes in the CS part, 
 $v$ and $a$ respectively become 
\begin{eqnarray} 
 v &=&  \alpha_{||} 
 + 
\hat{\alpha}_{||} 
\psi_1(z) 
\,, \qquad 
\alpha_{||} = \hat{\alpha}_{||} + V 
 \,, \\ 
 a &=& 
\hat{\alpha}_\perp  
\psi_0(z) 
 \,. 
\end{eqnarray}
  Substituting these expressions into Eq.(\ref{G3}), we calculate $\Gamma_3$ as follows:  
\begin{eqnarray} 
\Gamma_3 
&=&
 - \frac{3 N_c}{24\pi^2}  \int_{M^4} 
{\rm tr} \Bigg[
\langle 
 \psi_0 \dot{\psi}_1 
\rangle 
( \hat{\alpha}_\perp \hat{\alpha}_{||} - \hat{\alpha}_{||} \hat{\alpha}_{\perp} ) 
( d \alpha_{||} + \alpha_{||}^2) 
\nonumber \\ 
&& 
\hspace{60pt} 
+ 
\langle 
 \psi_0 \dot{\psi}_1 \psi_1 
\rangle 
\left\{ 
2 \alpha_{||} \hat{\alpha}_{||} \hat{\alpha}_\perp \hat{\alpha}_{||} 
+ 
( \hat{\alpha}_\perp \hat{\alpha}_{||} - \hat{\alpha}_{||} \hat{\alpha}_{\perp} ) d \hat{\alpha}_{||} 
+ 
( \hat{\alpha}_\perp {\alpha}_{||} - {\alpha}_{||} \hat{\alpha}_{\perp} ) \hat{\alpha}_{||}^2  
\right\} 
\nonumber \\ 
&&
\hspace{60pt} 
+ 
\langle 
 \psi_0 \dot{\psi}_1 \psi_1^2  
\rangle 
\left(
2 \hat{\alpha}_\perp \hat{\alpha}_{||}^3 
\right) 
+ 
\langle 
 \psi_0^2 \dot{\psi}_0 \psi_1 
 \rangle 
( 2 \hat{\alpha}_{||} \hat{\alpha}_\perp^3)  
\Bigg]  
 \nonumber \\ 
&=&
 - \frac{3 N_c}{24\pi^2}  \int_{M^4} 
{\rm tr} \Bigg[
\langle 
 \psi_0 \dot{\psi}_1 
\rangle 
( \hat{\alpha}_\perp \hat{\alpha}_{||} - \hat{\alpha}_{||} \hat{\alpha}_{\perp} ) 
( \hat{\alpha}_{||}^2 + D \hat{\alpha}_{||} + F_V )  
\nonumber \\ 
&& 
\hspace{60pt} 
+ 
\langle 
 \psi_0 \dot{\psi}_1 \psi_1 
\rangle 
\left\{ 
( D \hat{\alpha}_{||} + 2 \hat{\alpha}_{||}^2 )  
( \hat{\alpha}_\perp \hat{\alpha}_{||} - \hat{\alpha}_{||} \hat{\alpha}_{\perp} ) 
\right\} 
\nonumber \\ 
&&
\hspace{60pt} 
+ 
\langle 
 \psi_0 \dot{\psi}_1 \psi_1^2  
\rangle 
\left(
2 \hat{\alpha}_\perp \hat{\alpha}_{||}^3 
\right) 
+ 
\langle 
 \psi_0^2 \dot{\psi}_0 \psi_1 
 \rangle 
( 2 \hat{\alpha}_{||} \hat{\alpha}_\perp^3) \Bigg]  
\,, \nonumber \\ 
&=&
 - \frac{3 N_c}{24\pi^2} \int_{M^4} 
{\rm tr} \Bigg[
\langle 
 \psi_0 \dot{\psi}_1 
\rangle 
( \hat{\alpha}_\perp \hat{\alpha}_{||} - \hat{\alpha}_{||} \hat{\alpha}_{\perp} ) 
( - \hat{\alpha}_\perp^2 + \hat{F}_V )  
\nonumber \\ 
&& 
\hspace{60pt} 
+ 
\langle 
 \psi_0 \dot{\psi}_1 \psi_1 
\rangle 
\left\{ 
(  \hat{\alpha}_{||}^2 - \hat{\alpha}_{\perp}^2  -  F_V + \hat{F}_V )   
( \hat{\alpha}_\perp \hat{\alpha}_{||} - \hat{\alpha}_{||} \hat{\alpha}_{\perp} ) 
\right\} 
\nonumber \\ 
&&
\hspace{60pt} 
+ 
\langle 
 \psi_0 \dot{\psi}_1 \psi_1^2  
\rangle 
\left(
2 \hat{\alpha}_\perp \hat{\alpha}_{||}^3 
\right) 
+ 
\langle 
 \psi_0^2 \dot{\psi}_0 \psi_1 
 \rangle 
( 2 \hat{\alpha}_{||} \hat{\alpha}_\perp^3) \Bigg]  
\, \nonumber \\ 
&=&
 - \frac{3 N_c}{24\pi^2}  \int_{M^4} 
{\rm tr} \Bigg[
\langle - \frac{1}{2} 
 \psi_0 \dot{\psi}_1 \psi_1  
\rangle 
( \hat{\alpha}_\perp \hat{\alpha}_{||} - \hat{\alpha}_{||} \hat{\alpha}_{\perp} ) F_V 
\nonumber \\ 
&& 
\hspace{60pt} 
+ 
\langle 
2 \psi_0 \dot{\psi}_1 \psi_1 (1+\psi_1)
\rangle 
\hat{\alpha}_\perp  \hat{\alpha}_{||}^3 
\nonumber \\ 
&&
\hspace{60pt} 
+ 
\langle 
 2 \psi_0 \dot{\psi}_1 (1+ \psi_1 - \frac{1}{3} \psi_0^2 ) 
 \rangle 
\hat{\alpha}_{||} \hat{\alpha}_\perp^3 
\nonumber \\ 
&& 
\hspace{60pt}
+ 
\langle  
 \psi_0 \dot{\psi}_1 (1+ \psi_1)  
\rangle 
( \hat{\alpha}_\perp \hat{\alpha}_{||} - \hat{\alpha}_{||} \hat{\alpha}_{\perp} ) \hat{F}_V 
\Bigg]  
\,, \label{gamma3:symm}
\end{eqnarray}
where $\langle A \rangle = \int dz A(z)$ for an arbitrary function $A(z)$, 
and we have used an identity, 
\begin{equation} 
D \hat{\alpha}_{||} \equiv 
d \hat{\alpha}_{||} + V \hat{\alpha}_{||} + \hat{\alpha}_{||} V 
= - \hat{\alpha}_{||}^2 - \hat{\alpha}_{\perp}^2 - F_V + \hat{F}_V
\,, 
\end{equation}
and defined 
\begin{eqnarray}  
F_V &=& d V + V^2  
\,, \\ 
\hat{F}_V &=& \frac{\hat{F}_{L} + \hat{F}_{R}}{2}  
\,, \\ 
\hat{F}_{L,R} &=& 
\xi_{L,R}^{-1} \cdot F_{L,R} \cdot \xi_{L,R} 
\,, \\ 
F_{L,R} &=& d A_{L,R} + A_{L,R}^2 
\,. 
\end{eqnarray}

Moving on to four-dimensional Minkowski-space time and 
rewriting 1-forms in terms of hermitian fields,  
we obtain 
\begin{eqnarray} 
\Gamma_3 
&=&  \frac{N_c}{24\pi^2}  \int_4 
\Bigg\{ 
x_1 {\rm tr}[( \hat{\alpha}_\perp \hat{\alpha}_{||} - \hat{\alpha}_{||} \hat{\alpha}_{\perp} ) F_V ]
+ 
x_2   i {\rm tr}[ \hat{\alpha}_\perp  \hat{\alpha}_{||}^3 ] 
\nonumber \\ 
&& 
\hspace{60pt}
+   
x_3   i  {\rm tr} [\hat{\alpha}_{||}  \hat{\alpha}_{\perp}^3 ]
+ x_4 {\rm tr}[( \hat{\alpha}_\perp \hat{\alpha}_{||} - \hat{\alpha}_{||} \hat{\alpha}_{\perp} ) \hat{F}_V ]
\Bigg\} 
\,, \label{gamma3:def}
\end{eqnarray}
where 
\begin{eqnarray} 
x_1 &=& \langle  3 \psi_1 \psi_0 \dot{\psi}_1 \rangle 
\,, \\ 
x_2 &=& \langle  6 \psi_1 \psi_0 \dot{\psi}_1 (1 + \psi_1) \rangle 
\,, \\ 
x_3 &=& \langle 2 \psi_0 \dot{\psi}_1 ( - \psi_0^2 + 3 \psi_1 + 3)  \rangle 
\,, \\ 
x_4 &=& \langle  - 3  (\psi_1 + 1) \psi_0 \dot{\psi}_1 \rangle 
\,.  
\end{eqnarray}

The IP-odd terms in the HLS model are given 
in terms of $\hat{\alpha}_{||}, \hat{\alpha}_{\perp}$ as (See Eq.(\ref{HLS:anomaly-action})) 
\begin{eqnarray} 
  \Gamma_{\rm IP-odd}^{\rm HLS} 
  &=& \frac{N_c}{16 \pi^2} 
  \int_{M^4} 
  \Bigg\{  
( -4 c_1 - 4 c_2 )
\, i {\rm tr} [\hat{\alpha}_{\perp} \hat{\alpha}_{||}^3] 
\nonumber \\ 
&& 
\hspace{60pt}
+ 
(4 c_1 - 4 c_2) 
\, i{\rm tr} [\hat{\alpha}_{||} \hat{\alpha}_{\perp}^3] 
\nonumber \\ 
&& 
\hspace{60pt}
+ 
( - 2 c_3) \, 
 {\rm tr}[ (\hat{\alpha}_{\perp} \hat{\alpha}_{||} - \hat{\alpha}_{||} \hat{\alpha}_{\perp} ) 
F_V ]  
\nonumber \\ 
&& 
\hspace{60pt}
+ 
( - 2 c_4) \, 
 {\rm tr}[ (\hat{\alpha}_{\perp} \hat{\alpha}_{||} - \hat{\alpha}_{||} \hat{\alpha}_{\perp} ) 
\hat{F}_V ]  
  \Bigg\} 
  \,. 
\end{eqnarray} 
Comparing this form with $\Gamma_3$ in Eq.(\ref{gamma3:def}), 
we find  
\begin{eqnarray} 
c_1 &=& - \frac{1}{12} x_2 + \frac{1}{12} x_3 
\,, \\ 
c_2 &=&  
- \frac{1}{12} x_2 - \frac{1}{12} x_3 
\,, \\ 
c_3 &=& 
- \frac{1}{3} x_1 
\,,\\ 
c_4 &=& - \frac{1}{3} x_4 
\,, 
\end{eqnarray} 
which readily lead to Eqs.(\ref{c1})-(\ref{c4}).

\section{Alternative derivation of our results for IP-odd processes and its gauge/chiral invariance}  

\label{alternative}

In this appendix, 
to see that our formalism correctly incorporates contributions from
higher KK modes of the HLS gauge bosons in the IP-odd sector,   
we shall perform a low-energy expansion of the original forms of the vertex functions~\cite{Sakai:2005yt} 
for $\pi^0$-$\gamma^*$-$\gamma^*$, $\omega$-$\pi^0$-$\gamma^*$ and $\gamma^*$-$\pi^0$-$\pi^+$-$\pi^-$   
to be consistent with 
our integrating-out method as was done in Sec.~\ref{EMformfactor} (See Eq.(\ref{FVSS:2})). 
We also discuss a naive truncation and violation of gauge/chiral invariance.

\subsection{$\pi^0$-$\gamma^*$-$\gamma^*$ vertex function and $\pi^0$-$\gamma$ transition form factor}

We begin with the original form of the $\pi^0$-$\gamma^*$-$\gamma^*$ vertex function written in terms of infinite sum of 
vector meson exchanges~\cite{Sakai:2005yt}: 
\begin{equation} 
 \Gamma^{\mu\nu} [\pi^0, \gamma^*_\mu (q_1), \gamma_\nu^* (q_2)]  \Bigg|_{\rm SS}
=
\frac{e^2N_c}{24 \pi^2 F_\pi} 
\epsilon^{\mu\nu\alpha \beta} q_{1\alpha} q_{2 \beta} 
\cdot \sum_{k=1}^\infty \sum_{l=1}^\infty \left[ 
\frac{g_{\omega_k \rho_l \pi} g_{\rho_l} g_{\omega_k}}{(m_{\omega_k}^2 + q_1^2)(m_{\rho_l}^2 + q_2^2)} 
 + (q_1^2 \leftrightarrow q_2^2) 
\right] 
\,, \label{pi2g:SS}
\end{equation} 
where the coupling $g_{\omega_k \rho_l \pi} $ is defined in the SS model as  
\begin{equation} 
  {\cal L}_{\omega_k \rho_l \pi} =  - \frac{N_c}{8\pi F_\pi} g_{\omega_k \rho_l \pi} 
\sum_{a=1}^3  \epsilon^{\mu\nu\lambda\sigma} \partial_\mu (\omega_k)_\nu \partial_\lambda (\rho^a_l)_\sigma \cdot \pi^a
\,. 
\end{equation}
 Consistently with our method which integrates out higher KK modes into ${\cal O}(p^4)$ terms of the HLS Lagrangian,   
we expand Eq.(\ref{pi2g:SS}) as  
\begin{eqnarray} 
 \Gamma^{\mu\nu} [\pi^0, \gamma^*_\mu (q_1), \gamma_\nu^* (q_2)] \Bigg|_{\rm SS}
&=&
\frac{e^2N_c}{12 \pi^2 F_\pi} 
\epsilon^{\mu\nu\alpha \beta} q_{1\alpha} q_{2 \beta} 
\cdot \Bigg[ 
\left( 
\sum_{k=2}^\infty \sum_{l=2}^\infty  
\frac{g_{\omega_k \rho_l \pi} g_{\rho_l} g_{\omega_k}}{m_{\omega_k}^2m_{\rho_l}^2} \right) 
\nonumber \\ 
&& 
+ \frac{1}{4}\left( 2 
\sum_{k=2}^\infty 
\frac{g_{\omega_k \rho \pi} g_{\rho} g_{\omega_k}}{m_{\omega_k}^2m_{\rho}^2} \right) 
\{  
D_\omega(q_1^2) + D_\rho(q_2^2) +   (q_1^2 \leftrightarrow q_2^2) 
\}
\nonumber \\ 
&& 
+ \frac{1}{2} \left(
\frac{g_{\omega \rho \pi} g_\rho g_\omega}{m_\omega^2 m_\rho^2} 
\right) 
\{
 D_\omega(q_1^2)\cdot D_\rho(q_2^2) 
 + (q_1^2 \leftrightarrow q_2^2) 
\} 
\Bigg] 
\,, \label{pi2g:SS:expand}
\end{eqnarray} 
up to terms of ${\cal O}(q_{1,2}^2/m_{\rho_k, \omega_k}^2)$ ($k \ge 2$), which correspond to terms higher than ${\cal O}(p^4)$ in the Lagrangian. 
Using the sum rules~\cite{Sakai:2005yt}, 
\begin{equation} 
\sum_{k=1}^\infty 
\frac{g_{\omega_l \rho_k \pi} g_{\rho_k}}{m_{\rho_k}^2 } 
= 
 \sum_{k=1}^\infty 
\frac{g_{\omega_k \rho_l \pi} g_{\omega_k}}{m_{\omega_k}^2} 
= g_{\rho_l \pi \pi} 
 \,, \qquad
\sum_{k=1}^\infty 
\frac{g_{\rho_k \pi \pi} g_{\rho_k}}{m_{\rho_k}^2 } 
= 1 
\,, \label{sum3}
\end{equation} 
we have 
\begin{eqnarray} 
\sum_{k=2}^\infty \sum_{l=2}^\infty  
\frac{g_{\omega_k \rho_l \pi} g_{\rho_l} g_{\omega_k}}{m_{\omega_k}^2m_{\rho_l}^2}
&=& 
1 - \frac{2g_{\rho\pi\pi} g_\rho}{m_\rho^2} + \frac{g_{\omega \rho \pi} g_\rho g_\omega}{m_\omega^2 m_\rho^2}
\,, \label{Api2g:SS} \\ 
\sum_{k=2}^\infty 
\frac{g_{\omega_k \rho \pi} g_{\rho} g_{\omega_k}}{m_{\omega_k}^2m_{\rho}^2} 
&=&  
\frac{g_{\rho\pi\pi} g_\rho}{m_\rho^2} - \frac{g_{\omega \rho \pi} g_\rho g_\omega}{m_\omega^2 m_\rho^2}
\,. \label{Bpi2g:SS}
\end{eqnarray} 
  From Ref.~\cite{Sakai:2005yt} we read 
\begin{eqnarray}   
\frac{g_{\rho\pi\pi} g_\rho}{m_\rho^2} 
&=& \frac{1}{2} \frac{\langle \langle  \psi_1 \dot{\psi}_0   \rangle \rangle  \langle \psi_1 \rangle}{\langle \psi_1^2 \rangle} 
\,, \label{ratio:1}\\ 
\frac{ g_{\omega \rho \pi} g_\rho g_\omega}{m_\rho^2 m_\omega^2}  
&=& 
\frac{1}{2} \frac{\langle \langle  \psi_1^2 \dot{\psi}_0   \rangle \rangle  \langle \psi_1 \rangle^2}
{\langle \psi_1^2 \rangle^2} 
\,. \label{ratio:2}
\end{eqnarray} 
Substituting these into the right hand sides of Eqs.(\ref{Api2g:SS}) and (\ref{Bpi2g:SS}), we have  
 \begin{eqnarray} 
\sum_{k=2}^\infty \sum_{l=2}^\infty  
\frac{g_{\omega_k \rho_l \pi} g_{\rho_l} g_{\omega_k}}{m_{\omega_k}^2m_{\rho_l}^2} 
&=&
1 - \left[ 
\frac{\langle  \psi_1 \rangle \langle \langle \dot{\psi}_0 \psi_1 \rangle \rangle}
{\langle \psi_1^2 \rangle} 
- \frac{1}{2} 
\frac{\langle  \psi_1 \rangle^2 \langle \langle \dot{\psi}_0 \psi_1^2   \rangle \rangle}
{\langle \psi_1^2 \rangle^2} 
\right] 
\,, \label{Api2g-coffi1} \\ 
2 \sum_{k=2}^\infty 
\frac{g_{\omega_k \rho \pi} g_{\rho} g_{\omega_k}}{m_{\omega_k}^2m_{\rho}^2} 
&=& 
\frac{\langle \psi_1 \rangle \langle \langle \dot{\psi}_0 \psi_1  \rangle \rangle}
{\langle \psi_1^2 \rangle} 
- 
\frac{\langle \psi_1 \rangle^2  \langle \langle \dot{\psi}_0 \psi_1^2  \rangle \rangle}
{\langle \psi_1^2 \rangle^2}  
\,, \label{Bpi2g-coffi2} \\  
\frac{g_{\omega \rho \pi} g_\rho g_\omega}{m_\omega^2 m_\rho^2}  
&=& 
\frac{1}{2}\left[ 
\frac{\langle \psi_1 \rangle^2 \langle  \langle \dot{\psi}_0 \psi_1^2  \rangle \rangle}
{\langle \psi_1^2 \rangle^2} 
\right] 
\,. \label{Cpi2g-coffi1}
\end{eqnarray}  
The right hand sides are identical to 
$A^{\pi 2\gamma}_{_{\rm SS}}$-$C_{_{\rm SS}}^{\pi 2 \gamma}$ in 
Eqs.(\ref{Api2g})-(\ref{Cpi2g}): 
 \begin{eqnarray} 
\sum_{k=2}^\infty \sum_{l=2}^\infty  
\frac{g_{\omega_k \rho_l \pi} g_{\rho_l} g_{\omega_k}}{m_{\omega_k}^2m_{\rho_l}^2} 
&=&
A_{_{\rm SS}}^{\pi 2 \gamma} 
\,,   \label{Aa}   
\\ 
2 \sum_{k=2}^\infty 
\frac{g_{\omega_k \rho \pi} g_{\rho} g_{\omega_k}}{m_{\omega_k}^2m_{\rho}^2} 
&=& 
B_{_{\rm SS}}^{\pi 2 \gamma} 
\,,   \label{Bb} 
\\  
\frac{g_{\omega \rho \pi} g_\rho g_\omega}{m_\omega^2 m_\rho^2}  
&=& 
C_{_{\rm SS}}^{\pi 2 \gamma} 
\,,     
\end{eqnarray}  
and hence we arrive at the same result 
as that derived from our method integrating out higher KK modes (Eq.(\ref{pi2g:form2}) with Eqs.(\ref{Api2g})-(\ref{Cpi2g})): 
\begin{eqnarray} 
\Gamma^{\mu\nu} [\pi^0, \gamma^*_\mu (q_1), \gamma_\nu^* (q_2)] \Bigg|_{\rm SS}
&=& 
\frac{e^2N_c}{12\pi^2 F_\pi} 
\epsilon^{\mu\nu\alpha \beta} q_{1\alpha} q_{2 \beta} 
\Bigg[
A_{_{\rm SS}}^{\pi 2\gamma} + \frac{B_{_{\rm SS}}^{\pi 2\gamma}}{4}  
\left\{  
D_\rho(q_1^2) + D_\omega(q_1^2) + (q_1^2 \to q_2^2)  
\right\}  
\nonumber \\ 
&& 
+ \frac{C_{_{\rm SS}}^{\pi 2\gamma}}{2}
\left\{  
D_\rho(q_1^2) \cdot D_\omega(q_2^2) + (q_1^2 \leftrightarrow q_2^2)  
\right\} 
\Bigg]
\,. \label{pi2g:form:integ} 
\end{eqnarray} 
 For the transition form factor $F_{\pi^0 \gamma}$, we have 
\begin{equation} 
  F_{\pi^0 \gamma}(Q^2) \Bigg|_{\rm SS}
  = 
\left( A_{_{\rm SS}}^{\pi 2\gamma} +  \frac{B_{_{\rm SS}}^{\pi 2\gamma}}{2}
\right)
 + \left( \frac{B_{_{\rm SS}}^{\pi 2\gamma}}{4} + \frac{C_{_{\rm SS}}^{\pi 2\gamma}}{2} \right) [D_\rho(Q^2) + D_\omega(Q^2)] 
  \,. \label{Fpigamma:SS} 
\end{equation} 
Because the resultant form (\ref{pi2g:form:integ}) is the same as that obtained from our method 
which is manifestly gauge invariant by construction (See Eqs.(\ref{Amu:integ:trans}),(\ref{D}) and (\ref{action:Amu:integrate})),   
the low-energy theorem in Eq.(\ref{LET:pi2g}) is actually satisfied:  
\begin{equation}
 A_{_{\rm SS}}^{\pi 2 \gamma} +  B_{_{\rm SS}}^{\pi 2 \gamma} +  C_{_{\rm SS}}^{\pi 2 \gamma} = 1
 \,. %\label{LET:pi2g}
\end{equation}

 For a comparison, 
let us consider what would happen if one had naively truncated tower
of the HLS gauge bosons at the lowest level 
as in Eq.(\ref{action:naive}). 
  From Eqs.(\ref{Aa}) and (\ref{Bb}), one can easily see that 
such a naive truncation corresponds to simply neglecting higher KK modes,  
$A_{_{\rm SS}}^{\pi 2 \gamma}=B_{_{\rm SS}}^{\pi 2\gamma}=0$: 
\begin{equation} 
  F_{\pi^0 \gamma}(Q^2) \Bigg|_{\rm SS}^{\rm trun}
  = 
\frac{C_{_{\rm SS}}^{\pi 2\gamma}}{2} \left( \frac{m_\rho^2}{m_\rho^2 + Q^2} + 
\frac{m_\omega^2}{m_\omega^2 + Q^2}  \right)
  \,, \label{Fpigamma:SS:trun} 
\end{equation} 
with  $C_{_{\rm SS}}^{\pi 2\gamma} \simeq 0.5$ from Eq.(\ref{Cpi2g}). 
At $Q^2=0$ we have 
\begin{equation} 
  F_{\pi^0 \gamma}(0) \Bigg|_{\rm SS}^{\rm trun}
  = 
C_{_{\rm SS}}^{\pi 2\gamma} \simeq 0.5 \neq 1
  \,, 
\end{equation} 
which breaks the EM gauge symmetry. 
Note again that the naive truncation (\ref{Fpigamma:SS:trun}) is different from the $\rho/\omega$ meson 
dominance (\ref{GSW}) which is gauge invariant. 
  The violation of gauge symmetry can also be seen in the vertex function as 
\begin{eqnarray} 
\Gamma^{\mu\nu} [\pi^0, \gamma^*_\mu (q_1), \gamma_\nu^* (q_2)] \Bigg|^{\rm trun}_{\rm SS}
& \stackrel{q_1^2, q_2^2 \to 0}{\longrightarrow}& 
\frac{e^2N_c}{12\pi^2 F_\pi} 
\epsilon^{\mu\nu\alpha \beta} q_{1\alpha} q_{2 \beta} \cdot ( C_{_{\rm SS}}^{\pi 2\gamma}) 
\nonumber \\ 
&\neq& 
\frac{e^2N_c}{12\pi^2 F_\pi} 
\epsilon^{\mu\nu\alpha \beta} q_{1\alpha} q_{2 \beta}  
\,, \label{pi2g:form:truncate} 
\end{eqnarray} 
which contradicts with the low-energy theorem (\ref{LET:pi2g1}).

\subsection{$\omega$-$\pi^0$-$\gamma^*$ vertex function and  $\omega$-$\pi^0$ transition form factor}

We start with the form~\cite{Sakai:2005yt}: 
\begin{eqnarray} 
\Gamma^{\mu\nu} [\omega_\mu(p), \pi^0, \gamma^*_\nu (k)] \Bigg|_{\rm SS}
= 
\frac{e N_c}{8 \pi^2 F_\pi} 
\epsilon^{\mu\nu\alpha \beta} p_{\alpha} k_{\beta} 
\cdot \sum_{n=1}^\infty 
\frac{g_{\omega \rho_n \pi} g_{\rho_n}}{m_{\rho_n}^2 + k^2}
\,. 
\label{opig:SS}  
\end{eqnarray} 
We expand this expression 
to be consistent with our method integrating out higher KK modes into ${\cal O}(p^4)$ terms of the general HLS Lagrangian:    
\begin{eqnarray} 
\Gamma^{\mu\nu} [\omega_\mu(p), \pi^0, \gamma^*_\nu (k)] \Bigg|_{\rm SS}
&=& 
\frac{e N_c}{8 \pi^2 F_\pi} 
\epsilon^{\mu\nu\alpha \beta} p_{\alpha} k_{\beta} 
\cdot 
\Bigg[ 
\left( \sum_{n=2}^\infty 
\frac{g_{\omega \rho_n \pi} g_{\rho_n}}{m_{\rho_n}^2}
\right)  
+ \left( 
\frac{g_{\omega \rho \pi} g_\rho}{m_\rho^2} 
\right) 
D_\rho(k^2) 
\Bigg] 
\,, 
\label{opig:SS:expand}  
\end{eqnarray} 
up to terms of ${\cal O}(k^2/m_{\rho_n}^2)$ ($n \ge 2$) which correspond to terms higher than ${\cal O}(p^4)$ in the Lagrangian.  
Using the first sum rule displayed in Eq.(\ref{sum3}), we have 
\begin{eqnarray} 
\sum_{n=2}^\infty 
\frac{g_{\omega \rho_n \pi} g_{\rho_n}}{m_{\rho_n}^2}
=  
g_{\rho \pi\pi} - \frac{g_{\omega \rho \pi} g_\rho}{m_\rho^2} 
\,.  
\end{eqnarray} 
    From Ref.~\cite{Sakai:2005yt} we read 
\begin{eqnarray} 
g_{\rho\pi\pi} 
&=& \frac{1}{2 \sqrt{N_c G}} \sqrt{\frac{\langle \langle  \psi_0 \dot{\psi}_1 \rangle \rangle^2 }{ \langle  \psi_1^2 \rangle}} 
\,, \label{grhopipi:SS:1}\\ 
\frac{g_{\omega \rho \pi} g_\rho}{m_\rho^2} 
&=& 
\frac{1}{2 \sqrt{N_c G}}  
\frac{\langle \langle  \dot{\psi}_0 \psi_1^2 \rangle \rangle \langle  \psi_1  \rangle}{ \langle  \psi_1^2 \rangle^{3/2}} 
\,.  \label{ratio:3} 
\end{eqnarray} 
We then have 
\begin{eqnarray} 
\sum_{n=2}^\infty 
\frac{g_{\omega \rho_n \pi} g_{\rho_n}}{m_{\rho_n}^2}
=  
- \frac{1}{2\sqrt{N_c G}} 
\left[ 
\frac{\langle \langle \dot{\psi}_0 \psi_1 \rangle \rangle}{\sqrt{\langle \psi_1^2 \rangle}} 
- 
\frac{\langle  \psi_1 \rangle  \langle \langle \dot{\psi}_0 \psi_1^2 \rangle \rangle}{\langle \psi_1^2 \rangle^{3/2}} 
\right] 
\,.  \label{relate}
\end{eqnarray} 
 Comparing Eqs.(\ref{ratio:3}) and (\ref{relate}) with Eqs.(\ref{Bopg:5d}) and (\ref{Aopg:5d}), respectively, 
we find 
\begin{eqnarray} 
\sum_{n=2}^\infty 
\frac{g_{\omega \rho_n \pi} g_{\rho_n}}{m_{\rho_n}^2}
&=& 
A_{_{\rm SS}}^{\omega \pi \gamma} 
\,, \label{relation2} \\
\frac{g_{\omega \rho \pi} g_\rho}{m_\rho^2} 
&=& 
B_{_{\rm SS}}^{\omega \pi \gamma}
\,,
\end{eqnarray} 
and hence arrive at the same result as that derived from our integrating-out method 
(Eq.(\ref{opig:form2}) with Eqs.(\ref{Aopg:5d}) and (\ref{Bopg:5d})):  
\begin{eqnarray} 
\Gamma^{\mu\nu} [\omega_\mu(p), \pi^0, \gamma^*_\nu (k)]  \Bigg|_{\rm SS}
&=& 
\frac{e N_c}{8 \pi^2 F_\pi} 
\epsilon^{\mu\nu\alpha \beta} p_{\alpha} k_{\beta} 
\cdot \left[ 
A_{_{\rm SS}}^{\omega\pi \gamma} + B_{_{\rm SS}}^{\omega \pi\gamma}  D_\rho(k^2) 
\right]
\,. 
\end{eqnarray} 
 For the transition form factor, we have 
 \begin{eqnarray} 
  F_{\omega \pi^0} (q^2) \Bigg|_{\rm SS}
&=& 
\left( 
\frac{A_{_{\rm SS}}^{\omega\pi \gamma}}{A_{_{\rm SS}}^{\omega\pi \gamma}+B_{_{\rm SS}}^{\omega\pi \gamma}}\right) 
 + 
\left( \frac{B_{_{\rm SS}}^{\omega\pi \gamma}}{A_{_{\rm SS}}^{\omega\pi \gamma}+B_{_{\rm SS}}^{\omega\pi \gamma}}
\right) \frac{m_\rho^2}{m_\rho^2 - q^2} 
\,, \nonumber \\ 
&=& 
\left( 
1- {\tilde r}_{_{\rm SS}}\right) 
 + 
{\tilde r}_{_{\rm SS}} \frac{m_\rho^2}{m_\rho^2 - q^2} 
\,. \label{Fopi:SS}
\end{eqnarray}

 A naive truncation as in Eq.(\ref{action:naive}), which corresponds to setting $A_{_{\rm SS}}^{\omega\pi \gamma}=0$,  
would lead to the same form of $F_{\omega\pi^0}$ as that of the $\rho$ meson dominance (\ref{Fopi:rhoD}), 
although $g_{\omega \pi \gamma}$ in Eq.(\ref{g:opig}) yields the value   
about $1/(1.53) \simeq 2/3$ times smaller (See footnote~\ref{foot}.). 
Unlike the case of the pion EM and $\pi^0$-$\gamma$ transition form factors, 
the violation of gauge symmetry is not manifest 
in the $\omega$-$\pi^0$ transition form factor $F_{\omega \pi^0}$ 
since there is no low-energy theorem for this process.

\subsection{$\gamma^*$-$\pi^0$-$\pi^+$-$\pi^-$ vertex function}

The original form~\cite{Sakai:2005yt} can be expanded 
consistently with our integrating-out method: 
 \begin{eqnarray} 
&& 
\Gamma_{\mu} [\gamma^*_\mu (p), \pi^0(q_0), \pi^+(q_+), \pi^-(q_-)] \Bigg|_{\rm SS} 
\nonumber \\ 
&=&
- \frac{e N_c}{12\pi^2 F_\pi} 
\epsilon_{\mu\nu\alpha \beta} q_0^\nu q_+^\alpha q_-^\beta  
\sum_{k=1}^\infty \sum_{l=1}^\infty 
\Bigg[ 
\frac{g_{\omega_k \rho_l \pi} g_{\rho_l \pi\pi} g_{\omega_k}}{(m_{\omega_k}^2 + p^2)(m_{\rho_l}^2 + (q_+ + q_-)^2)}
+ (q_+ \leftrightarrow q_0) + (q_- \leftrightarrow q_0)  
\Bigg] 
\, 
\nonumber \\
&=& 
- \frac{e N_c}{12\pi^2 F_\pi^3} 
\epsilon_{\mu\nu\alpha \beta} q_0^\nu q_+^\alpha q_-^\beta  
\Bigg[ 
\left( 3 F_\pi^2
\sum_{k=2}^\infty \sum_{l=2}^\infty  
\frac{g_{\omega_k \rho_l \pi} g_{\rho_l \pi\pi} g_{\omega_k}}{m_{\omega_k}^2m_{\rho_l}^2}
\right) 
+ 
\left(
3 F_\pi^2
 \sum_{l=2}^\infty  
\frac{g_{\omega \rho_l \pi} g_{\rho_l \pi\pi} g_{\omega}}{m_{\omega}^2m_{\rho_l}^2}
\right)
\cdot D_\omega(p^2)
\nonumber\\ 
&& 
+ 
\frac{1}{3} \left(3 F_\pi^2
 \sum_{k=2}^\infty  
\frac{g_{\omega_k \rho \pi} g_{\rho \pi\pi} g_{\omega_k}}{m_{\omega_k}^2m_{\rho}^2}
\right)
\cdot \{ D_\rho((q_+ + q_-)^2)  + (q_+ \leftrightarrow q_0) + (q_- \leftrightarrow q_0)   \} 
\nonumber \\ 
&& 
+ 
\frac{1}{3} \left( 3 F_\pi^2
\frac{g_{\omega \rho \pi} g_{\rho \pi\pi} g_{\omega}}{m_{\omega}^2m_{\rho}^2}
\right) 
D_\omega(p^2) \cdot \{ D_\rho((q_+ + q_-)^2)  + (q_+ \leftrightarrow q_0) + (q_- \leftrightarrow q_0)   \} 
\Bigg] 
\,, 
\label{g3pi:SS:expand}
\end{eqnarray} 
where we have neglected terms of ${\cal O}(p^2,q_{\pm,0}^2/m_{\rho_k,\omega_k}^2)$ ($k \ge 2$) which correspond 
to terms higher than ${\cal O}(p^4)$ in the Lagrangian.  
  Using the sum rules in Eq.(\ref{sum3}) and~\cite{Sakai:2005yt}  
\begin{equation} 
\sum_{l=1}^\infty
\frac{g_{\rho_l\pi\pi}^2}{m_{\rho_l}^2} 
= \frac{1}{3 F_\pi^2}
\,, \label{sum4} 
\end{equation} 
we have 
\begin{eqnarray} 
3 F_\pi^2
\sum_{k=2}^\infty \sum_{l=2}^\infty  
\frac{g_{\omega_k \rho_l \pi} g_{\rho_l \pi\pi} g_{\omega_k}}{m_{\omega_k}^2m_{\rho_l}^2}
&=&
1 - 3 \left( \frac{g_{\rho\pi\pi}^2 F_\pi^2}{m_\rho^2} \right) 
\left( 
  2 - \frac{g_{\omega \rho \pi} g_\omega}{ g_{\rho\pi\pi} m_\omega^2} 
\right) 
\,, \label{Ag3pi:ref:SS} 
\\ 
3 F_\pi^2
 \sum_{l=2}^\infty  
\frac{g_{\omega \rho_l \pi} g_{\rho_l \pi\pi} g_{\omega}}{m_{\omega}^2m_{\rho_l}^2}
&=& 
3 F_\pi^2
 \sum_{k=2}^\infty  
\frac{g_{\omega_k \rho \pi} g_{\rho \pi\pi} g_{\omega_k}}{m_{\omega_k}^2m_{\rho}^2}
= 
3 \left( \frac{g_{\rho\pi\pi}^2 F_\pi^2}{m_\rho^2} \right) 
\left(1 - \frac{g_{\omega \rho \pi} g_\omega}{ g_{\rho\pi\pi} m_\omega^2} 
\right) 
\,, \label{Bg3pi:ref:SS} 
\end{eqnarray} 
    From Ref.~\cite{Sakai:2005yt} we read off 
\begin{eqnarray} 
  \frac{g_{\rho\pi\pi}^2 F_\pi^2}{m_\rho^2} 
&=&  
\frac{1}{4} \frac{ \langle \langle  \dot{\psi}_0  \psi_1 \rangle \rangle  \langle  \psi_1 (1-\psi_0^2)  \rangle  }
{\langle \psi_1^2 \rangle}
\,, \label{ksrf2} \\ 
\frac{g_{\omega \rho \pi} g_\omega}{ g_{\rho\pi\pi} m_\omega^2} 
&=& 
\frac{ \langle \psi_1 \rangle   \langle \langle  \dot{\psi}_0  \psi_1^2 \rangle \rangle}
{\langle  \psi_1^2  \rangle    \langle \langle \psi_0 \dot{\psi}_1 \rangle \rangle}
\,. 
\end{eqnarray}  
 Putting these into the right hand sides of Eqs.(\ref{Ag3pi:ref:SS}) and (\ref{Bg3pi:ref:SS}), we have 
 \begin{eqnarray} 
3 F_\pi^2
\sum_{k=2}^\infty \sum_{l=2}^\infty  
\frac{g_{\omega_k \rho_l \pi} g_{\rho_l \pi\pi} g_{\omega_k}}{m_{\omega_k}^2m_{\rho_l}^2}
&=&
 1 - \frac{3}{4} \Bigg[ 
 \frac{\langle  \psi_1 \rangle \langle \langle  \dot{\psi}_0 \psi_1 (1-\psi_0^2) \rangle \rangle}{\langle \psi_1^2 \rangle}
+ 
 \frac{\langle \langle \dot{\psi}_0 \psi_1 \rangle \rangle 
\langle \psi_1(1-\psi_0^2)  \rangle}{\langle \psi_1^2 \rangle}
- 
 \frac{\langle \langle \dot{\psi}_0 \psi_1^2 \rangle \rangle 
\langle \psi_1 \rangle \langle  \psi_1(1-\psi_0^2)  \rangle}
{\langle \psi_1^2 \rangle^2} 
 \Bigg]  
\,,  
\nonumber \\ 
\\ 
3 F_\pi^2
 \sum_{l=2}^\infty  
\frac{g_{\omega \rho_l \pi} g_{\rho_l \pi\pi} g_{\omega}}{m_{\omega}^2m_{\rho_l}^2}
&=& 
\frac{3}{4} \Bigg[ 
 \frac{\langle  \psi_1 \rangle 
\langle \langle  \dot{\psi}_0 \psi_1 (1-\psi_0^2)  \rangle\rangle}
{\langle  \psi_1^2 \rangle}
- 
 \frac{\langle \langle \dot{\psi}_0 \psi_1^2  \rangle \rangle 
\langle \psi_1 \rangle 
\langle  \psi_1(1-\psi_0^2)  \rangle}
{\langle \psi_1^2 \rangle^2} 
 \Bigg] 
\,, 
\\ 
3 F_\pi^2
 \sum_{k=2}^\infty  
\frac{g_{\omega_k \rho \pi} g_{\rho \pi\pi} g_{\omega_k}}{m_{\omega_k}^2m_{\rho}^2}
&=& 
\frac{3}{4} \Bigg[ 
 \frac{\langle \langle \dot{\psi}_0 \psi_1  \rangle \rangle 
\langle \psi_1(1-\psi_0^2)  \rangle}
{\langle \psi_1^2 \rangle}
- 
 \frac{\langle \langle \dot{\psi}_0 \psi_1^2  \rangle \rangle 
\langle \psi_1 \rangle 
\langle \psi_1(1-\psi_0^2)  \rangle}
{\langle \psi_1^2 \rangle^2} 
 \Bigg] 
\,, 
\\
3 F_\pi^2
\frac{g_{\omega \rho \pi} g_{\rho \pi\pi} g_{\omega}}{m_{\omega}^2m_{\rho}^2}
&=& 
\frac{3}{4} \left[ 
  \frac{\langle \langle \dot{\psi}_0 \psi_1^2 \rangle \rangle 
\langle \psi_1 \rangle 
\langle \psi_1(1-\psi_0^2)  \rangle}
{\langle \psi_1^2 \rangle^2} 
\right]  
\,. 
\end{eqnarray} 
The right hand sides are identical to $A_{_{\rm SS}}^{\gamma 3\pi} $-$D_{_{\rm SS}}^{\gamma 3\pi} 
$ in Eqs.(\ref{Ag3pi})-(\ref{Dg3pi}):  
\begin{eqnarray} 
3 F_\pi^2
\sum_{k=2}^\infty \sum_{l=2}^\infty  
\frac{g_{\omega_k \rho_l \pi} g_{\rho_l \pi\pi} g_{\omega_k}}{m_{\omega_k}^2m_{\rho_l}^2}
&=&
A_{_{\rm SS}}^{\gamma 3\pi} 
\,, \label{Ag3pi:SS} 
\\ 
3 F_\pi^2
 \sum_{l=2}^\infty  
\frac{g_{\omega \rho_l \pi} g_{\rho_l \pi\pi} g_{\omega}}{m_{\omega}^2m_{\rho_l}^2}
&=& 
B_{_{\rm SS}}^{\gamma 3\pi} 
\,, %\label{Bg3pi:ref} 
\\ 
3 F_\pi^2
 \sum_{k=2}^\infty  
\frac{g_{\omega_k \rho \pi} g_{\rho \pi\pi} g_{\omega_k}}{m_{\omega_k}^2m_{\rho}^2}
&=& 
C_{_{\rm SS}}^{\gamma 3\pi} 
\,.  \label{Cg3pi:SS}  
\\ 
3 F_\pi^2
\frac{g_{\omega \rho \pi} g_{\rho \pi\pi} g_{\omega}}{m_{\omega}^2m_{\rho}^2}
&=& 
D_{_{\rm SS}}^{\gamma 3\pi} 
\,, 
\end{eqnarray} 
and hence we arrive at the same result as that of our method (Eq.(\ref{g3pi:form2}) with Eqs.(\ref{Ag3pi})-(\ref{Dg3pi})):  
 \begin{eqnarray} 
&& 
\Gamma_\mu [\gamma^*_\mu (p), \pi^0(q_0), \pi^+(q_+), \pi^-(q_-)] \Bigg|_{\rm SS} 
\nonumber \\ 
&=& 
- \frac{e N_c}{12\pi^2 F_\pi^3} 
\epsilon_{\mu\nu\alpha \beta} q_0^\nu q_+^\alpha q_-^\beta  
\Bigg[
A_{_{\rm SS}}^{\gamma 3\pi} + B_{_{\rm SS}}^{\gamma 3\pi} \cdot D_\omega(p^2) 
+ 
\frac{C_{_{\rm SS}}^{\gamma 3 \pi}}{3}  \cdot 
\Bigg\{  
D_\rho((q_++q_-)^2) 
+ 
D_\rho((q_-+q_0)^2) 
+ 
D_\rho((q_0+q_+)^2) 
\Bigg\}  
\nonumber \\ 
&& 
+ \frac{D_{_{\rm SS}}^{\gamma 3\pi}}{3} \cdot 
 D_\omega(p^2) 
 \cdot 
 \Bigg\{ 
D_\rho((q_++q_-)^2) 
+ 
D_\rho((q_-+q_0)^2) 
+ 
D_\rho((q_0+q_+)^2) 
\Bigg\}  
\Bigg]
\,. 
\label{g3pi:SS:reduced}
\end{eqnarray} 
 Since the resultant form is equivalent to that obtained from our method  which is manifestly gauge invariant by 
construction (See Eqs.(\ref{Amu:integ:trans}),(\ref{D}) and (\ref{action:Amu:integrate})), 
the low-energy theorem  (\ref{LET:g3pi}) is actually satisfied: 
\begin{equation} 
A_{_{\rm SS}}^{\gamma 3\pi} +B_{_{\rm SS}}^{\gamma 3\pi} + C_{_{\rm SS}}^{\gamma 3\pi} + D_{_{\rm SS}}^{\gamma 3\pi} = 1
\,. \label{LET:g3pi2}
\end{equation}

In contrast, a naive truncation as in Eq.(\ref{action:naive}), 
which corresponds to taking $A_{_{\rm SS}}^{\gamma 3 \pi}=B_{_{\rm SS}}^{\gamma 3 \pi}= C_{_{\rm SS}}^{\gamma 3 \pi}=0$ in Eqs.(\ref{Ag3pi:SS})-(\ref{Cg3pi:SS}),  
would provide us with  
 \begin{eqnarray}  
&&\Gamma_{\mu} [\gamma^*_\mu (p), \pi^0(q_0), \pi^+(q_+), \pi^-(q_-)] \Bigg|^{\rm trun}_{\rm SS} 
\nonumber \\ 
&& 
=
- \frac{e N_c}{36\pi^2 F_\pi^3} 
\epsilon_{\mu\nu\alpha \beta} q_0^\nu q_+^\alpha q_-^\beta  
\, 
D_{_{\rm SS}}^{\gamma 3\pi} \, 
\frac{m_\omega^2}{m_\omega^2 - p^2} 
\left[ 
\frac{m_\rho^2}{m_\rho^2 - (q_++q_-)^2}
+ (q_+ \leftrightarrow q_0) 
+ (q_- \leftrightarrow q_0) 
\right] 
\,, 
\label{g3pi:form:truncate}
\end{eqnarray} 
with $D_{_{\rm SS}}^{\gamma 3 \pi} \simeq 1.5$ from Eq.(\ref{A-Dvals:g3pi}). 
At the low-energy limit $p^2, q_{\pm,0}^2 \to 0$, we have  
 \begin{eqnarray}  
\Gamma_{\mu} [\gamma^*_\mu (p), \pi^0(q_0), \pi^+(q_+), \pi^-(q_-)] \Bigg|^{\rm trun}_{\rm SS} 
&\stackrel{p^2, q_{\pm,0}^2 \to 0}{\longrightarrow} &
- \frac{e N_c}{12\pi^2 F_\pi^3} 
\epsilon_{\mu\nu\alpha \beta} q_0^\nu q_+^\alpha q_-^\beta  
( D_{_{\rm SS}}^{\gamma 3\pi} )  
\nonumber \\ 
&\neq& 
- \frac{e N_c}{12\pi^2 F_\pi^3} 
\epsilon_{\mu\nu\alpha \beta} q_0^\nu q_+^\alpha q_-^\beta  
\,, \label{g3pi:form:truncate:LE}
\end{eqnarray} 
which contradicts with the low-energy theorem (\ref{LET:g3pi}) 
and hence breaks the EM gauge symmetry.  
It should be noted again that the $\rho/\omega$ truncation (\ref{g3pi:form:truncate}) is different from 
the $\rho/\omega$ meson dominance (\ref{rho-omega-D:g3pi}) which is gauge invariant.

\end{document}